\documentclass[review]{elsarticle}

\usepackage{graphicx}
\usepackage{epstopdf}
\usepackage{dcolumn}
\usepackage{bm}
\usepackage{epsfig}
\usepackage{booktabs}
\usepackage{subfigure}
\usepackage{graphics}
\usepackage{amssymb}
\usepackage{amsmath}
\usepackage{array}
\usepackage{color}
\usepackage{natbib}
\usepackage{caption}
\usepackage{lineno,hyperref}

\begin{document}

\begin{frontmatter}

\title{Phase-field-based lattice Boltzmann model for immiscible incompressible N-phase flows}
\author[mymainaddress]{Xiaolei Yuan}
\author[mythirdaryaddress]{Hong Liang}
\author[mymainaddress,mysecondaryaddress]{Zhenhua Chai}
\author[mymainaddress,mysecondaryaddress]{Baochang Shi\corref{mycorrespondingauthor}}
\cortext[mycorrespondingauthor]{Corresponding author}
\ead{shibc@hust.edu.cn}

\address[mymainaddress]{School of Mathematics and Statistics, Huazhong University of Science and Technology, Wuhan 430074, China}
\address[mysecondaryaddress]{Hubei Key Laboratory of Engineering Modeling and Scientific Computing, Huazhong University of Science and Technology, Wuhan
    430074, China}
\address[mythirdaryaddress]{Department of Physics, Hangzhou Dianzi University, Hangzhou 310018, China}

\begin{abstract}
In this paper, we develop an efficient lattice Boltzmann (LB) model for simulating immiscible incompressible $N$-phase flows $(N \geq 2)$ based on the Cahn-Hilliard phase field theory. In order to facilitate the design of LB model and reduce the calculation of the gradient term, the governing equations of the $N$-phase system are reformulated, and they satisfy the conservation of mass, momentum and the second law of thermodynamics. In the present model, $(N-1)$ LB equations are employed to capture the interface, and another LB equation is used to solve the Navier-Stokes (N-S) equations, where a new distribution function for the total force is delicately designed to reduce the calculation of the gradient term. The developed model is first validated by two classical benchmark problems, including the tests of static droplets and the spreading of a liquid lens, the simulation results show that the current LB model is accurate and efficient for simulating incompressible $N$-phase fluid flows. To further demonstrate the capability of the LB model, two numerical simulations, including dynamics of droplet collision for four fluid phases and dynamics of droplets and interfaces for five fluid phases,
are performed to test the developed model. The results show that the present model can successfully handle complex interactions among $N$ ($N \geq 2$) immiscible incompressible flows.
\end{abstract}


\end{frontmatter}

\section{\label{sec:level1}Introduction}

$N$-phase flow problems ($N \geq 2$) are ubiquitous in nature and engineering, such as the emulsion and foam formation in the microfluidic devices, where one or more fluid phases are dispersed in another continuous phase \cite{Utada2005,Guzowski2015,Taylor1934,Dong2014}, the enhanced oil recovery, the geological CO$_{2}$ sequestration in depleted oil/gas reservoirs \cite{Dong2014,Boyer2010,Liang1}, as well as many daily phenomena such as rain drops, spraying of pesticides, bubbles in water, etc. In many cases, the number of fluid phases is greater than or equal to $3$, and $N$-phase flow problems are usually accompanied by the droplet/bubble generation, coalescence, and breakup. There are roughly three types of methods for studying these complicated $N$-phase flow problems, namely theoretical approach, experimental approach and numerical simulation. Due to limitations of theoretical and experimental methods, numerical simulation plays an important role in the study of $N$-phase flows. However, numerical simulation of such $N$-phase problems ($N \geq 3$) is still very challenging because of the inherent nonlinearities, complex topological changes and the complexity of moving interfaces. Due to the importance and complexity of the $N$-phase problem ($N \geq 3$), our main focus in the present paper will be on situations involving more than two fluid phases.

Compared to the numerous researches of two-phase fluid flows \cite{Hohenberg1977,Anderson1998,Lowengrub1998,Chen2002,Badalassi2003,Shen2013,Liang2019}, there are fewer theoretical and numerical studies on fluid flows that involves three or more phases. Nonetheless, some efforts have been made to develop efficient numerical methods for dealing with $N$-phase flows, including the volume of fluid (VOF) method \cite{Hirt1981,Gueyffier1999}, level set method \cite{Osher1988}, front tracking method \cite{Unverdi1992}, smoothed particle hydrodynamics method \cite{Gingold1977,Lucy1977}, diffuse interface method (phase field method) \cite{Rayleigh1892,Garcke1999,Kim2004,Boyer2006,Boyer2010,Kim2007}, and lattice Boltzmann (LB) method \cite{Lamura1999,Li2007,Chen2000,Nekovee2000,Halliday2007,Leclaire2013,Wohrwag2018,Yu2019,Liang1,Shi}. In this work, we will focus on the phase-field-based LB method for immiscible incompressible $N$-phase flows ($N \geq 3$).

In the phase field method, the thickness of the interface between the two immiscible flows is supposed to be very small but nonzero. Then different phases can be characterized by phase variables (order parameters) which vary continuously across thin interfacial layers. One can derive a set of governing equations for the whole computational domain from the free energy, where the order parameters satisfy an advection-diffusion equation (usually the Cahn-Hilliard equation), and this equation is coupled with the Navier-Stokes equations. With the Cahn-Hilliard phase field approach, Boyer \emph{et al.} \cite{Boyer2006} proposed a diffuse interface model for the study of three immiscible component incompressible viscous flows. The effects of different forms for the bulk free energy have been investigated. Kim \emph{et al.} \cite{Kim2007} proposed a phase field model for the immiscible incompressible ternary fluid flows with interfaces. In these models, the pairwise surface tensions $\sigma_{ij}$ are decomposed into three positive phase-specific surface-tension coefficients. However, this decomposition will encounter some difficulties when $N \geq 4$, as the number of pairwise surface tensions $\frac{N(N-1)}{2}$ would be greater than the number of mixing energy density coefficients $N$, which leads to an overdetermined system \cite{Kim2009}. To overcome this shortcomings, Kim \emph{et al.} \cite{Kim2009} proposed the generalized continuous surface tension force model for multi-component fluid flows. Boyer and his collaborator \cite{Boyer2014} have proposed a generalized model for the $N$-phase mixtures using the consistency principle which can be described as that the $N$-phase model exactly coincide with the classical two-phase model when only two phases are present in the system. Recently, Dong \cite{Dong2014} have derived a physical formulation and a numerical algorithm for the mixture of N ($N \geq 2$) immiscible incompressible fluids in the thermodynamics framework. In his model, the mixture velocity is the volume-averaged velocity, and the mixing energy density coefficients involved in the $N$-phase model can be determined based on the pairwise surface tensions among the N fluids.


As a mesoscopic level method, the LB method has made rapid progress since its appearance in the late 1980s due to its simplicity, scalability on parallel computers, and ease to handle complex geometries \cite{Kruger,Chai0}. Based on different physical perspectives, the existing LB models for multiphase flows can be generally classified into four categories: color-gradient model \cite{Gunstensen1991}, pseudo-potential model \cite{Shan1994}, free-energy model \cite{Swift1995}, and phase-field-based model \cite{He0,Wang2015,Liang,Fakhari2017,Chai3}. These methods have gained great success in the study of two-phase flow problems, and they have been extended to simulate immiscible ternary fluids. For instance, Lamura \emph{et al.} \cite{Lamura1999} proposed a LB model to simulate oil-water-surfactant mixtures based on a Ginzburg-Landau free energy. However, this model can only be used to simulate ternary flows where an amphiphile phase is located at oil-water interface, and cannot be extended to arbitrary ternary flows. Chen \emph{et al.} \cite{Chen2000} presented a LB model for amphiphilic ternary fluid dynamics, which can be regarded as an extension of the pseudo-potential model. His model suffers from the inherent defects in the pseudo-potential model, such as high spurious velocities and lack of flexibility in adjusting surface tension \cite{Yu2019}. Leclaire \emph{et al.} \cite{Leclaire2013} developed a LB model for the simulation of multiple immiscible fluids based on the improved color-gradient model, where three subcollision operators are also applied. We note that thermodynamic consistency, which is distinctly important in the physical formulation, are lack in the above three LB models. Recently, an alternative LB ternary model was proposed by Liang \emph{et al.} \cite{Liang1} based on the Cahn-Hilliard phase field theory, which provides a firm physical foundation on the dynamics of the interfaces among three fluids. Shi \emph{et al.} \cite{Shi} extended the LB flux solver which was originally proposed for simulating incompressible flows of binary fluids based on two-component Cahn-Hilliard model to three-component fluid flows, while their models are limited to fluids with a very small density difference (no more than $1.25$). Although Shi \emph{et al.} \cite{Shi} pointed out that the model can be extended to $N$-phase $(N \geq 4)$ fluid flows by introducing the free energy $W(\phi,\nabla \phi)=\int_{\Omega}\left[0.25\sum_{i=1}^N \phi_i^2(1-\phi_i^2)+\frac{\epsilon^2}{2}\sum_{i=1}^N |\nabla \phi_i|^2 \right]\, d \mathbf{x}$, where $\phi_i$ is the order parameter and $\epsilon$ is a small parameter denoting the interface thickness, we note that this free energy expression is unreasonable as it does not involve the interaction between different phases. In fact, this interaction is related to the multiple pairwise surface tensions. Besides, the author does not give a numerical example when $N \geq 4$. Therefore, all the mentioned LB models are limited to the case of $N \leq 3$.

To our best knowledge, there is still no available work on the construction of LB model for fluid flows with $N \geq 4$. In this work, we develop an efficient LB model for simulating immiscible incompressible $N$-phase flows $(N \geq 2)$ based on the Cahn-Hilliard phase field theory. The proposed model has several distinct features. Firstly, the governing equations of the $N$-phase system proposed by Dong \cite{Dong2014} are reformulated, and they also conserve the mass, momentum and second law of thermodynamics. In such a sense, the reformulated system is thermodynamically consistent. Secondly, the velocity in this model is the volume-averaged mixture velocity and is divergence free. Thirdly, the mixing energy density coefficients involve the interaction between different phases and can be determined by the pairwise surface tensions among the N fluids. Fourthly, in the present model, $(N-1)$ LB equations are employed to capture the interface, and another LB equation is used to solve the N-S equations. Fifthly, we introduce a new distribution function for the total force to reduce the calculation of the gradient term, and the governing equations can be recovered correctly from the present LB model through Chapmann-Enskog analysis.

The rest of present paper is organized as follows. In Sec. \uppercase\expandafter{\romannumeral2}, we describe the physical formulation of $N$-phase flows, and a phase-field-based LB model for immiscible incompressible $N$-phase flows is given in Sec. \uppercase\expandafter{\romannumeral3}. In Sec. \uppercase\expandafter{\romannumeral4}, we use two classic numerical examples to validate our model. It is found that the numerical results agree well with analytical solutions. In Sec. \uppercase\expandafter{\romannumeral5}, some numerical applications for $N$-phase fluid flows ($N \geq 4$) are conducted. Finally, some conclusions are given in Sec. \uppercase\expandafter{\romannumeral6}.

\section{Physical formulation} \label{sec:method}

In this section, we will describe the physical formulation of immiscible incompressible $N$-phase flows ($N\geq 2$) based on the
mass conservation, momentum conservation, the second law of thermodynamics, and Galilean invariance. We refer to \cite{Dong2014}
for detailed derivations of this system.

Consider an incompressible system consisting of N immiscible Newtonian fluids. Let $\Omega$ denote the flow domain, $\tilde{\rho}_i$ ($1\leq i \leq N$) represent constant densities of these $N$ pure fluids, $V$ is an arbitrary control volume with the mass $M$, $V_i$ is the volume of fluid $i$ before mixing with the mass $M_i$, and $\tilde{\mu}_i$ expresses their constant dynamic viscosities. Here we assume that there is no volume loss or increase during the $N$-phase mixing process, which leads to
\begin{equation}
V=V_1+V_2+ \cdots+V_N.
\end{equation}
For convenience, we introduce the following auxiliary parameters
\begin{equation}
\tilde{\gamma}_i=\frac{1}{\tilde{\rho}_i}, 1\leq i \leq N;\quad
\Gamma=\sum_{i=1}^{N} \tilde{\gamma}_i; \label{eq:N1}
\end{equation}
$\phi_i$ ($1\leq i \leq N-1$) are employed to denote ($N-1$)
independent order parameters, and $-1\leq \phi_i \leq 1$. $\phi=(\phi_1, \phi_2, \cdots, \phi_{N-1})$ represents the vector of $(N-1)$ phase field functions, $c_i$ and $\rho_i$ ($1\leq i \leq N$) are used to represent the volume fraction and density of fluid $i$ within the mixture, and $\rho$ is introduced to denote the average density of the mixture. Then they satisfy the following relations
\cite{Dong2014,Dong2015,Dong2017}
\begin{equation}
c_i=\frac{V_i}{V}=\frac{M_i/{\tilde{\rho}_i}}{M_i/\rho_i}=\frac{\rho_i}{\tilde{\rho}_i}=\tilde{\gamma}_i \rho_i, 1 \leq
i\leq N; \quad \sum_{i=1}^{N}c_i=\frac{\sum_{i=1}^N V_i}{V}=1; \quad \rho=\sum_{i=1}^{N}\rho_i.
\label{eq:N2}
\end{equation}

The mass conservation of $N$-phase flows can be given by the following ($N-1$) mass balance equations \cite{Dong2014,Dong2015,Dong2017}
\begin{equation}
\frac{\partial \varphi_i(\phi)}{\partial t}+\mathbf{u}\cdot \nabla
\varphi_i(\phi)=-\nabla \cdot \mathbf{J}_{ai}, \quad 1 \leq i\leq N-1,
\label{eq:N3}
\end{equation}
where $\varphi_i(\phi)$ is defined as $\varphi_i \equiv \rho_i-\rho_N$ ($1 \leq i\leq N-1$), and $\mathbf{u}$ is the volume averaged mixture velocity and can be rigorously proved that it is divergence free
\cite{Dong2014}
\begin{equation}
\nabla \cdot \mathbf{u}=0. \label{eq:N4}
\end{equation}
$\mathbf{J}_{ai}$ is the relative differential flux between fluids $i$ and $N$, and it can be written as
\begin{equation}
\mathbf{J}_{ai}=-\tilde{m}_i(\phi) \nabla C_i, \quad 1 \leq i\leq
N-1, \label{eq:N5}
\end{equation}
based on the second law of thermodynamics, where $\tilde{m}_i(\phi) \geq 0$ is the mobility. $C_i$ ($\leq i\leq N-1$) are ($N-1$) effective chemical potentials determined by the linear system
\begin{equation}
\sum_{j=1}^{N-1}\frac{\partial \varphi_j}{\partial
\phi_i}C_j=\frac{\partial W(\phi, \nabla \phi)}{\partial \phi_i}-\nabla \cdot \frac
{\partial W(\phi, \nabla \phi)}{\partial({\nabla \phi_i})},\quad 1 \leq i\leq N-1,
\label{eq:N6}
\end{equation}
where $W(\phi, \nabla \phi)$ represents the free energy density function.

The momentum conservation can be described by \cite{Dong2014}
\begin{equation}
\rho(\frac{\partial \mathbf{u}}{\partial t}+\mathbf{u}\cdot \nabla
\mathbf{u})+\tilde{\mathbf{J}}\cdot \nabla \mathbf{u}=-\nabla
p+\nabla \cdot \mathbf{S}, \label{eq:N7}
\end{equation}
where $\tilde{\mathbf{J}}$ is the flux and satisfies
\begin{equation}
\tilde{\mathbf{J}}=\sum_{i=1}^{N-1}\left(
1-\frac{N}{\Gamma}\tilde{\gamma}_i\right)\mathbf{J}_{ai}.
\label{eq:N8}
\end{equation}
$p$ is the pressure, and $\mathbf{S}$ represents a trace-free stress tensor which is determined as
\begin{equation}
\mathbf{S}=\mu(\phi)\mathbf{D}(\mathbf{u})-\sum_{i=1}^{N-1}\nabla
\phi_i \otimes \frac{\partial W(\phi, \nabla \phi)}{\partial \nabla
\phi_i},
 \label{eq:N9}
\end{equation}
based on the second law of thermodynamics, where
$\mathbf{D(u)}=\nabla \mathbf{u}+\nabla \mathbf{u}^{T}$, $\mu(\phi)=\sum_{k=1}^N \tilde{\mu}_k c_k(\phi)$ is the viscosity, $\otimes$ denotes the tensor product.

With these constitutive relations, we obtain the following phase field system for $N$-phase flows \cite{Dong2014}
\begin{subequations}
\begin{align}
&\nabla \cdot \mathbf{u}=0,\label{eq:N10a} \\
&\rho\left(\frac{\partial \mathbf{u}}{\partial t}+\mathbf{u}\cdot
\nabla \mathbf{u}\right)+\tilde{\mathbf{J}}\cdot \nabla
\mathbf{u}=-\nabla p+\nabla \cdot
[\mu(\phi)\mathbf{D(u)}]-\sum_{i=1}^{N-1}\nabla \cdot \left(\nabla
\phi_i \otimes \frac{\partial W}{\partial \nabla \phi_i}\right),\label{eq:N10b} \\
&\sum_{j=1}^{N-1}\frac{\partial \varphi_i}{\partial \phi_j}\left(
\frac{\partial \phi_j}{\partial t}+\mathbf{u} \cdot \nabla
\phi_j\right)=\nabla \cdot[\tilde{m}_i(\phi)\nabla C_i], \quad 1
\leq i\leq N-1,\label{eq:N10c}
\end{align}
\label{eq:N10}
\end{subequations}
based on the mass conservation [Eq. (\ref{eq:N3})] and momentum conservation [Eq. (\ref{eq:N7})]. Once the free energy density $W(\phi,\nabla \phi)$ and the functions $\varphi_i(\phi)$ ($1 \leq i\leq N-1$) are specified, we can compute all the other quantities in this model. Following \cite{Dong2014}, we chose $W(\phi,\nabla \phi)$ and $\varphi_i(\phi)$ as
\begin{equation}
\begin{split}
W(\phi,\nabla \phi)&=\sum_{i,j}^{N-1}\frac{\lambda_{ij}}{2}\nabla \phi_i\cdot \nabla\phi_j+\frac{\beta^2}{\eta^2}H(\phi) \\
&=\sum_{i,j}^{N-1}\frac{\lambda_{ij}}{2}\nabla \phi_i\cdot \nabla \phi_j+\frac{\beta^2}{\eta^2}\sum_{k=1}^{N}c_k^2(1-c_k)^2,
\label{eq:N11}
\end{split}
\end{equation}
\begin{equation}
\varphi_i(\phi)=\rho_i-\rho_N=\frac{1}{2}(\tilde{\rho}_i-\tilde{\rho}_N)+\frac{1}{2}(\tilde{\rho}_i+\tilde{\rho}_N)\phi_i,\quad 1 \leq i\leq N-1,
\label{eq:N12}
\end{equation}
where the constant $\beta^2$ represents the characteristic energy scale, and the constant $\eta>0$ denotes the characteristic scale of the interfacial thickness. The constants $\lambda_{ij}$ ($1\leq i,j\leq N-1$) denote the mixing energy density coefficients, and they can be determined by the pairwise surface tensions $\sigma_{kl}$ ($1\leq k<l\leq N$),
\begin{equation}
\sum_{i=1}^{N-1}(L_i^{kl})^2 \lambda_{ii}+\sum_{j=1}^{N-1}\sum_{i=1}^{j-1}2L_i^{kl}L_j^{kl}\lambda_{ij}=\frac{9}{2}\frac{\eta^2}{\beta^2}\sigma_{kl}^2,\quad 1\leq k<l\leq N,
\label{eq:N14}
\end{equation}
where $\lambda_{ij}$ should be symmetry, and $L_i^{kl}$ ($1\leq i,j\leq N-1$) are determined as
\begin{equation}
L_i^{kN}=\left\{
\begin{array}
{l}1,\quad \quad \, if \ i=k, \\
\frac{\tilde{\rho}_N}{\tilde{\rho}_i+\tilde{\rho}_N}, if \ i \neq k.
\end{array}
\right.
1 \leq k<l=N,
\label{eq:N15}
\end{equation}
or
\begin{equation}
L_i^{kl}=\left\{
\begin{array}{l}
\frac{\tilde{\rho}_k}{\tilde{\rho}_k+\tilde{\rho}_N},\ \ \quad \, if \ i=k, \\
-\frac{\tilde{\rho}_l}{\tilde{\rho}_l+\tilde{\rho}_N}, \, \quad if \ i=l, \\
0, \qquad \qquad otherwise,
\end{array}
\right.
1 \leq k<l<N.
\label{eq:N16}
\end{equation}
With the expressions (\ref{eq:N11}) and (\ref{eq:N12}), Eqs. (\ref{eq:N10a})-(\ref{eq:N10c}) will reduce to
\begin{subequations}
\begin{align}
&\nabla \cdot \mathbf{u}=0,\label{eq:N13a} \\
&\rho\left(\frac{\partial \mathbf{u}}{\partial t}+\mathbf{u}\cdot
\nabla \mathbf{u}\right)+\tilde{\mathbf{J}}\cdot \nabla
\mathbf{u}=-\nabla p+\nabla \cdot
[\mu(\phi)\mathbf{D(u)}]-\sum_{i,j=1}^{N-1}\nabla \cdot \left(\lambda_{ij}\nabla\phi_i\nabla\phi_j\right)+\mathbf{G}(\mathbf{x},t),\label{eq:N13b} \\
&\frac{\partial\phi_i}{\partial t}+\mathbf{u}\cdot \nabla \phi_i=m_i \nabla^2 C_i, \quad 1
\leq i\leq N-1,\label{eq:N13c}
\end{align}
\label{eq:N13}
\end{subequations}
where we have taken into account an external body force $\mathbf{G}(\mathbf{x},t)$ in the momentum equation, and the constant $m_i$ ($m_i>0$) is the mobility.

For the convenience of calculation, the parameters $\rho_i$, $c_i$, $C_i$, $\tilde{\mathbf{J}}$, $\rho$ can be rewritten as
\begin{equation}
\left\{
\begin{array}{l}
\rho_i=\frac{1}{\Gamma}+\sum_{j=1}^{N-1} \left(\delta_{ij}-\frac{ \tilde{\gamma}_j}{\Gamma} \right) \left[\frac{1}{2}(\tilde{\rho}_j-\tilde{\rho}_N)+\frac{1}{2}(\tilde{\rho}_j+\tilde{\rho}_N)\phi_j \right],\quad 1 \leq i \leq N,\\
c_i(\phi)=\tilde{\gamma}_i \rho_i=\frac{\tilde{\gamma}_i}{\Gamma}+\sum_{j=1}^{N-1} \left(\tilde{\gamma}_i \delta_{ij}-\frac{\tilde{\gamma}_i \tilde{\gamma}_j}{\Gamma} \right) \left[\frac{1}{2}(\tilde{\rho}_j-\tilde{\rho}_N)+\frac{1}{2}(\tilde{\rho}_j+\tilde{\rho}_N)\phi_j \right],\quad 1 \leq i \leq N,\\
C_i(\phi)=\frac{\delta W}{\delta \phi_i}=-\sum_{j=1}^{N-1} \lambda_{ij} \nabla^2 \phi_j+\frac{\beta^2}{\eta^2}h_i(\phi), \quad 1 \leq i \leq N-1,\\
\tilde{\mathbf{J}}=-\sum_{i=1}^{N-1}\left(1-\frac{N}{\Gamma} \tilde{\gamma}_i \right) \frac{\tilde{\rho}_i+\tilde{\rho}_N}{2} m_i \nabla \left[-\sum_{j=1}^{N-1}\lambda_{ij}\nabla^{2}\phi_j+\frac{\beta^2}{\eta^2}h_i(\phi) \right], \\
\rho=\sum_{k=1}^N \rho_k=\frac{N}{\Gamma}+\sum_{i=1}^{N-1}\left(1-\frac{N}{\Gamma}\tilde{\gamma}_i \right)\left[\frac{1}{2}(\tilde{\rho}_i-\tilde{\rho}_N)+\frac{1}{2}(\tilde{\rho}_i+\tilde{\rho}_N)\phi_j \right],
\end{array}
\right.
\label{eq:N18}
\end{equation}
where $\delta_{ij}$ is the Kronecker delta function, and $h_i(\phi)$ is given as
\begin{equation}
h_i({\phi})=\frac{\partial H}{\partial \phi_i}=\frac{\tilde{\rho}_i+\tilde{\rho}_N}{2}\sum_{k=1}^{N} \left( \tilde{\gamma}_k \delta_{ki}-\frac{\tilde{\gamma}_k \tilde{\gamma}_i}{\Gamma} \right) c_k(1-c_k)(1-2c_k), \quad 1 \leq i \leq N-1.
\label{eq:N17}
\end{equation}

Note that the $N$-phase system consisting of Eqs. (\ref{eq:N13a})-(\ref{eq:N13c}) is complicated due to excessive gradient calculation in momentum equation, and it is not easy to solve this system directly using LB method. In order to facilitate the design of an efficient LB model, we introduce the following system of equations
\begin{subequations}
\begin{align}
&\nabla \cdot \mathbf{u}=0,\label{eq:N19a} \\
&\frac{\partial (\rho \mathbf{u})}{\partial t}+\nabla \cdot (\rho \mathbf{uu})=-\nabla p+\nabla \cdot
[\mu(\phi)\mathbf{D(u)}]-\nabla \cdot (\mathbf{\tilde{J}u})+\mathbf{F}_s+\mathbf{G},\label{eq:N19b} \\
&\frac{\partial\phi_i}{\partial t}+\nabla \cdot (\phi_i \mathbf{u})=m_i \nabla^2 C_i, \quad 1
\leq i\leq N-1,\label{eq:N19c}
\end{align}
\label{eq:N19}
\end{subequations}
where $\mathbf{F}_s$ is the surface tension which can be given as $\mathbf{F}_s=\sum_{i=1}^{N-1}C_i \nabla\phi_i$. Next, we will show that Eq. (\ref{eq:N13b}) and Eq.(\ref{eq:N19b}) are equivalent.

Consider the term $-\sum_{i,j=1}^{N-1}\nabla \cdot \left(\lambda_{ij}\nabla\phi_i\nabla\phi_j\right)$. For $1 \leq i=j \leq N-1$,
\begin{equation}
\begin{split}
-\lambda_{ii}\nabla \cdot (\nabla\phi_i\nabla\phi_i)&=-\lambda_{ii}\nabla_{\beta}(\nabla_{\alpha}\phi_i \nabla_{\beta}\phi_i)\\
&=-\lambda_{ii}[\nabla_{\beta}(\nabla_{\alpha}\phi_i) \nabla_{\beta}\phi_i+\nabla_{\alpha}\phi_i \nabla_{\beta}^2 \phi_i]\\
&=-\lambda_{ii}[\nabla_{\alpha}(\frac{1}{2}\nabla_{\beta}\phi_i \nabla_{\beta}\phi_i)+\nabla_{\alpha}\phi_i \nabla_{\beta}^2 \phi_i]\\
&=-\lambda_{ii}[\nabla(\frac{1}{2} \left|\nabla\phi_i \right|^2)+\nabla\phi_i \nabla^2 \phi_i],
\end{split}
\label{eq:N20}
\end{equation}
where the first term $-\lambda_{ii}\nabla(\frac{1}{2} \left|\nabla\phi_i \right|^2)$ can be absorbed into the pressure gradient term in momentum equation.
For $1 \leq i<j \leq N-1$, we have
\begin{equation}
\begin{split}
&-[\lambda_{ij} \nabla\cdot(\nabla \phi_i\nabla \phi_j)+\lambda_{ji} \nabla\cdot(\nabla \phi_j\nabla \phi_i)]\\
=&-[\lambda_{ij} \nabla_{\beta}(\nabla_{\alpha} \phi_i\nabla_{\beta} \phi_j)+\lambda_{ji} \nabla_{\beta}(\nabla_{\alpha} \phi_j\nabla_{\beta} \phi_i)]\\
=&-\lambda_{ij} [\nabla_{\beta}(\nabla_{\alpha} \phi_i)\nabla_{\beta} \phi_j+\nabla_{\alpha}\phi_i\nabla_{\beta}^2 \phi_j+\nabla_{\beta}(\nabla_{\alpha} \phi_j)\nabla_{\beta} \phi_i+\nabla_{\alpha}\phi_j\nabla_{\beta}^2 \phi_i]\\
=&-\lambda_{ij} \{ \nabla_{\alpha}[(\nabla_{\beta}\phi_i)(\nabla_{\beta}\phi_j)]+\nabla_{\alpha}\phi_i\nabla_{\beta}^2 \phi_j+\nabla_{\alpha}\phi_j\nabla_{\beta}^2 \phi_i \}\\
=&-\lambda_{ij} \nabla[(\nabla\phi_i)\cdot(\nabla\phi_j)]-\lambda_{ij}[\nabla\phi_i \nabla^2 \phi_j+\nabla\phi_j \nabla^2 \phi_i],
\end{split}
\label{eq:N21}
\end{equation}
where the relation $\lambda_{ij}=\lambda_{ji}$ has been used, and the first term $-\lambda_{ij} \nabla[(\nabla\phi_i)\cdot(\nabla\phi_j)]$ can also be absorbed into the pressure gradient term in momentum equation.

Note that the surface tension $\mathbf{F}_s$ can be transformed as follows
\begin{equation}
\begin{split}
\mathbf{F}_s &=\sum_{i=1}^{N-1}C_i \nabla\phi_i \\
&=\sum_{i=1}^{N-1} \left[ -\sum_{j=1}^{N-1} \lambda_{ij} \nabla^2 \phi_j+\frac{\beta^2}{\eta^2}h_i(\phi) \right] \nabla\phi_i \\
&=-\sum_{i,j=1}^{N-1}\lambda_{ij} \nabla^2 \phi_j \nabla\phi_i-\sum_{i=1}^{N-1}\frac{\beta^2}{\eta^2}h_i(\phi) \nabla\phi_i \\
&=-\sum_{i,j=1}^{N-1}\lambda_{ij} \nabla^2 \phi_j \nabla\phi_i-\sum_{i=1}^{N-1}\frac{\beta^2}{\eta^2}\frac{\partial H}{\partial \phi_i}\nabla\phi_i\\
&=-\sum_{i,j=1}^{N-1}\lambda_{ij} \nabla^2 \phi_j \nabla\phi_i-\sum_{i=1}^{N-1}\frac{\beta^2}{\eta^2}\nabla H,
\end{split}
\label{eq:N22}
\end{equation}
where the last term $-\sum_{i=1}^{N-1}\frac{\beta^2}{\eta^2}\nabla H$ can also be absorbed into the pressure gradient term. In contrast to Eqs. (\ref{eq:N20})-(\ref{eq:N22}), we conclude that $-\sum_{i,j=1}^{N-1}\nabla \cdot \left(\lambda_{ij}\nabla\phi_i\nabla\phi_j\right)$ and $\mathbf{F}_s$ are equivalent except for the terms which are absorbed into $-\nabla p$. In addition, with the relation of $\tilde{\mathbf{J}}\cdot \nabla \mathbf{u}=\nabla \cdot (\tilde{\mathbf{J}}\mathbf{u})-(\nabla \cdot \tilde{\mathbf{J}})\mathbf{u}$ and the mass conservation equation $\frac{\partial \rho}{\partial t}+\mathbf{u}\cdot \nabla \rho=-\nabla \cdot \tilde{\mathbf{J}}$ \cite{Dong2014}, Eq. (\ref{eq:N13b}) can be converted to Eq. (\ref{eq:N19b}).

Thus, we have proved that the two immiscible $N$-phase systems [Eqs. (\ref{eq:N13a})-(\ref{eq:N13c}) and Eqs. (\ref{eq:N19a})-(\ref{eq:N19c})] are equivalent. But Eqs. (\ref{eq:N19a})-(\ref{eq:N19c}) are much simpler from the perspective of LB model construction. Since the chemical potential $C_i$ is a quantity that must be calculated, we only need to calculate gradient terms for $(N-1)$ times when calculating the surface tension $\mathbf{F}_s$, which is much simpler than calculating the term $-\sum_{i,j=1}^{N-1}\nabla \cdot (\lambda_{ij} \nabla\phi_i \nabla \phi_j)$, especially when $N$ is relatively large. Simultaneously, the reformulated $N$-phase system Eqs. (\ref{eq:N19a})-(\ref{eq:N19c}) can also satisfy the conservations of mass, momentum and the second law of thermodynamics. Therefore, next we will construct the LB model based on the system of Eqs. (\ref{eq:N19a})-(\ref{eq:N19c}).

\section{LB model for immiscible incompressible N-phase flows}

In this section, we will present a LB model for simulating the incompressible $N$-phase flows. The governing equations Eqs. (\ref{eq:N19a})-(\ref{eq:N19c}) can be regarded as the coupling between the incompressible N-S equations and the $(N-1)$ C-H equations. Next, we will take two sets of LB models to solve the N-S equations and the C-H equations separately.

\subsection{LB model for the Navier-Stokes equations}
To obtain the evolution equation, we integrate the following discrete velocity Boltzmann equation
\begin{equation}
\frac{\partial f_k}{\partial t}+\mathbf{c}_k \cdot \nabla
f_k=\Omega_k+G_k \label{eq:20}
\end{equation}
along a characteristic line $\mathbf{c}_k$ over a time interval $\Delta t$ \cite{Luo,Du}, then we get
\begin{equation}
f_k(\textbf{x}+\textbf{c}_k \Delta t,t+\Delta t)-f_k(\textbf{x},t) =
\int_0^{\Delta t} \Omega_k(\textbf{x}+\textbf{c}_k t',t+t') d
t'+\int_0^{\Delta t} G_k(\textbf{x}+\textbf{c}_k  t',t+ t') d t'.
\label{eq:21}
\end{equation}
where $f_k(\mathbf{x},t)$ denotes particle distribution function for fluids $i$ with velocity $\mathbf{c}_k$ at position $\mathbf{x}$ and time $t$, $G_k(\mathbf{x},t)$ represents the force term, $\Omega_k(\mathbf{x},t)$ is the collision operator which can be approximated by
\begin{equation}
\Omega_k=-\frac{1}{\tau'}(f_k-f_k^{eq}),
\end{equation}
where $\tau'$ denotes the relaxation time and $f_k^{eq}(\mathbf{x},t)$ represents the equilibrium distribution function.

The trapezoidal rule is used to integrate the collision term, then Eq. (\ref{eq:21}) becomes
\begin{equation}
f_k(\textbf{x}+\textbf{c}_k \Delta t,t+\Delta t)-f_k(\textbf{x},t)
=\frac{\Delta t}{2} \left[\Omega_k(\textbf{x}+\textbf{c}_k \Delta
t,t+\Delta t)+\Omega_k(\textbf{x},t)\right] +\int_0^{\Delta t}
G_k(\textbf{x}+\textbf{c}_k t',t+t') d t'. \label{eq:21b}
\end{equation}

Let $\bar{f}_k=f_k-\frac{\Delta t}{2}\Omega_k$, we can transform Eq. (\ref{eq:21b}) into
\begin{equation}
\bar{f}_k(\textbf{x}+\textbf{c}_k \Delta t,t+\Delta
t)-\bar{f}_k(\textbf{x},t) =-\frac{1}{\tau_g}\left[
\bar{f}_k(\mathbf{x},t)-f_k^{eq}(\mathbf{x},t)\right]
+\int_0^{\Delta t} G_k(\textbf{x}+\textbf{c}_k t',t+t') d t',
\label{eq:21bb}
\end{equation}
where $\tau_g=\frac{2\tau'+\Delta t}{2\Delta t}$ denotes the
dimensionless relaxation time, and $\bar f_k$ satisfies $\sum_k \bar
f_k=\sum_k f_k$, and $\sum_k \mathbf{c}_k \bar f_k=\sum_k
\mathbf{c}_k f_k$.

Through Taylor expansion of $G_k(\textbf{x}+\textbf{c}_k t',t+t')$ and neglecting the terms of order
$O(\Delta t^2)$, the last term in the right hand of Eq.
(\ref{eq:21bb}) can be transformed into \cite{Du}
\begin{equation}
\begin{split}
\int_0^{\Delta t} G_k(\textbf{x}+\textbf{c}_k \Delta t',t+\Delta t')
d t' &=\int_0^{\Delta t} \left(G_k(\mathbf{x},t)+t' D_k
G_k(\mathbf{x},t)\right)dt' \\
&=\Delta t G_k(\mathbf{x},t)+\frac{\Delta t^2}{2}D_k
G_k(\mathbf{x},t),
\end{split}
\label{eq:22}
\end{equation}
where $D_k=\partial t+\mathbf{c}_k \cdot \nabla$.

The LB evolution equation with the BGK collision operator for the
N-S equations can be written as
\begin{equation}
\bar{f}_k(\textbf{x}+\textbf{c}_k \Delta t,t+\Delta
t)-\bar{f}_k(\textbf{x},t) =-\frac{1}{\tau_g}\left[
\bar{f}_k(\mathbf{x},t)-f_k^{eq}(\mathbf{x},t)\right] +\Delta t
\left[G_k(\mathbf{x},t)+\frac{\Delta t}{2}D_k G_k(\mathbf{x},t) \right].
\label{eq:23}
\end{equation}
For the term $D_k G_k(\mathbf{x},t)$ in Eq. (\ref{eq:23}), one can take different discretization schemes to deal with it. When the up-wind scheme is used, the evolution equation becomes
\begin{equation}
\begin{split}
\bar{f}_k(\textbf{x}+\textbf{c}_k \Delta t,t+\Delta
t)-\bar{f}_k(\textbf{x},t) =&-\frac{1}{\tau_g}\left[
\bar{f}_k(\mathbf{x},t)-f_k^{eq}(\mathbf{x},t)\right]+\Delta t
\left[G_k(\mathbf{x},t)\right.\\
& \left.+\frac{G_k(\textbf{x}+\textbf{c}_k \Delta t,t+\Delta t)-G_k(\textbf{x},t)}{2} \right].
\end{split}
\end{equation}

To remove the implicitness, we introduce a modified particle distribution function,
\begin{equation}
g_k(\mathbf{x},t)=\bar f_k(\mathbf{x},t)-\frac{\Delta t}{2}G_k(\mathbf{x},t).
\label{eq:23a}
\end{equation}
With some simple operations, the explicit evolution equation can be derived,
\begin{equation}
g_k(\textbf{x}+\textbf{c}_k \Delta t,t+\Delta t)-g_k(\textbf{x},t)
=-\frac{1}{\tau_g}
\left(g_k(\mathbf{x},t)-g_k^{eq}(\mathbf{x},t)\right)+\Delta
t(1-\frac{1}{2\tau_g})G_k, \label{eq:24}
\end{equation}
where $g_k^{eq}$ denotes the new equilibrium distribution function which satisfies $g_k^{eq}=f_k^{eq}$.

To recover the N-S equations (\ref{eq:N19a})-(\ref{eq:N19b}) through Chapman-Enskog analysis, the equilibrium distribution function $g_k^{eq}$ is delicately designed as
\begin{equation}
g_k^{eq}=\left\{
\begin{array}{l}
\rho_0+\frac{p}{c_s^2}(\omega_k-1)+{\rho} s_k(\mathbf{u}), \;\;\;k=0\\
\frac{p}{c_s^2}\omega_k+{\rho} s_k(\mathbf{u}),
\;\;\;\;\;\;\;\;\;\;\;\;   k \ne 0,
\end{array}
\right. \label{eq:25}
\end{equation}
with
\begin{equation}
s_k({\mathbf{u}})=\omega_k \left[ \frac{\mathbf{c}_k \cdot
\mathbf{u}}{c_s^2}+\frac{(\mathbf{c}_k \cdot
\mathbf{u})^2}{2c_s^4}-\frac{\mathbf{u} \cdot \mathbf{u}}{2c_s^2}
\right],
\end{equation}
where $\rho_0$ is a constant, $\omega_k$ and $\mathbf{c}_k$ denote the weighting coefficient and the discrete velocity, and $c_s$ represents the speed of sound. We would like to point that the present model is based on the D$d$Q$q$ lattice with $q$ velocity directions in $d$-dimensional space. The values of $\mathbf{c}_k$ and $\omega_k$ depend on the lattice model used. In D$1$Q$3$ model,
$\{\mathbf{c}_k\}=c(0,1,-1)$, $\omega_0=2/3$, $\omega_{1,2}=1/6$,
$c_s=c/\sqrt{3}$, where $c=\Delta x / \Delta t$ , with $\Delta x$
and $\Delta t$ representing the spacing and time step, respectively;
In the D$2$Q$9$ model, $\omega_k$ is chosen as $\omega_0=4/9$,
$\omega_{1-4}=1/9$, $\omega_{5-8}=1/36$, $c_s=c/\sqrt{3}$, and
$\mathbf{c}_k$ is defined as
$$\{\mathbf{c}_k\}=c\left( \begin{array}{rrrrrrrrr}
0&1&0&-1&0&1&-1&-1&1\\
0&0&1&0&-1&1&1&-1&-1 \end{array} \right);$$ In the D$3$Q$15$ model,
 $\omega_0=2/9$, $\omega_{1-6}=1/9$,
$\omega_{7-14}=1/72$, $c_s=c/\sqrt{3}$, and $\mathbf{c}_i$ is defined as
$$\{\mathbf{c}_k\}=c\left( \begin{array}{rrrrrrrrrrrrrrr}
0&1&0&0&-1&0&0&1&1&1&-1&-1&-1&-1&1\\
0&0&1&0&0&-1&0&1&1&-1&1&-1&-1&1&-1\\
0&0&0&1&0&0&-1&1&-1&1&1&-1&1&-1&-1 \end{array} \right).$$

In order to accurately recover the macroscopic equations, the design of the force distribution function is very important. In the present model, the distribution function for the total force is given by
\begin{equation}
G_k=\omega_k\left \{ \mathbf{u} \cdot \nabla \rho+\frac{\mathbf{c}_k \cdot \mathbf{F}}{c_s^2}+\frac{(\mathbf{c}_k
\mathbf{c}_k-c_s^2\mathbf{I}):\left[\mathbf{u F}+\mathbf{Fu}+c_s^2 \mathbf{u} \nabla \rho +c_s^2(\nabla \rho)\mathbf{u}+\frac{\tilde{\mathbf{J}}\mathbf{u}}{\Delta t (\tau_g-0.5)}\right]}{2c_s^4}\right \}, \label{eq:N20_1}
\end{equation}
where $\mathbf{F}$ is the total force and is expressed as
\begin{equation}
\mathbf{F}=\mathbf{F}_s+\mathbf{G}.
\end{equation}

It can be rigorously proved that the present LB model can correctly recover Eqs. (\ref{eq:N19a})-(\ref{eq:N19b}) through the Chapman-Enskog analysis (see \ref{app:sec1} for the details) with the following fluid kinematic viscosity
\begin{equation}
\nu=c_s^2(\tau_g-0.5)\Delta t.
\label{eq:N21_1}
\end{equation}

In the LB method, the macroscopic quantities, e.g. fluid velocity $\mathbf{u}$ and pressure $p$, can be calculated from the moment of the distribution function \cite{Liang,Liang1},
\begin{equation}
\mathbf{u}=\frac{1}{{\rho}} \left( \sum_i \mathbf{c}_i g_i+0.5\Delta
t \mathbf{F} \right), \label{eq:N22_1}
\end{equation}
\begin{equation}
p=\frac{c_s^2}{1-\omega_0} \left[ \sum_{i\ne 0} g_k +\frac{\Delta
t}{2}\mathbf{u}\cdot \nabla \rho+{\rho} s_0(\mathbf{u}) \right
]. \label{eq:N23}
\end{equation}



\emph{Remark 1.} The distribution function for the total force can also be designed as
\begin{equation}
G_k=\omega_k\left \{ \mathbf{u} \cdot \nabla \rho+\frac{\mathbf{c}_k \cdot \mathbf{F}}{c_s^2}+\frac{(\mathbf{c}_k
\mathbf{c}_k-c_s^2\mathbf{I}):\left[\mathbf{u F}+\mathbf{Fu}+c_s^2 \mathbf{u} \nabla \rho +c_s^2(\nabla \rho)\mathbf{u}\right]}{2c_s^4}\right \},
\end{equation}
where $\mathbf{F}$ is redefined as
\begin{equation}
\mathbf{F}=-\nabla \cdot (\tilde{\mathbf{J}}\mathbf{u})+\mathbf{F}_s+\mathbf{G}.
\end{equation}
We would like to point out that the first force distribution function [Eq. (\ref{eq:N20_1})] is better because we don't have to calculate the gradient term $-\nabla \cdot (\tilde{\mathbf{J}} \mathbf{u})$. Therefore, in the following simulations, we will adopt Eq. (\ref{eq:N20_1}) as the force distribution function.


\subsection{LB model for the Cahn-Hilliard equations}
To complete the modeling of incompressible $N$-phase flows, the N-S equations should be coupled with the system of $(N-1)$ C-H equations. Here the evolution equations for the C-H equations can be written as
\begin{equation}
\begin{split}
h_k^i(\textbf{x}+\textbf{c}_k \Delta t,t+\Delta t)-h_k^i(\textbf{x},t)=&
-\frac{1}{\tau_{i}} \left[ h_k^i(\textbf{x},t)-h_k^{i,eq}(\textbf{x},t)
\right ]+\Delta t R_k^i(\textbf{x},t), \quad 1 \leq i \leq N-1,
\end{split}
\label{eq:N24}
\end{equation}
where $h_k^i(\textbf{x},t)$ denotes the distribution function of order
parameter $\phi_i$, $\tau_{i}$ represents the non-dimensional relaxation time for $i$-phase which is
related to the mobility, $h_k^{i,eq}(\textbf{x},t)$ is the local
equilibrium distribution function, which is introduced as
\cite{Fakhari2010,Huang,Chai}
\begin{equation}
h_k^{i,eq}(\textbf{x},t)=\left \{
\begin{array}{l}
\phi_i+(\omega_k-1)\eta_i C_i, \;\;\;k=0\\
\omega_k \eta_i C_i+\omega_k \frac{\textbf{c}_k \cdot \phi_i
\textbf{u}}{c_s^2}, \;\;\;\;   k \ne 0,
\end{array}
\right. \label{eq:N25}
\end{equation}
where $\eta_i$ denotes an adjustable parameter that controls the mobility for $i$-fluid.
$R_k(\textbf{x},t)$ is the source term and is defined as \cite{Liang}
 \begin{equation}
 {R_k}=\left(1-\frac{1}{2\tau_i} \right) \frac{\omega_k \mathbf{c}_k \cdot \partial_t \phi
 \textbf{u}}{c_s^2}.
 \label{eq:N26}
 \end{equation}
In our simulations, the temporal derivative in Eq.
(\ref{eq:N26}) is calculated by the first-order Eulerian scheme, i.e.,
\begin{equation}
\partial_t \phi \textbf{u}|_{(\mathbf{x},t)}=\frac{\phi
\textbf{u}|_{(\mathbf{x},t)}-\phi \textbf{u}|_{(\mathbf{x},t-\Delta
t)} }{\Delta t}.
 \label{eq:N27}
\end{equation}

In the present model, the order parameter $\phi_i$ is
calculated by taking the zeroth moment of the order distribution function,
\begin{equation}
\phi_i=\sum_k h_k^i,
\label{eq:N28}
\end{equation}
Then, according to mass conservation, the density $\rho$ can be projected on the basis of $\phi_i$,
\begin{equation}
\rho=\sum_{k=1}^N \rho_k=\frac{N}{\Gamma}+\sum_{i=1}^{N-1}\left(1-\frac{N}{\Gamma}\tilde{\gamma}_i \right)\left[\frac{1}{2}(\tilde{\rho}_i-\tilde{\rho}_N)+\frac{1}{2}(\tilde{\rho}_i+\tilde{\rho}_N)\phi_j \right].
\label{eq:N29}
\end{equation}

It is shown using the Chapman-Enskog analysis \cite{Liang} that the C-H equations [Eqs. (\ref{eq:N19c})] can be recovered with second-order accuracy and the mobility can be determined by
\begin{equation}
m_i=\eta_i c_s^2 (\tau_i-0.5)\Delta t. \label{eq:N30}
\end{equation}


In addition, to compute the first-order spatial derivatives and the Laplacian operators in the force term
$G_k$, surface tension $\mathbf{F}_s$ and chemical potentials $C_i$ $1 \leq i \leq N-1$,
the following isotropic schemes are adopted if not specified \cite{Zu}:
\begin{subequations}
\begin{equation}
\nabla \zeta(\mathbf{x},t)=\sum_{i \ne 0}\frac{ \omega_{i}
\mathbf{c}_{i} [\zeta(\mathbf{x}+\mathbf{c}_{i} \Delta t,t)-\zeta(\mathbf{x}-\mathbf{c}_{i} \Delta t,t)]}{2c_s^2
\Delta t}, \label{eq:N31a}
\end{equation}
\begin{equation}
\nabla^2 \zeta(\mathbf{x},t)=\sum_{i\ne 0} \frac{\omega_i[\zeta
(\mathbf{x}+\mathbf{c}_i \Delta t,t)-2\zeta(\mathbf{x},t)+\zeta
(\mathbf{x}-\mathbf{c}_i \Delta t,t)]}{c_s^2
\Delta t^2}, \label{eq:N31b}
\end{equation}
\label{eq:N31}
\end{subequations}
where $\zeta$ is any macroscopic quantity. This scheme is referred to as isotropic central scheme, and it not only has a second-order accuracy in space, but also conserve the global mass of a $N$-phase system \cite{Liang}.

\section{Model validation}
In this section, we employ two classical benchmark problems, including tests of static droplets and the spreading of a liquid lens between the other two components, to demonstrate the capability and accuracy of the LB model. Some detailed comparisons of numerical results and the analytical solutions are conducted. Here we indicate that in all the simulations, the D$2$Q$9$ lattice structure is adopted, and the grid resolution test has been conducted. It is demonstrated that the grids used in the present work are accurate enough to give grid-independent results.

\subsection{Static droplets}
\begin{figure}[ht]
\subfigure[]{ \label{fig0_0_1}
\includegraphics[width=0.315\textwidth]{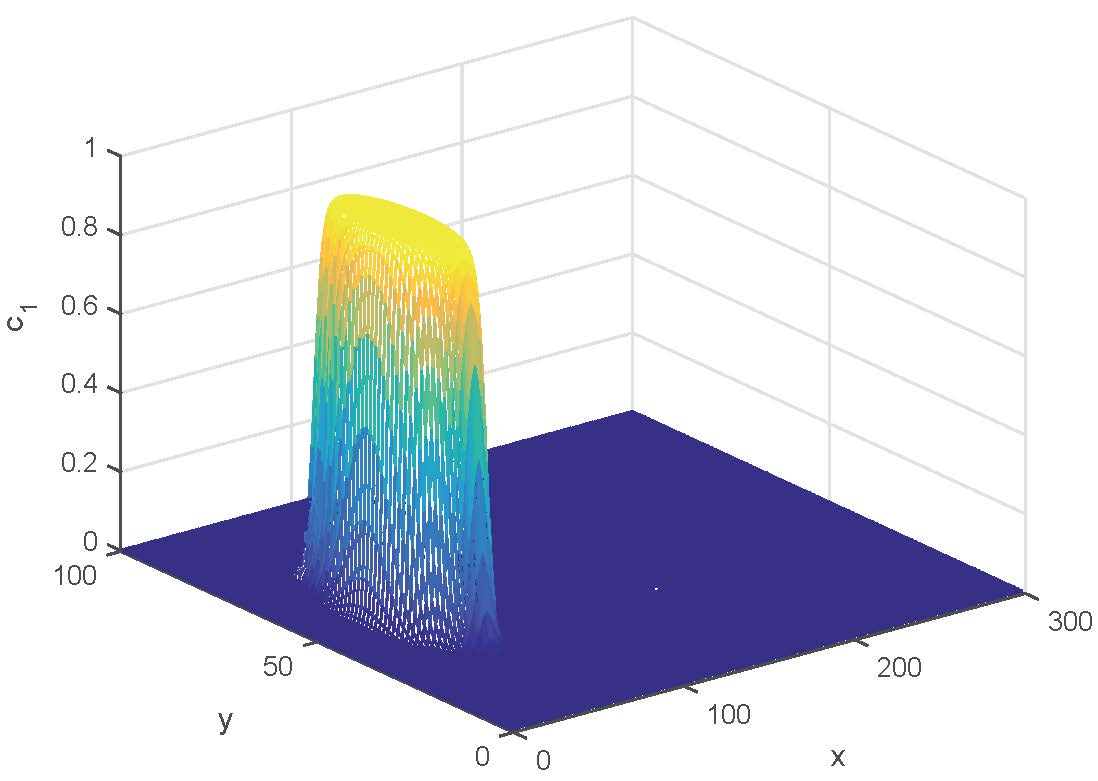}}
\subfigure[]{ \label{fig0_0_2}
\includegraphics[width=0.315\textwidth]{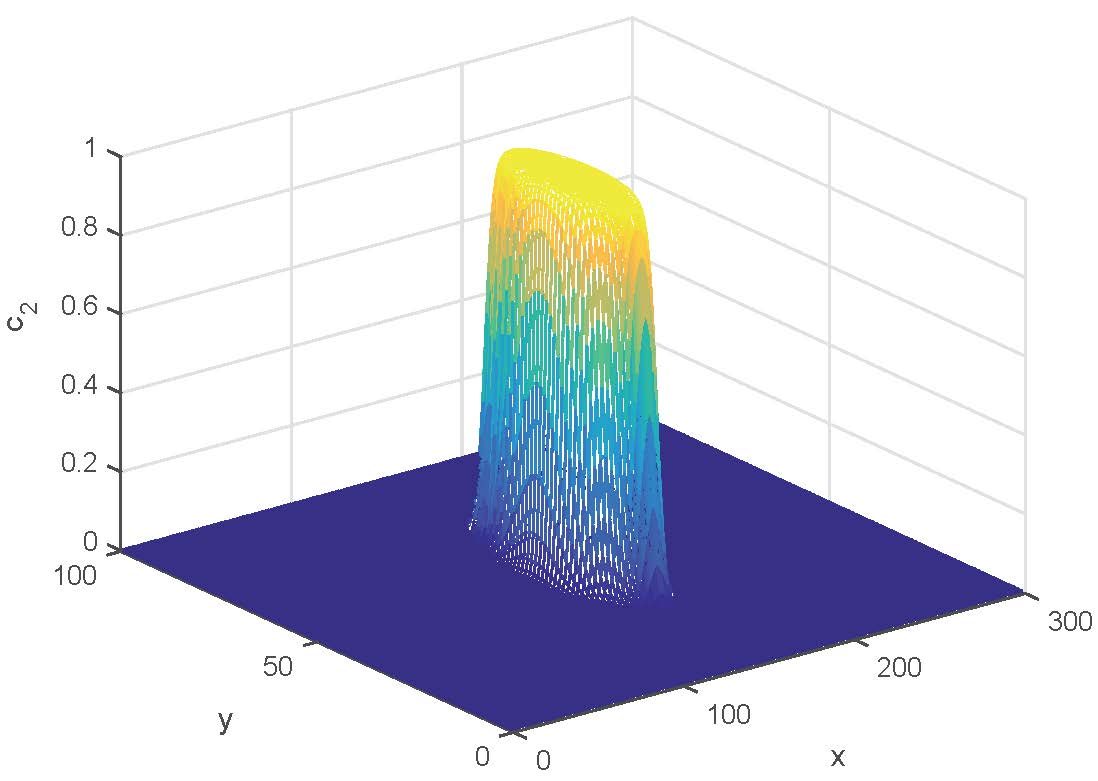}}
\subfigure[]{ \label{fig0_0_3}
\includegraphics[width=0.315\textwidth]{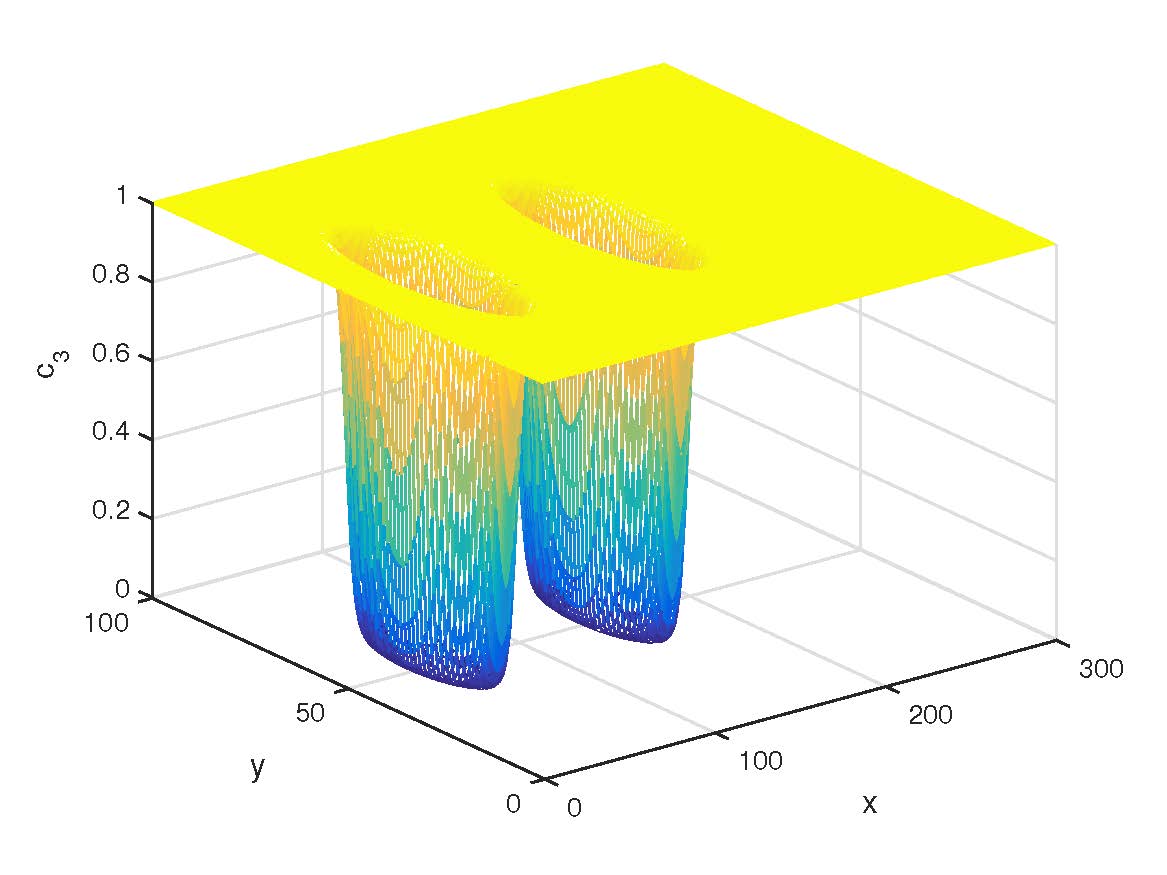}}
\caption{(Color online) The steady distributions of three volume fractions ($c_1$, $c_2$, $c_3$) obtained by the present LB model.}
\label{fig0_0_a}
\end{figure}

\begin{figure}[ht]
\centering
\includegraphics[width=0.7\textwidth]{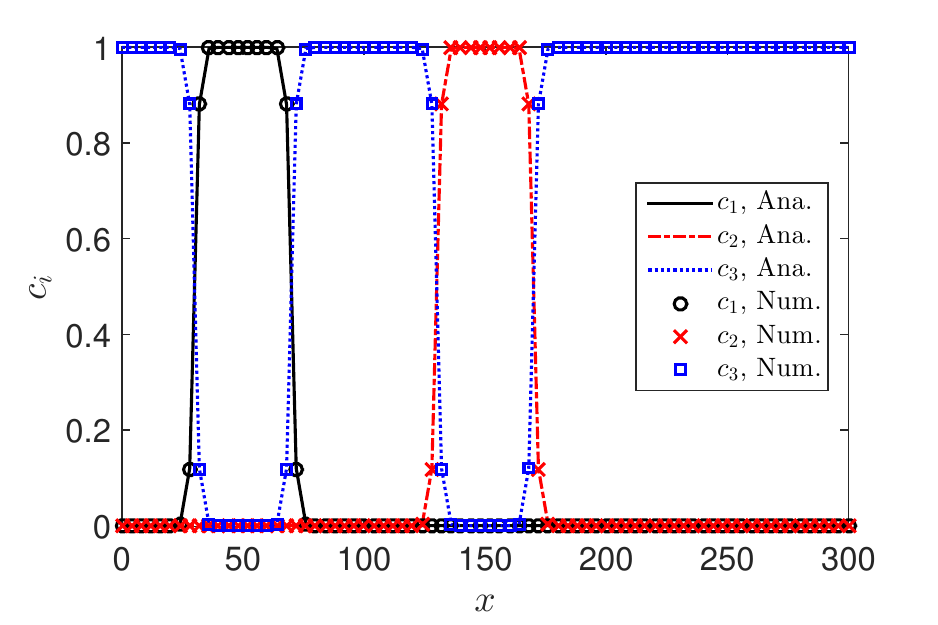}
\caption{(Color online) Volume fractions along the centerline ($x=50$, $0 \leq y \leq 300$), where Ana. denotes analytical results and Num. denotes numerical results.}
\label{fig0_0_b}
\end{figure}

A basic test of static droplets is first performed to verify the present LB model. Initially, two droplets with radius $R=20$ are placed in a lattice domain of $NX \times NY=300\times 100$ and the periodic boundary condition is applied at all boundaries. The initial volume fractions are given by
\begin{subequations}
\begin{align}
&c_1(x,y)=0.5+0.5\tanh
\frac{R-\sqrt{(x-x_{c_1})^2+(y-y_{c_1})^2}}{\sqrt2 \eta}, \\
&c_2(x,y)=0.5+0.5\tanh
\frac{R-\sqrt{(x-x_{c_2})^2+(y-y_{c_2})^2}}{\sqrt2 \eta}, \\
&c_3(x,y)=1.0-c_1(x,y)-c_2(x,y),
\end{align}
\label{eq:N31_a}
\end{subequations}
where $(x_{c_1},y_{c_1})$ and $(x_{c_2},y_{c_2})$ represent the coordinate of circular droplet and are fixed as $(x_{c_1},y_{c_1})=(50,50)$, $(x_{c_2},y_{c_2})=(150,50)$. In all the simulations, $c=\delta x=\delta t=1.0$. Some other parameters are given as $\rho_1 : \rho_2  : \rho_3=20:1:5$, $\tau_g=\tau_1=\tau_2=0.8$, $\sigma_{12}=\sigma_{13}=\sigma_{23}=0.01$, $m_1=m_2=0.001$. When choosing $\beta$ and $\eta$ in the free energy density expression, we follow such a relation
\begin{equation}
\beta=\sqrt{3\sqrt{2}\sigma_{min} \eta},
\label{eq:N36}
\end{equation}
where $\sigma_{min}=\min {\sigma_{ij}}$ is the minimum value among the $\frac{1}{2}N(N-1)$ surface tensions \cite{Dong2014}. Here $\eta$ is set to $\eta=\sqrt{2}$, then $\beta=\sqrt{0.06}$. Fig. \ref{fig0_0_a} shows the steady distributions of volume fractions obtained by the present LB model. It can be observed that the present LB model can accurately preserve the profiles of volume fractions with initial configurations for this $3$-phase problem. In order to give quantitative comparisons, we also plot the volume fractions along the centerline ($x=50$, $0 \leq y \leq 300$) in Fig. \ref{fig0_0_b}. From this figure, one can observe that the numerical results of the volume fractions agree well with the theoretical results.

\begin{figure}[ht]
\subfigure[]{ \label{fig0_1_1}
\includegraphics[width=0.5\textwidth]{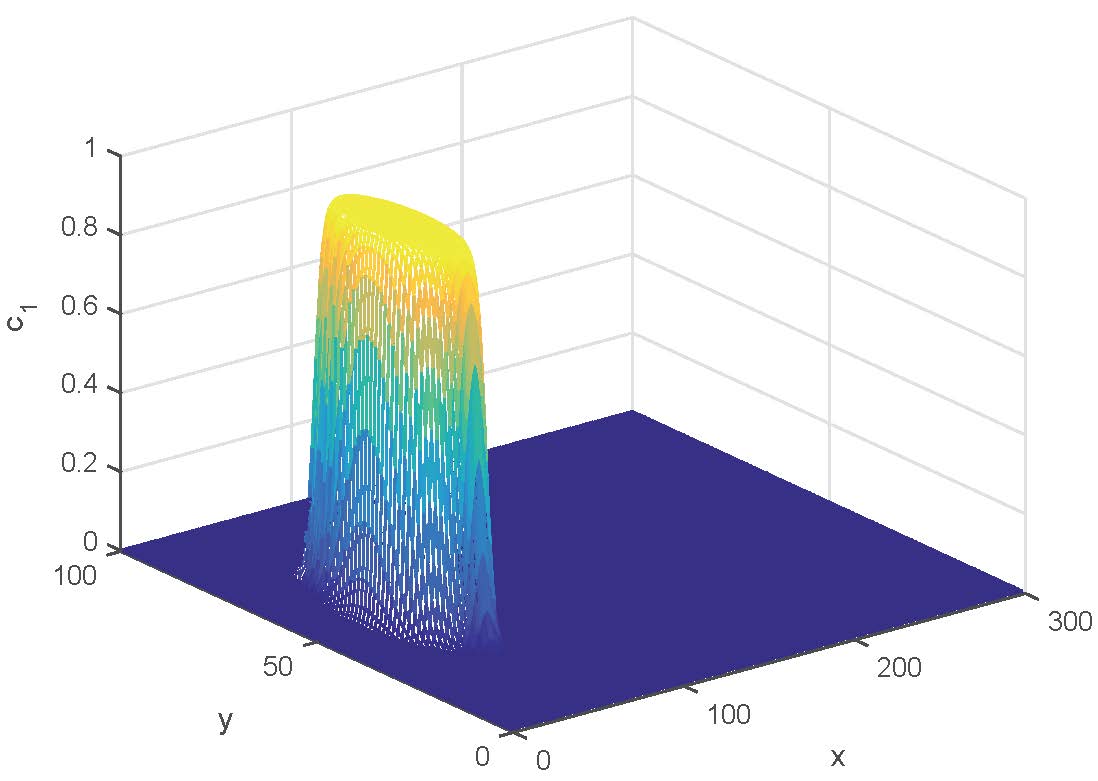}}
\subfigure[]{ \label{fig0_1_2}
\includegraphics[width=0.5\textwidth]{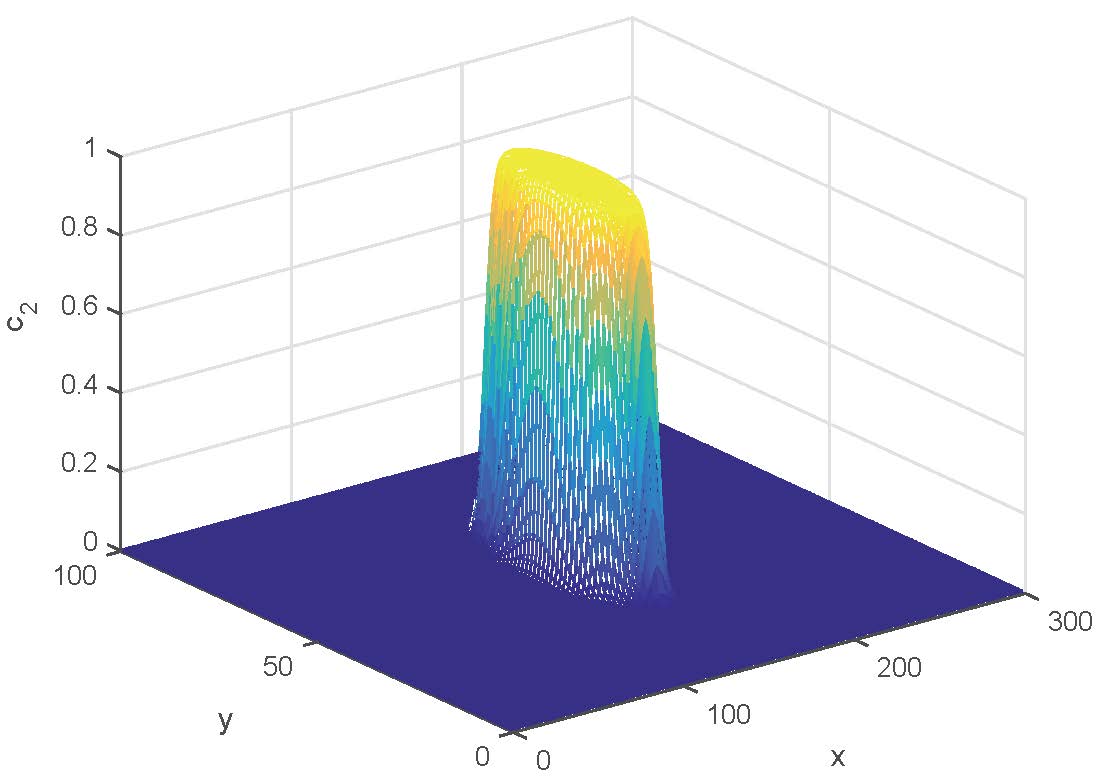}}
\subfigure[]{ \label{fig0_1_3}
\includegraphics[width=0.5\textwidth]{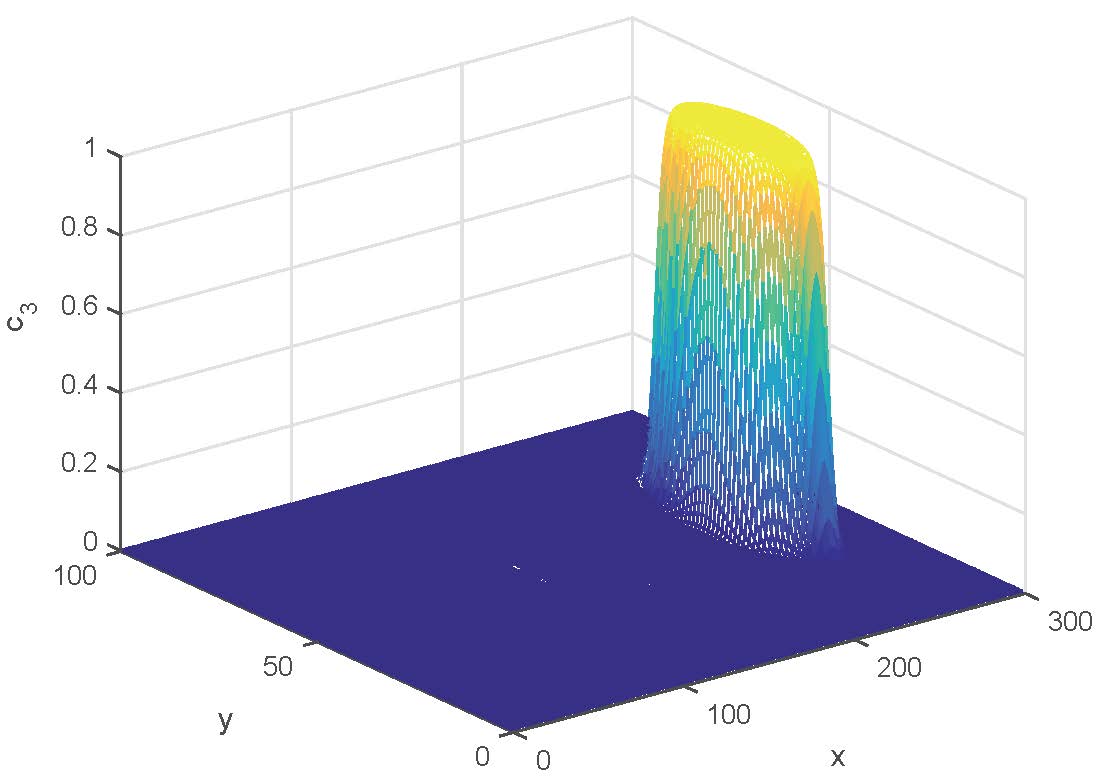}}
\subfigure[]{ \label{fig0_1_4}
\includegraphics[width=0.5\textwidth]{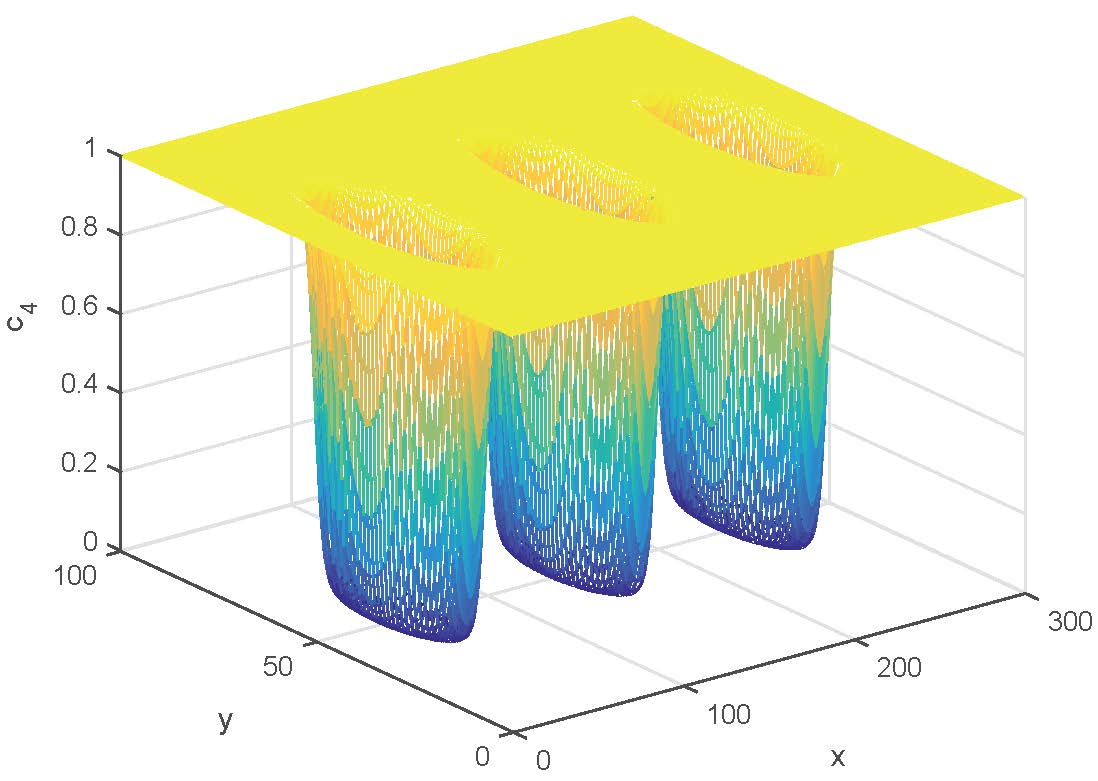}}
\caption{(Color online) The steady distributions of three volume fractions ($c_1$, $c_2$, $c_3$, $c_4$) obtained by the present LB model.}
\label{fig0_1_a}
\end{figure}

Besides, we also consider a $4$-phase problem of static droplets. In this test, three droplets with radius $R=20$ are placed in the same domain of $NX \times NY=300\times 100$. The initial volume fractions are expressed as
\begin{subequations}
\begin{align}
&c_1(x,y)=0.5+0.5\tanh
\frac{R-\sqrt{(x-x_{c_1})^2+(y-y_{c_1})^2}}{\sqrt2 \eta}, \\
&c_2(x,y)=0.5+0.5\tanh
\frac{R-\sqrt{(x-x_{c_2})^2+(y-y_{c_2})^2}}{\sqrt2 \eta}, \\
&c_3(x,y)=0.5+0.5\tanh
\frac{R-\sqrt{(x-x_{c_3})^2+(y-y_{c_3})^2}}{\sqrt2 \eta}, \\
&c_4(x,y)=1.0-c_1(x,y)-c_2(x,y)-c_3(x,y),
\end{align}
\label{eq:N31_b}
\end{subequations}
with $(x_{c_1},y_{c_1})=(50,50)$, $(x_{c_2},y_{c_2})=(150,50)$ and $(x_{c_3},y_{c_3})=(250,50)$. Some other parameters are given as $\rho_1 : \rho_2  : \rho_3:\rho4=20:1:10:5$, $\tau_g=\tau_1=\tau_2=\tau_3=0.8$, $\sigma_{12}=\sigma_{13}=\sigma_{14}=\sigma_{23}=\sigma_{24}=\sigma{34}=0.01$, $m_1=m_2=m_3=0.001$. The rest of the parameters are set as previous case. As shown in Fig. \ref{fig0_1_a}, the present LB model can also accurately preserve the profiles of volume fractions with initial configurations for $4$-phase problems. To further give a detailed comparison, we also plot the volume fractions along the center line and compare the results between analytical results and numerical results. The results show that the developed LB model is accurate enough to simulate multiphase problems.

\begin{figure}[ht]
\centering
\includegraphics[width=0.7\textwidth]{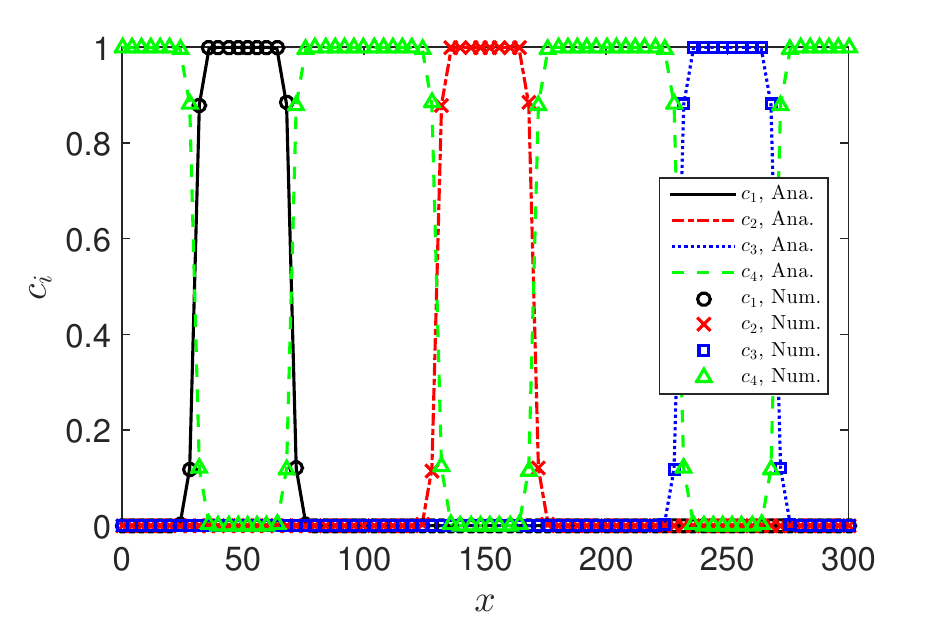}
\caption{(Color online) Volume fractions along the centerline ($x=50$, $0 \leq y \leq 300$), where Ana. denotes analytical results and Num. denotes numerical results.}
\label{fig0_1_b}
\end{figure}

\begin{figure}[ht]
\centering
\includegraphics[width=0.7\textwidth]{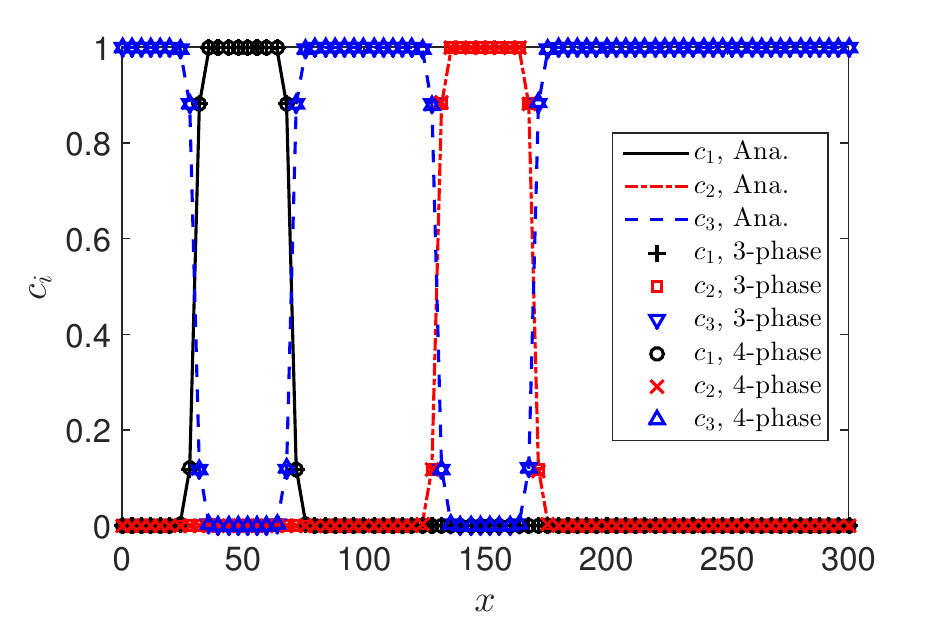}
\caption{(Color online) Comparisons of volume fractions obtained from our $3$-phase model and $4$-phase model with one phase disappeared as well as theoretical values along the centerline ($x=50$, $0 \leq y \leq 300$).}
\label{fig0_2}
\end{figure}

Next, we will show that our LB model for $N$-phase flows can satisfy the reduction-consistent property. Here, the reduction-consistent property means that when one phase of the $N$-phase system disappears, the designed model can still be used to simulate the remaining ($N-1$)-phase system. Consider the case where the volume fraction of the third phase is $0$ in the previous $4$-phase example. The remaining parts can be viewed as a $3$-phase system. For convenience, the $4$th-phase in the former case is marked as the $3$th-phase. Fig. \ref{fig0_2} depicts the comparisons of volume fractions obtained from our $3$-phase model, $4$-phase model with one phase disappeared, and theoretical values along the centerline. It can be seen that the results from our $3$-phase model and $4$-phase model with one phase disappeared are almost identical, and they are also in good agreement with the theoretical values. Thus, we conclude that our LB model can satisfy the reduction-consistent property.

\subsection{Spreading of a liquid lens}

\begin{figure}[ht]
\centering
\includegraphics[width=0.5\textwidth]{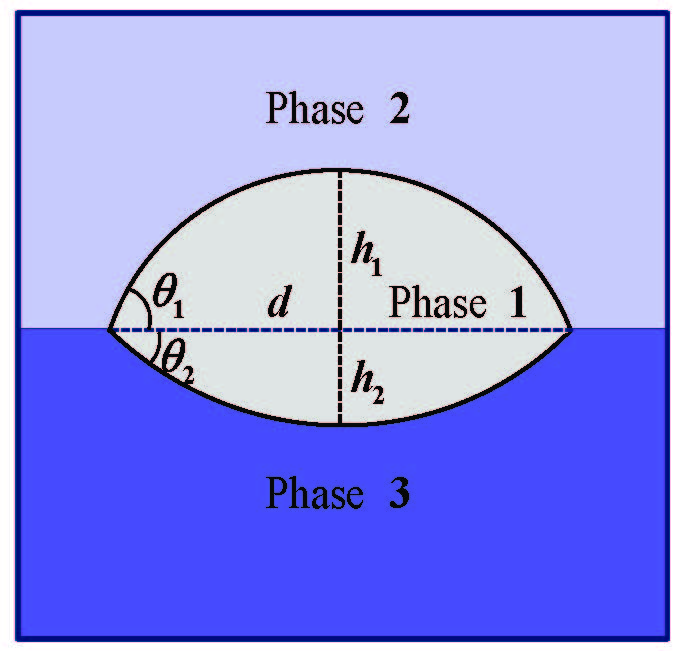}
\caption{(Color online) Schematic of the lens shape at the equilibrium state.}
\label{fig1_1}
\end{figure}

Here we consider a droplet located at the interface between the other two immiscible fluids. Under the influence of surface tension, the droplet will form a lens when the system reaches equilibrium. According to Neumann's law \cite{Rawlinson}, the following relationships are satisfied between the contact angles and the surface tensions at the equilibrium state,
\begin{equation}
\cos(\theta_1)=\frac{\sigma_{12}^{2}+\sigma_{23}^{2}-\sigma_{13}^{2}}{2\sigma_{12} \sigma_{23}},\quad
\cos(\theta_2)=\frac{\sigma_{13}^{2}+\sigma_{23}^{2}-\sigma_{12}^{2}}{2\sigma_{13} \sigma_{23}},
\label{eq:N32}
\end{equation}
where $\theta_i$ ($i=1, 2$) denote the contact angles as shown in Fig. {\ref{fig1_1}}. The relationship between the
lens area $A$, its length $d$ (the distance between two triple junctions) and the contact angles is
\begin{equation}
A=\left(\frac{d}{2} \right)^2 \sum_{i=1}^2 \frac{1}{\sin(\theta_i)}\left(\frac{\theta_i}{\sin \theta_i}-\cos \theta_i \right),
\label{eq:N33}
\end{equation}
After some simple geometric transformations, we can derive the following expression
\begin{equation}
h_i=\left(\frac{d}{2} \right)\frac{1-\cos \theta_i}{\sin \theta_i}, \quad i=1,2.
\label{eq:N34}
\end{equation}

Initially, a circular droplet with the radius $R=30$ is placed in the middle of the computational domain with $NX \times NY=150 \times 150$. The initial volume fractions are given by
\begin{subequations}
\begin{align}
&c_1(x,y)=0.5+0.5\tanh
\frac{R-\sqrt{(x-x_c)^2+(y-y_c)^2}}{\sqrt2 \eta}, \\
&c_2(x,y)=\max\left[ 0.5+0.5\tanh \frac{y-y_c}{\sqrt 2 \eta}-c_1(x,y),0 \right],\\
&c_3(x,y)=1.0-c_1(x,y)-c_2(x,y),
\end{align}
\label{eq:N35}
\end{subequations}
where $(x_c,y_c)$ is the coordinate of circular droplet.

\begin{figure}[ht]
\subfigure[]{ \label{fig1_2_1}
\includegraphics[width=0.3\textwidth]{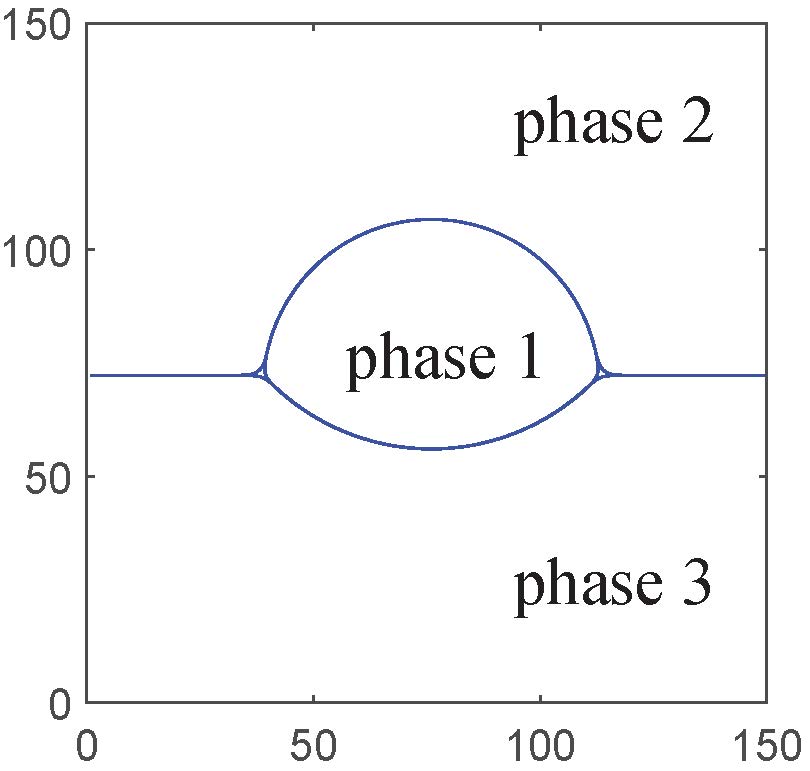}}
\subfigure[]{ \label{fig1_2_2}
\includegraphics[width=0.3\textwidth]{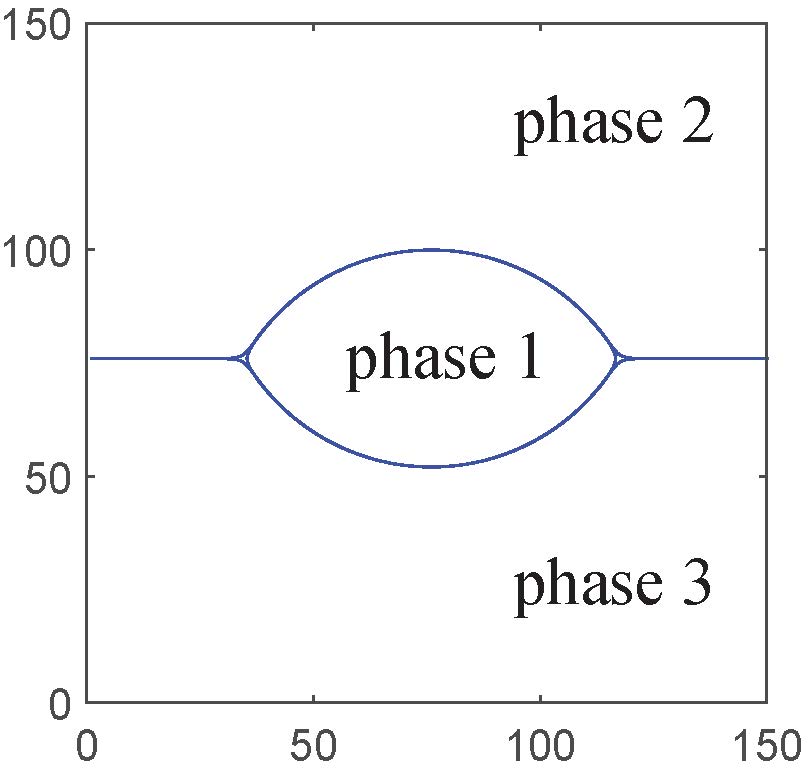}}
\subfigure[]{ \label{fig1_2_3}
\includegraphics[width=0.3\textwidth]{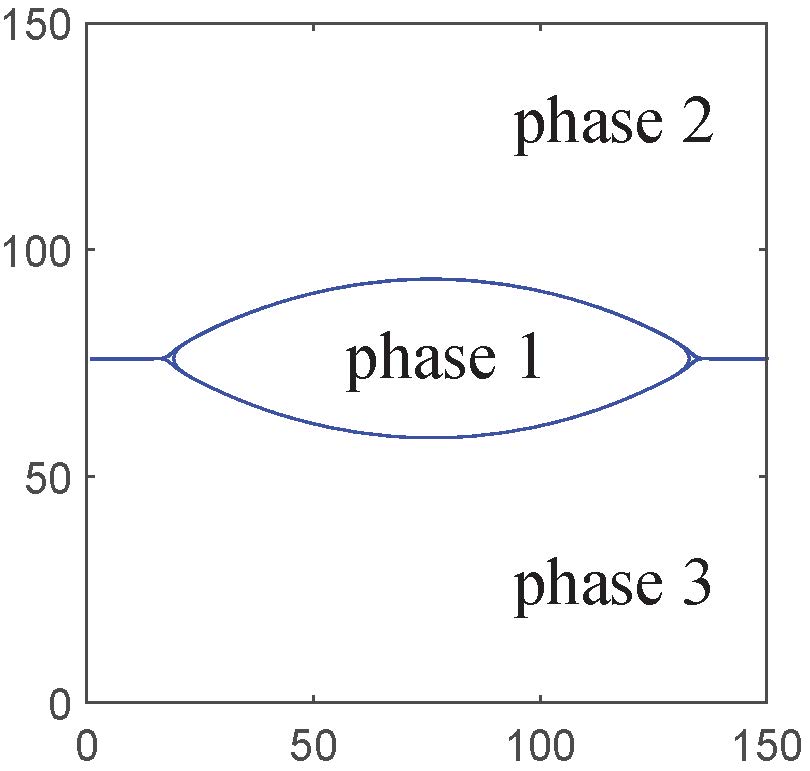}}
\caption{(Color online) Schematic of the lens shape at the equilibrium state.}
\label{fig1_2}
\end{figure}

In this test, the periodic boundary conditions are employed in the $x$-direction, and the halfway bounce-back boundary conditions are enforced on the top and bottom walls. The density ratio of the three fluids is set to $\rho_1 : \rho_2  : \rho_3=10:1:5$.   Some other physical parameters are given as $\tau_g=\tau_1=\tau_2=0.8$, $\sigma_{12}:\sigma_{13}:\sigma_{23}=1: \frac{4}{3}:1$, $1:1:1$, and $0.6:0.6:1$, $m_1=m_2=0.1$, $\eta=\sqrt{2}$, $\beta=\sqrt{0.06}$. It is shown in Fig. {\ref{fig1_2}} that the droplets will undergo some deformations due to the effect of surface tension, and eventually form distinct equilibrium configurations. The shapes of the liquid interface in this work are in good agreement with the previous results \cite{Smith,Boyer2006}. In order to give quantitative results, we also measured the contact angles $\theta_1$ and $\theta_2$ and compared them with the analytical solutions (see Table {\ref{Tab_1}}), where the numerical solutions $\theta_1$ and $\theta_2$ are calculated by
\begin{equation}
\theta_1=2\arctan \left(\frac{2h_1}{d} \right),\quad
\theta_2=2\arctan \left(\frac{2h_2}{d} \right).
\end{equation}
From Table {\ref{Tab_1}}, one can observe that the numerical contact angles $\theta_1$ and $\theta_2$ are in good agreement with analytical solutions, where the maximum relative errors are no more than $1.5\%$. In addition, we also give a comparison between the numerical solutions and the analytical solutions of the length $d$ and the height $(h_1, h_2)$, see Table {\ref{Tab_2}}. It is shown that the maximum relative errors of $d$, $h_1$ and $h_2$ shall not exceed $1.8\%$, which means that the present LB model can accurately capture the interface of the incompressible three-phase flow without considering gravity.

\begin{table}[tbp]
\caption{The equilibrium contact angles $\theta_1$ and $\theta_2$ with different surface tension ratios.} \label{Tab_1}
\centering
\begin{tabular}{lcccccc}
\hline\hline
Surface tension &  \multicolumn{2}{c}{Present LB model}  & \multicolumn{2}{c}{Analytical solutions} &\multicolumn{2}{c}{Relative errors} \\
\cline{2-3} \cline{4-5} \cline{6-7}
$(\sigma_{12}:\sigma_{13}:\sigma_{23})$  & $\theta_1$ &  $\theta_2$  & $\theta_1$ &  $\theta_2$  & $\theta_1$ &  $\theta_2$\\
\midrule[1pt]
$1: \frac{4}{3} :1$ & $84.6^{\circ}$ & $50.0^{\circ}$ & $83.5^{\circ}$ & $52.5^{\circ}$ & $1.2\%$ & $0.2\%$  \\
$1:    1        :1$ & $60.7^{\circ}$ & $60.7^{\circ}$ & $60.0^{\circ}$ & $60.0^{\circ}$ & $1.2\%$ & $1.2\%$  \\
$0.6:   0.6     :1$ & $34.1^{\circ}$ & $34.1^{\circ}$ & $34.0^{\circ}$ & $34.0^{\circ}$ & $1.5\%$ & $1.5\%$  \\
\hline\hline
\end{tabular}
\end{table}

\begin{table}[tbp]
\caption{The equilibrium length $d$ and height $(h_1,h_2)$ with different surface tension ratios.} \label{Tab_2}
\centering
\begin{tabular}{lccccccccc}
\hline\hline
Surface tension &  \multicolumn{3}{c}{Present LB model}  & \multicolumn{3}{c}{Analytical solutions} &\multicolumn{3}{c}{Relative errors} \\
\cline{2-4} \cline{5-7} \cline{8-10}
$(\sigma_{12}:\sigma_{13}:\sigma_{23})$  & $d$ & $h_1$& $h_2$  & $d$ & $h_1$& $h_2$  & $d$ & $h_1$& $h_2$  \\
\midrule[1pt]
$1: \frac{4}{3} :1$ & $73.0$ & $33.2$ & $16.3$ &  $73.8$ & $33.0$ & $16.5$& $1.1\%$ & $0.6\%$ & $1.2\%$  \\
$1:    1        :1$ & $82.0$ & $23.6$ & $23.6$ &  $80.7$ & $23.3$ & $23.3$& $1.7\%$ & $1.3\%$ & $1.3\%$  \\
$0.6:   0.6     :1$ & $114.0$ & $17.5$ & $17.5$ &  $113.9$ & $17.2$ & $17.2$& $0.1\%$ & $1.8\%$ & $1.8\%$  \\
\hline\hline
\end{tabular}
\end{table}

\begin{figure}[ht]
\subfigure[]{ \label{lens1_31}
\includegraphics[width=0.315\textwidth]{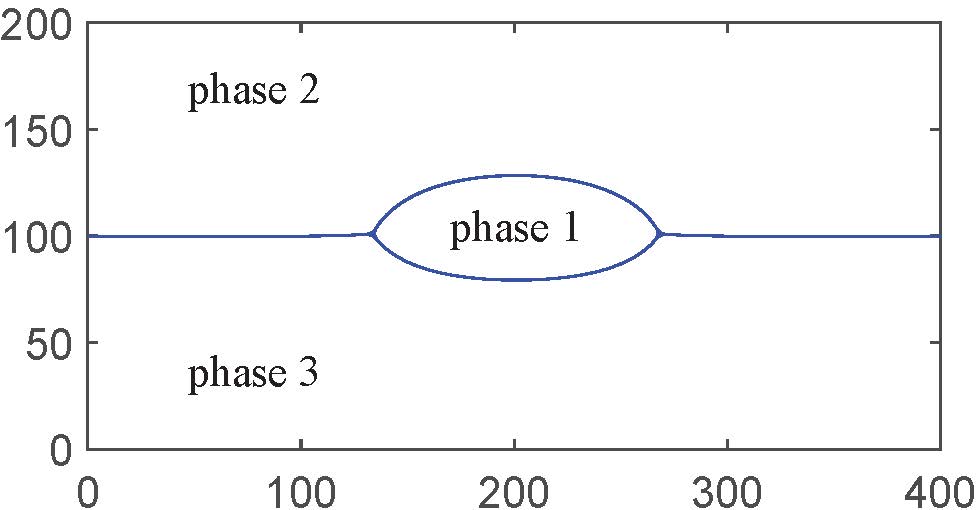}}
\subfigure[]{ \label{lens1_32}
\includegraphics[width=0.315\textwidth]{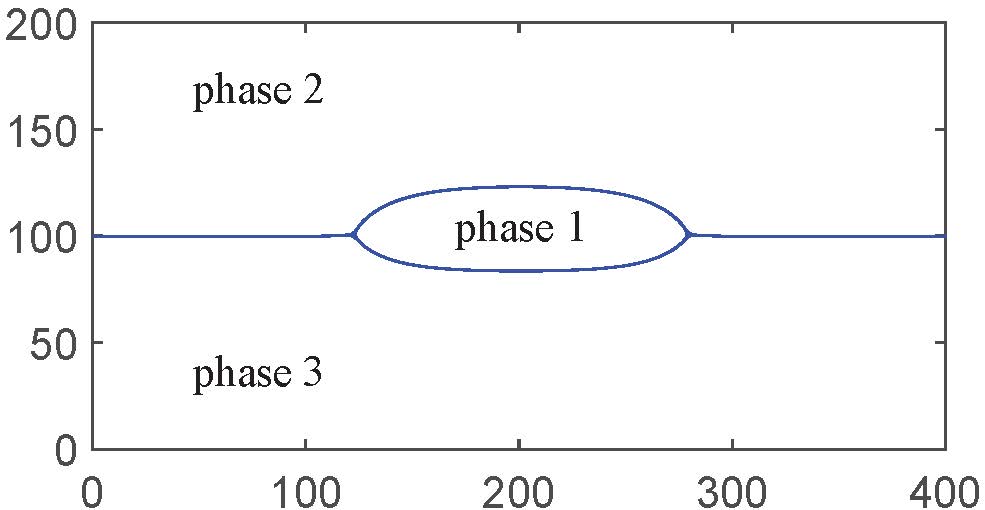}}
\subfigure[]{ \label{lens1_33}
\includegraphics[width=0.315\textwidth]{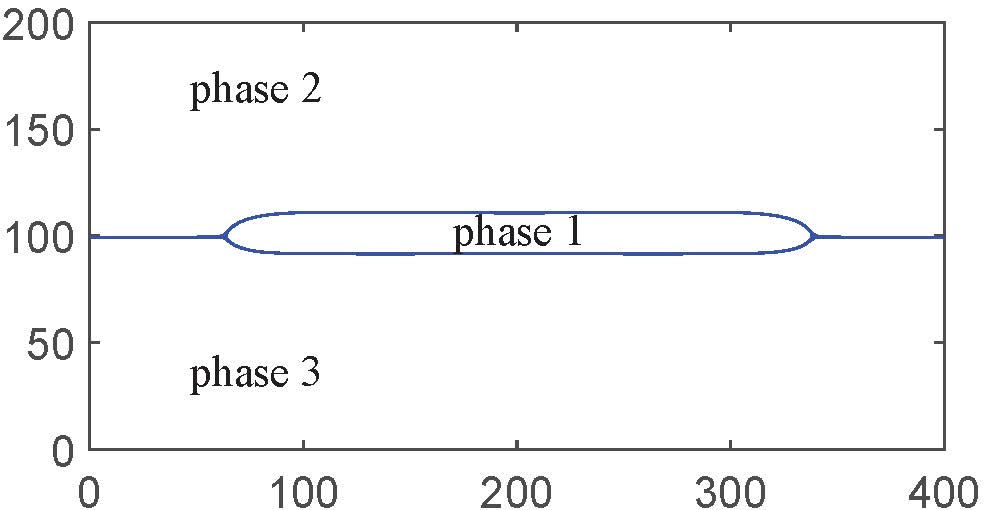}}
\caption{(Color online) The equilibrium configurations of the droplet with different gravity forces: (a) $g=5 \times 10^{-6}$, (b) $1 \times 10^{-5}$, (c) $g=5 \times 10^{-5}$.}
\label{fig1_3}
\end{figure}

\begin{figure}[ht]
\centering
\includegraphics[width=0.7\textwidth]{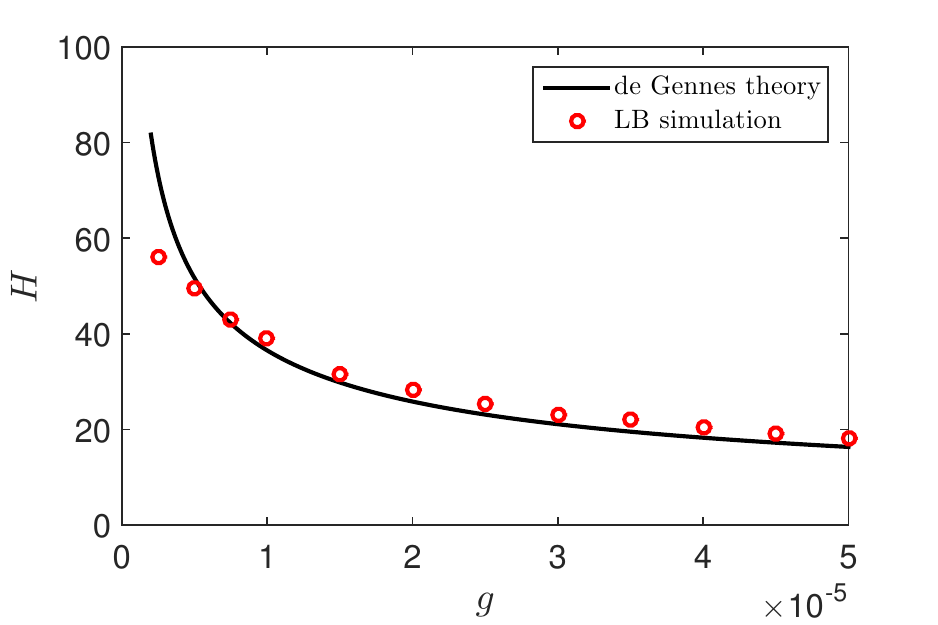}
\caption{(Color online) Comparison of the asymptotic puddle thickness $H$ as a function of gravity $g$ between the present results and the de Gennes theory.}
\label{fig1_4}
\end{figure}

Furthermore, now we consider the effect of gravity on the equilibrium configurations. To generate the gravitational effects, a
body force, $\mathbf{G}=(0,\rho g)$, is added to the momentum equation, where $g$ is the gravitational acceleration. In this test, we vary the value of the gravitational acceleration while fixing all the other physical parameters. Due to the large deformation of the interface, the computational domain is set to $NX \times NY=400 \times 200$. The physical parameters are given as $\rho_1 : \rho_2  : \rho_3=3:1:6$, $\tau_g=\tau_1=\tau_2=0.8$, $\sigma_{12}:\sigma_{13}:\sigma_{23}=1:1:1$, $m_1=m_2=0.1$. As shown in Fig. \ref{lens1_31}, when the gravity is relatively small and surface tension is dominant, the droplet forms a lens on the water surface. As the gravity increases, the droplet becomes flatter [see Fig. \ref{lens1_32}]. When the gravity is large enough and dominates over the surface tension, it will forms a puddle [see Fig. \ref{lens1_33}]. The current results are qualitatively consistent with the de Gennes theory \cite{Gennes}. To give a quantitative comparison, we measured asymptotic thickness $H$ as a function of gravity and compared it with the de Gennes theory, where the asymptotic puddle thickness can be given as \cite{Gennes,Langmuir}
\begin{equation}
H=\sqrt{\frac{2(\sigma_{12}+\sigma_{13}-\sigma_{23})}{\frac{\rho_1}{\rho_3}(\rho_3-\rho_1)g}}.
\label{eq:N37}
\end{equation}
Fig. {\ref{fig1_4}} depicts the asymptotic thickness as a function of gravity between the present results and the de Gennes theory. One can observe that the present results agree well with the de Gennes theory when the gravity is large ($g \geq 0.5$ for this case). However, when the gravity is small, a large discrepancy between the numerical results and theoretical results, because the expression ({\ref{eq:N37}}) is not valid for the case that $g$ is too small \cite{Dong2014}.

\begin{figure}[ht]
\subfigure[]{ \label{lens1_5_1}
\includegraphics[width=0.5\textwidth]{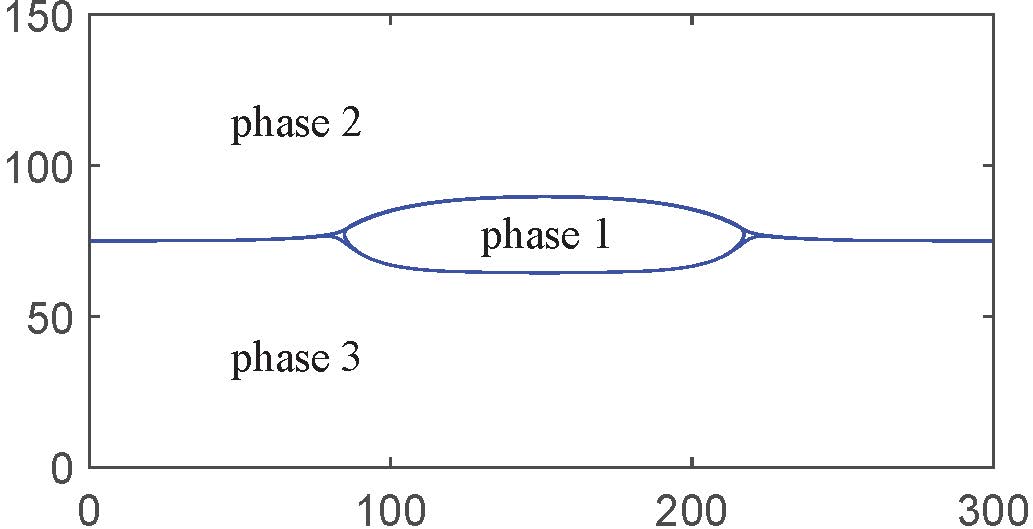}}
\subfigure[]{ \label{lens1_5_2}
\includegraphics[width=0.5\textwidth]{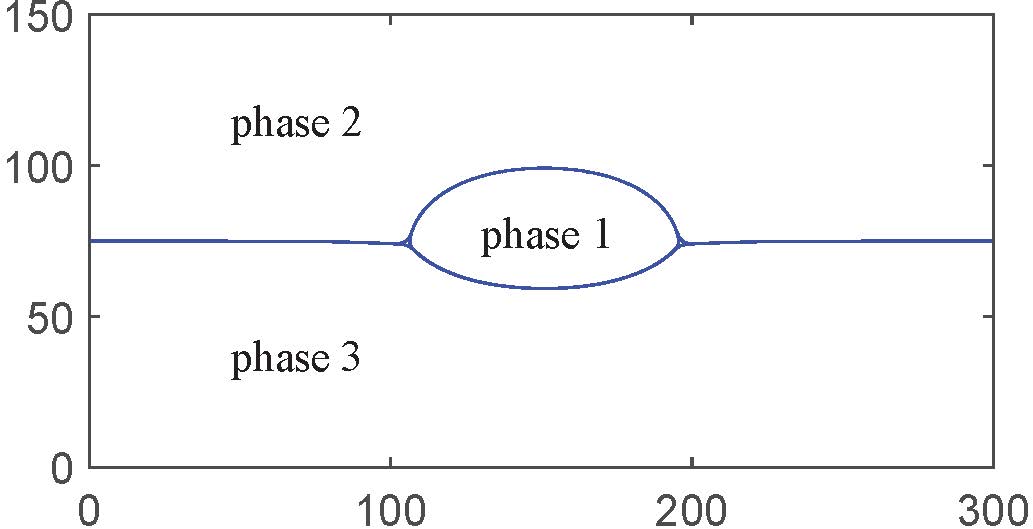}}
\caption{(Color online) The equilibrium configurations of the droplet with different surface tensions: (a) $\sigma_{13}=0.004$, (b) $\sigma_{13}=0.014$. Here the gravitational acceleration is fixed to $g=1 \times 10^{-5}$}.
\label{fig1_5}
\end{figure}

\begin{figure}[ht]
\centering
\includegraphics[width=0.7\textwidth]{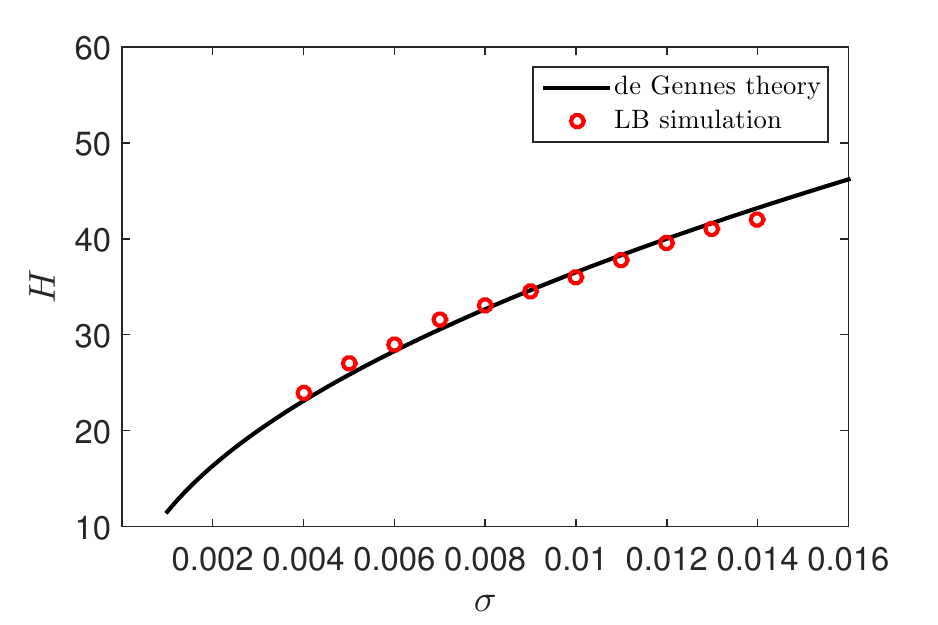}
\caption{(Color online) Comparison of the asymptotic puddle thickness $H$ as a function of surface tension $\sigma_{13}$ between the present results and the de Gennes theory.}
\label{fig1_6}
\end{figure}

In addition, we also consider the effect of surface tension on the equilibrium configurations. In this test, we fix the gravitational acceleration $g$ and the surface tensions $\sigma_{12}$, $\sigma_{23}$ to $g=1 \times 10^{-5}$, $\sigma_{12}=\sigma_{23}=0.01$ and vary the surface tension $\sigma_{13}$. The computational domain is set to $NX \times NY=300 \times 150$. The other parameters are the same as the previous test. As shown in Fig. \ref{fig1_5}, the droplet thickness strongly depend on the surface tension. As the surface tension increases, the droplet thickness gradually increases. To quantitatively describe the relationship between asymptotic thickness and surface tension, we give a comparison of the results between the current simulation and the de Gennes theory \cite{Gennes} (see Fig. {\ref{fig1_6}}). From Fig. {\ref{fig1_6}}, one can conclude that $H$ and $\sigma_{13}$ increase simultaneously, and the simulation results are in good agreement with the de Gennes theory.

Therefore, all the results of this section, especially qualitative and quantitative comparisons with de Gennes theory, show that the current LB model is accurate and efficient for simulating the motion of incompressible multiphase systems.


\section{Numerical applications}
To further demonstrate the capabilities and performance of the LB model we proposed for $N$-phase flows ($N \geq 2$), two numerical applications involving four to five fluid phases are carried out. These applications include the dynamics of droplet collision for four fluid phases and dynamics of droplets and interfaces for five fluid phases.
The simulation results show that the present LB model can be used to study the interactions among multiple types of fluid interfaces.
\subsection{Dynamics of droplet collision for four fluid phases}

\begin{figure}[ht]
\centering
\includegraphics[width=0.3\textwidth]{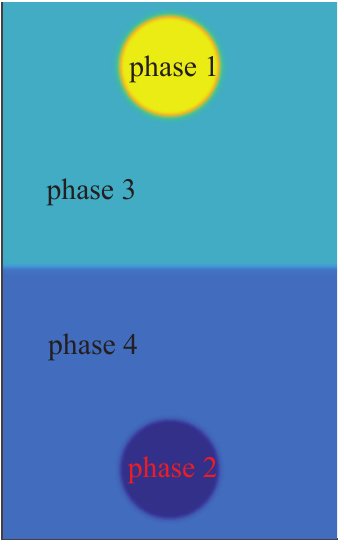}
\caption{(Color online) Initial configuration of droplet collision for four fluid phases.}
\label{fig2_1a}
\end{figure}

\begin{figure}[ht]
\centering
\subfigure[]{ \label{fig2_1b}
\includegraphics[width=1.2in]{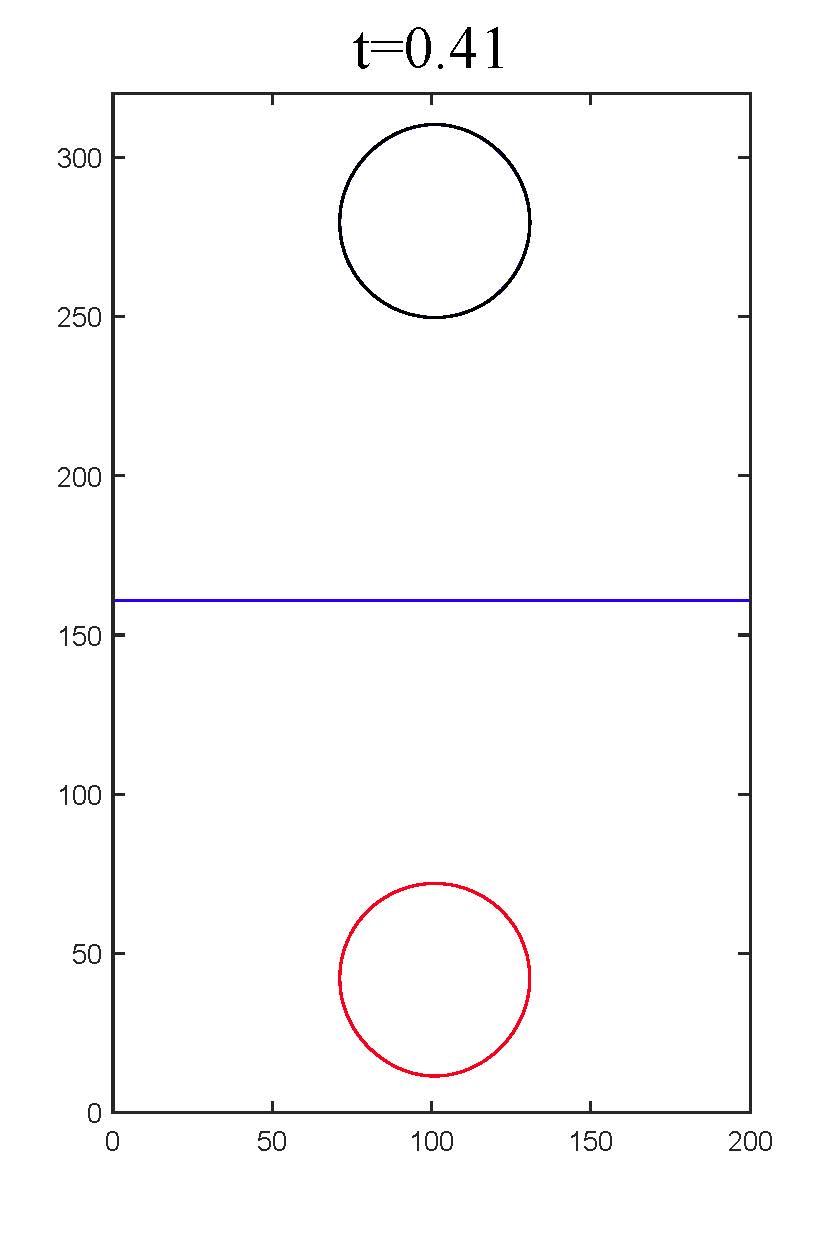}}
\subfigure[]{ \label{fig2_1c}
\includegraphics[width=1.2in]{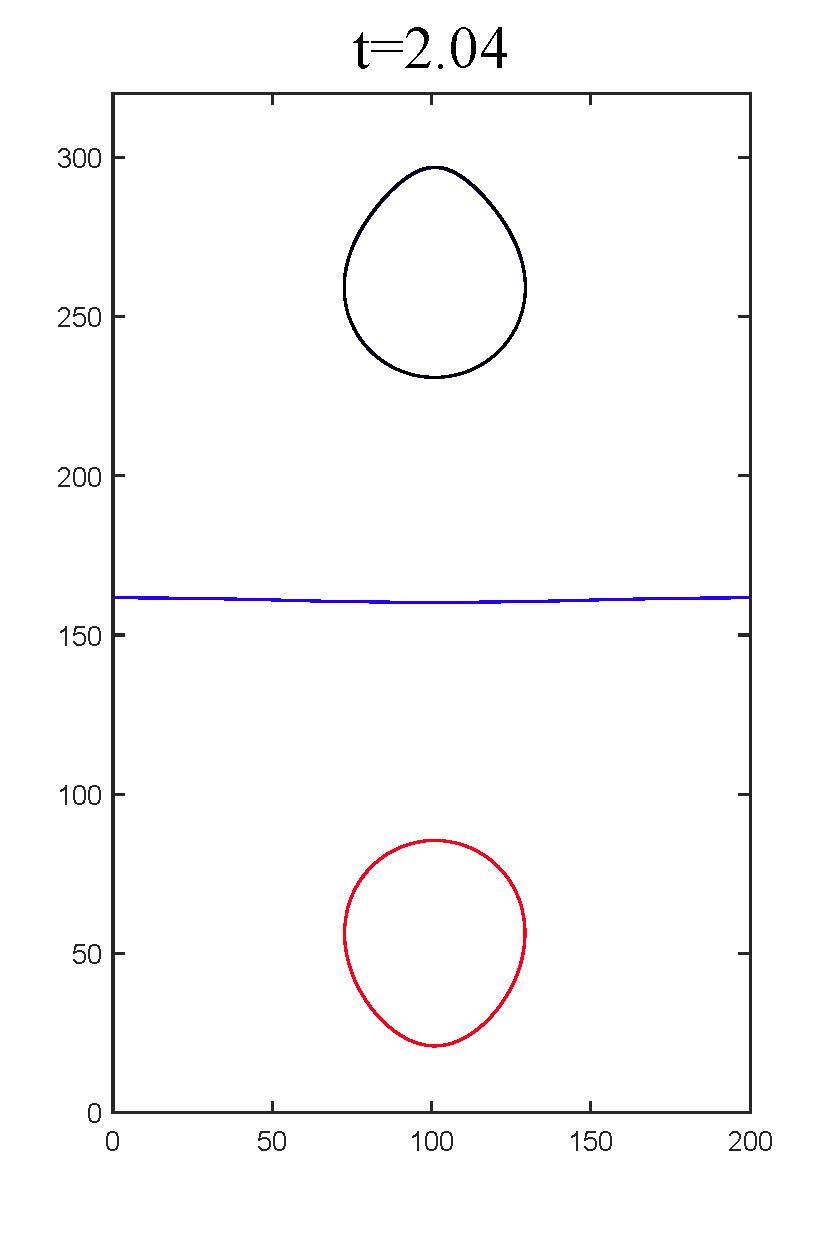}}
\subfigure[]{ \label{fig2_1d}
\includegraphics[width=1.2in]{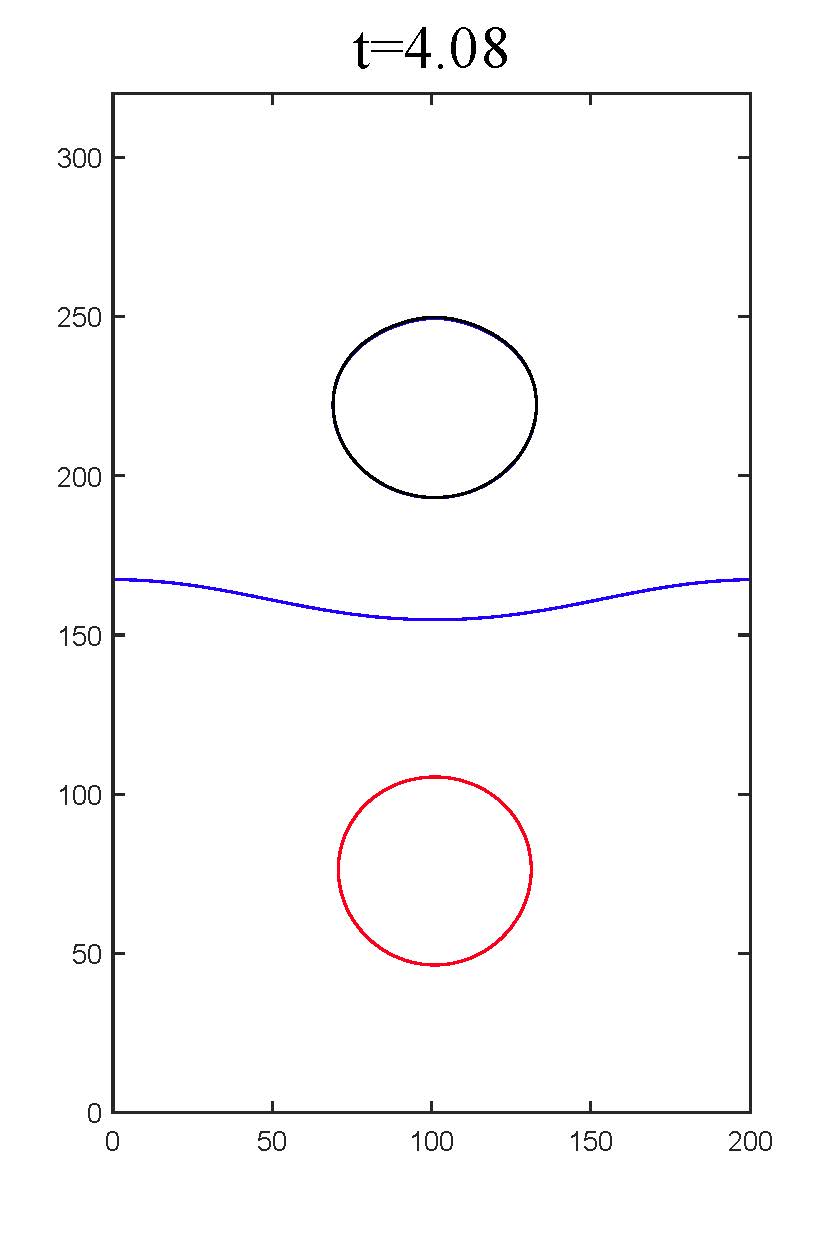}}\\
\subfigure[]{ \label{fig2_1e}
\includegraphics[width=1.2in]{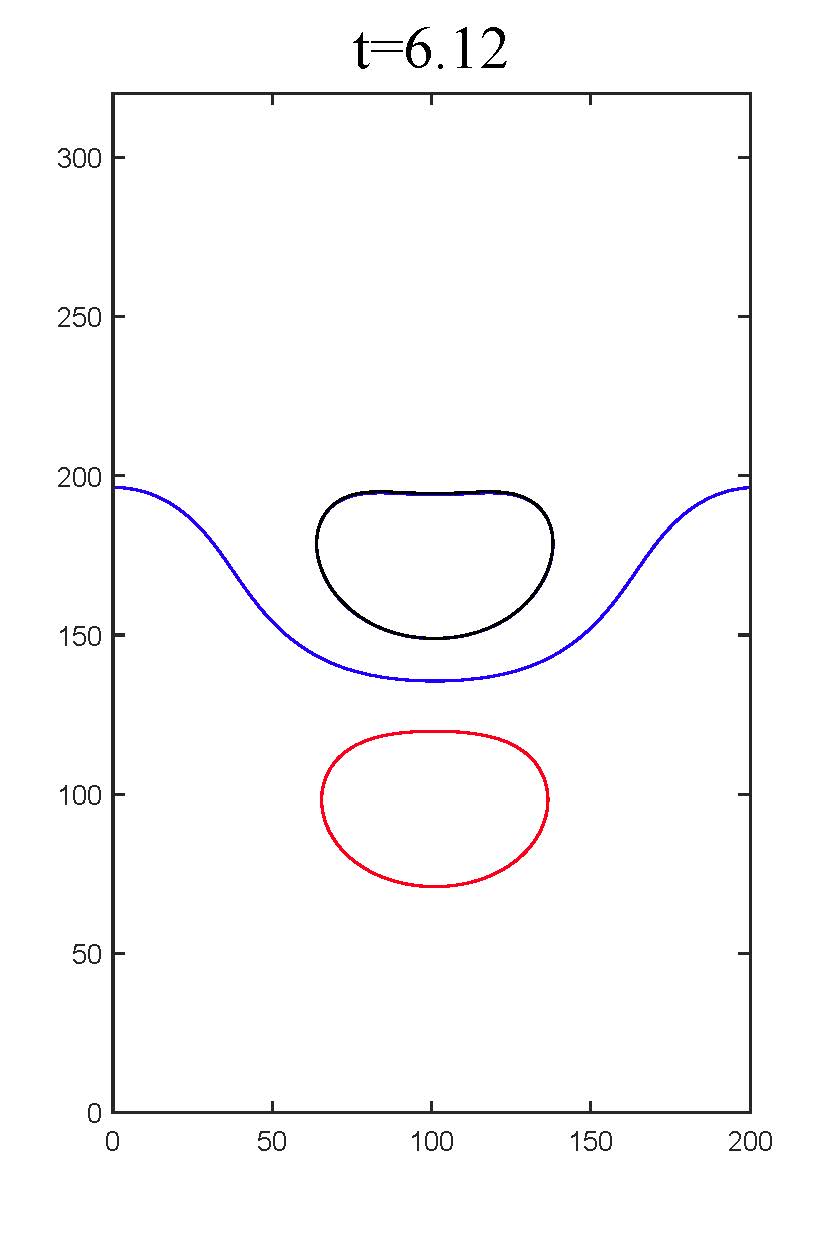}}
\subfigure[]{ \label{fig2_1f}
\includegraphics[width=1.2in]{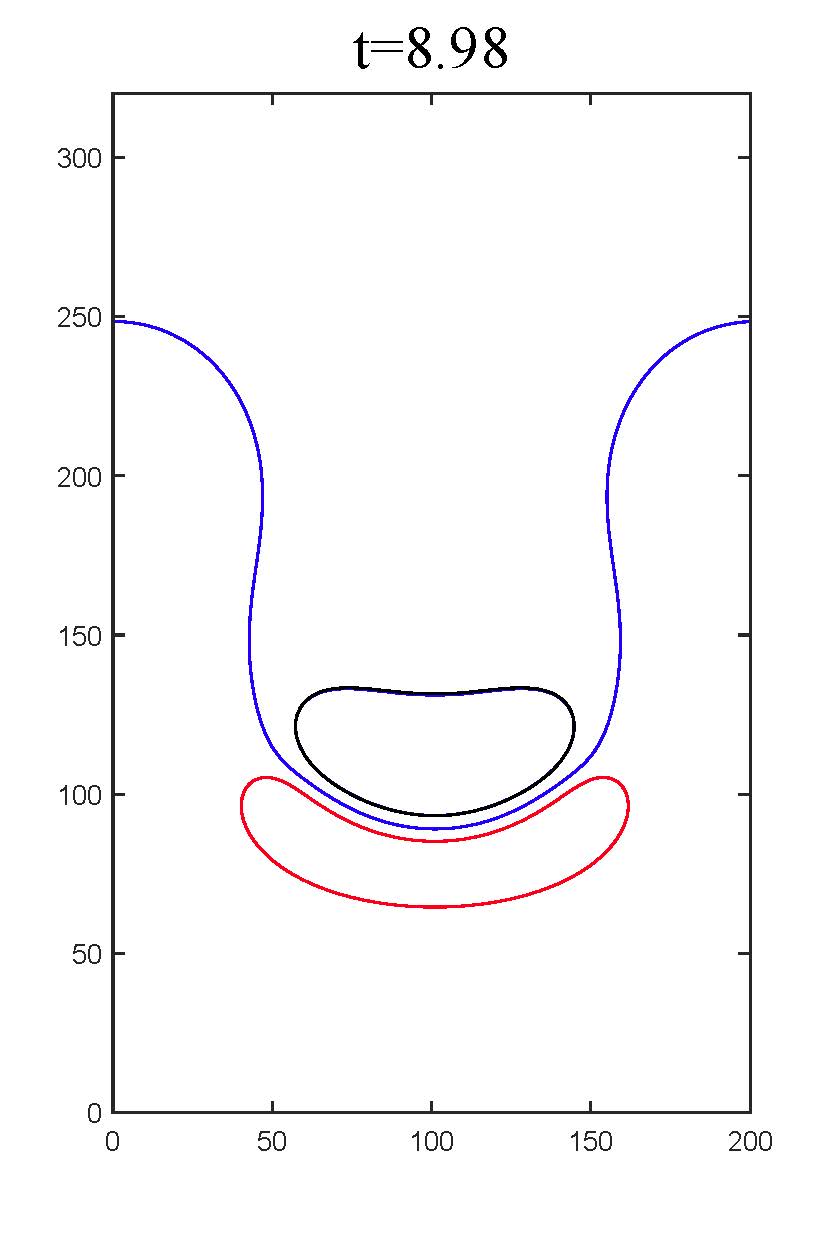}}
\subfigure[]{ \label{fig2_1g}
\includegraphics[width=1.2in]{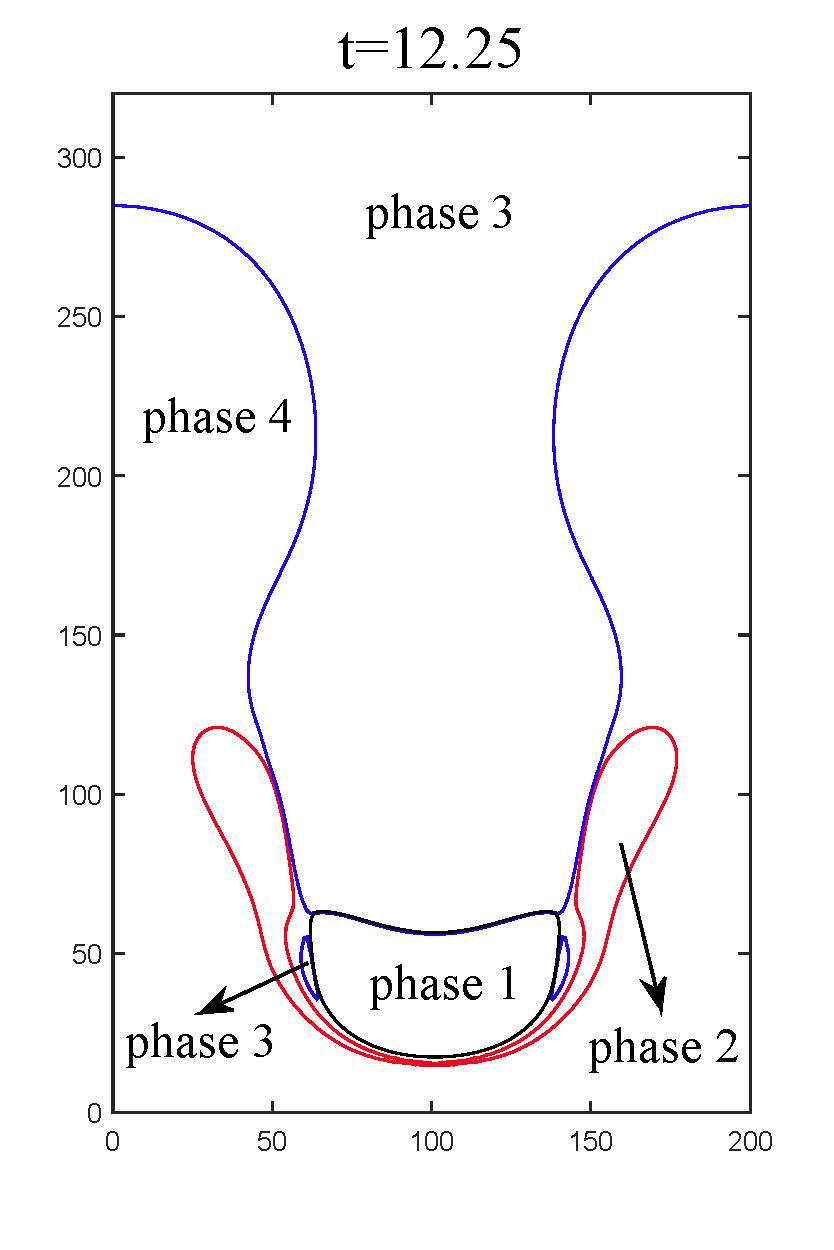}}\\
\caption{(Color online) Evolution of the fluid interfaces with $\sigma_{13}=0.01$. The time has been non-dimensionalized through $t=\frac{t_l}{\sqrt{\frac{D}{g}}}$, where $t_l$ denotes the time in the lattice unit.}
\label{fig2_1}
\end{figure}

In this subsection, we study a dynamic problem involving four fluid phases as another test for validating the developed LB method. This phenomenon is very common in engineering and industrial problems, as well as in nature, such as in spraying processes, in micro-reactors, and in kitchen \cite{Qian2009,Hinterbichler,Dong2014}. The problem configuration is shown in Fig. \ref{fig2_1a}, where the upper part of the domain is filled with phase $3$, and the lower part is filled with phase $4$. One droplet, referred to as phase $1$, is located inside phase $3$. Simultaneously, a bubble of the same size, denoted as phase $2$, is located in phase $4$. Then, the droplet and the bubble start to move from rest under gravity (buoyancy). In this simulation, the computational domain is chosen as $NY\times NX=320\times200$. Periodic boundary conditions are applied in the horizonal direction, and the halfway bounce-back boundary conditions are enforced on the top and bottom walls. The initial volume fractions are given by
\begin{subequations}
\begin{align}
&c_1(x,y)=0.5+0.5\tanh
\frac{R-\sqrt{(x-x_{c_1})^2+(y-y_{c_1})^2}}{\sqrt2 \eta}, \\
&c_2(x,y)=0.5+0.5\tanh
\frac{R-\sqrt{(x-x_{c_2})^2+(y-y_{c_2})^2}}{\sqrt2 \eta}, \\
&c_3(x,y)=\left(0.5+0.5\tanh\frac{y-y_0}{\sqrt2 \eta} \right)\times[1.0-c_1(x,y)]\\
&c_4(x,y)=1.0-c_1(x,y)-c_2(x,y)-c_3(x,y),
\end{align}
\label{eq:N38}
\end{subequations}
where $(x_{c_1},y_{c_1})$ and $(x_{c_2},y_{c_2})$ are the center of two droplets, and $y_0=\frac{1}{2}NY$ is the initial surface position between phase $3$ and phase $4$. Both of phase $1$ and phase $2$ have a diameter of $D=60$. Different values of surface tension $\sigma_{13}$ are investigated. The other parameters are fixed as $\rho_1 : \rho_2  : \rho_3: \rho_4=6:1:3:2$, $\sigma_{12}=\sigma_{14}=\sigma_{23}=\sigma_{24}=\sigma_{34}=0.01 $, $\tau_g=\tau_1=\tau_2=\tau_3=0.8$, $m_1=m_2=m_3=0.1$, $\eta=\sqrt{2}$, $\beta=\sqrt{0.06}$, $g=10^{-5}$, $(x_{c1},y_{c1})=(160,280)$, $(x_{c2},y_{c2})=(160,40)$. Fig. {\ref{fig2_1}} shows the time sequence of evolution of the fluid interfaces with $\sigma_{13}=0.01$, where the solid line represents the contour levels of the volume fraction $c_i=0.5$. As time evolves, the significant deformations of phase $1$ and phase $2$ are generated, and the interface between the fluids $3$ and $4$ is curved due to the effect of the droplets on both sides (Fig. {\ref{fig2_1e}}). Next, the films are formed between phase $1$ and phase $3$ and between phase $2$ and phase $3$, respectively (Fig. {\ref{fig2_1f}}). And due to the disturbance caused by the movement of the droplets and bubbles, the interface of the two continuous phases produces a phenomenon similar to Rayleigh-Taylor (RT) instability. That is to say, the heavy and light fluids (phase $3$ and phase $4$) penetrate into each other and the penetration length increases with time, which then leads to the formation of the spike. As depicted in Fig. {\ref{fig2_1g}}, at $t=12.25$, the interface between different fluids becomes more complicated. The bubble is squeezed into a strip, and two small drops of phase $3$ are trapped between the droplet (phase $1$) and the bubble (phase $3$).

\begin{figure}[ht]
\centering
\subfigure[]{ \label{fig2_2b}
\includegraphics[width=1.2in]{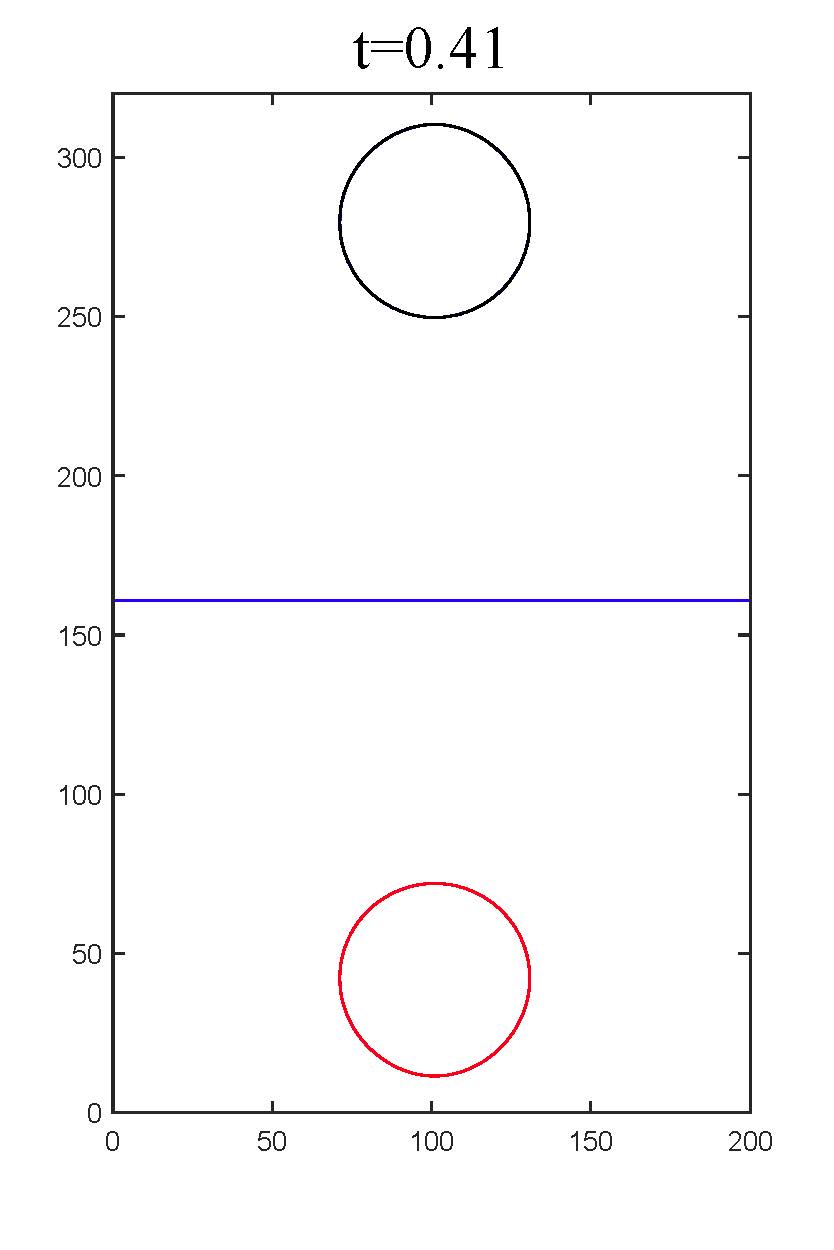}}
\subfigure[]{ \label{fig2_2c}
\includegraphics[width=1.2in]{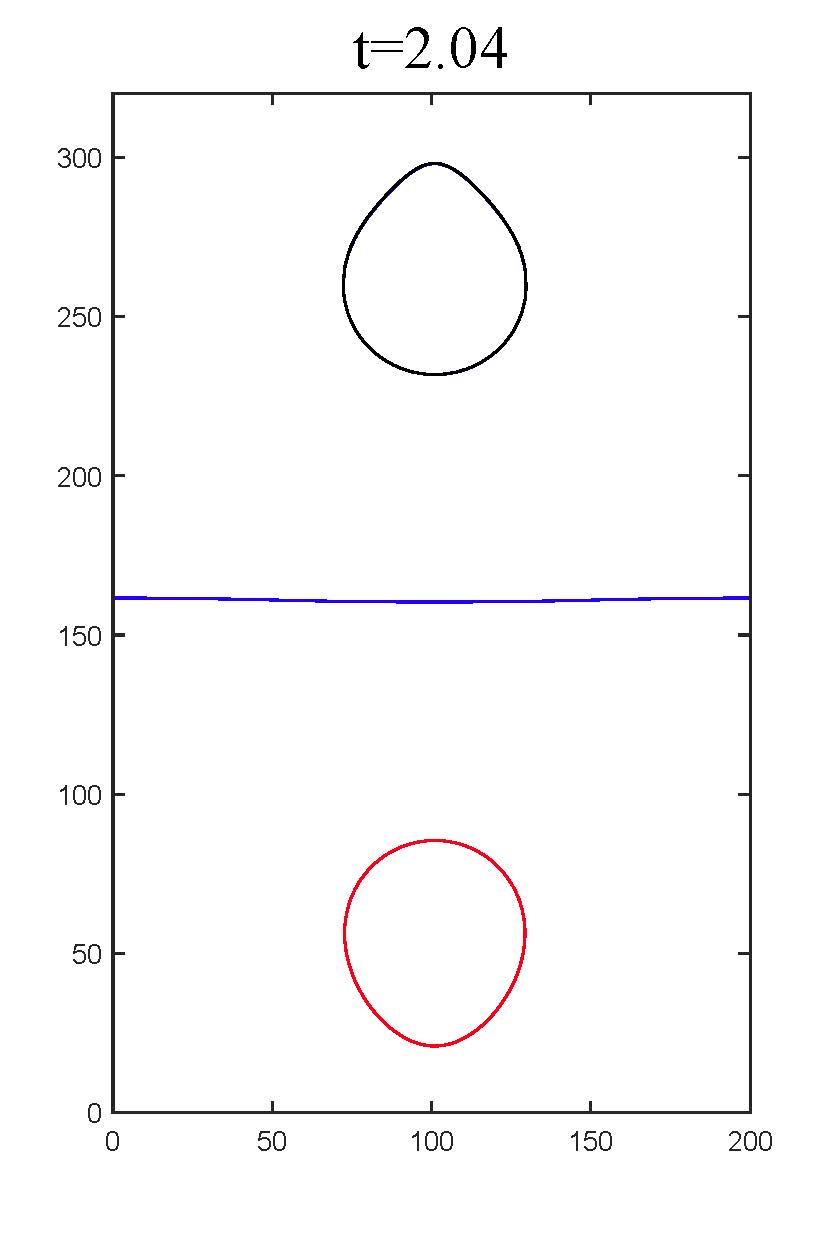}}
\subfigure[]{ \label{fig2_2d}
\includegraphics[width=1.2in]{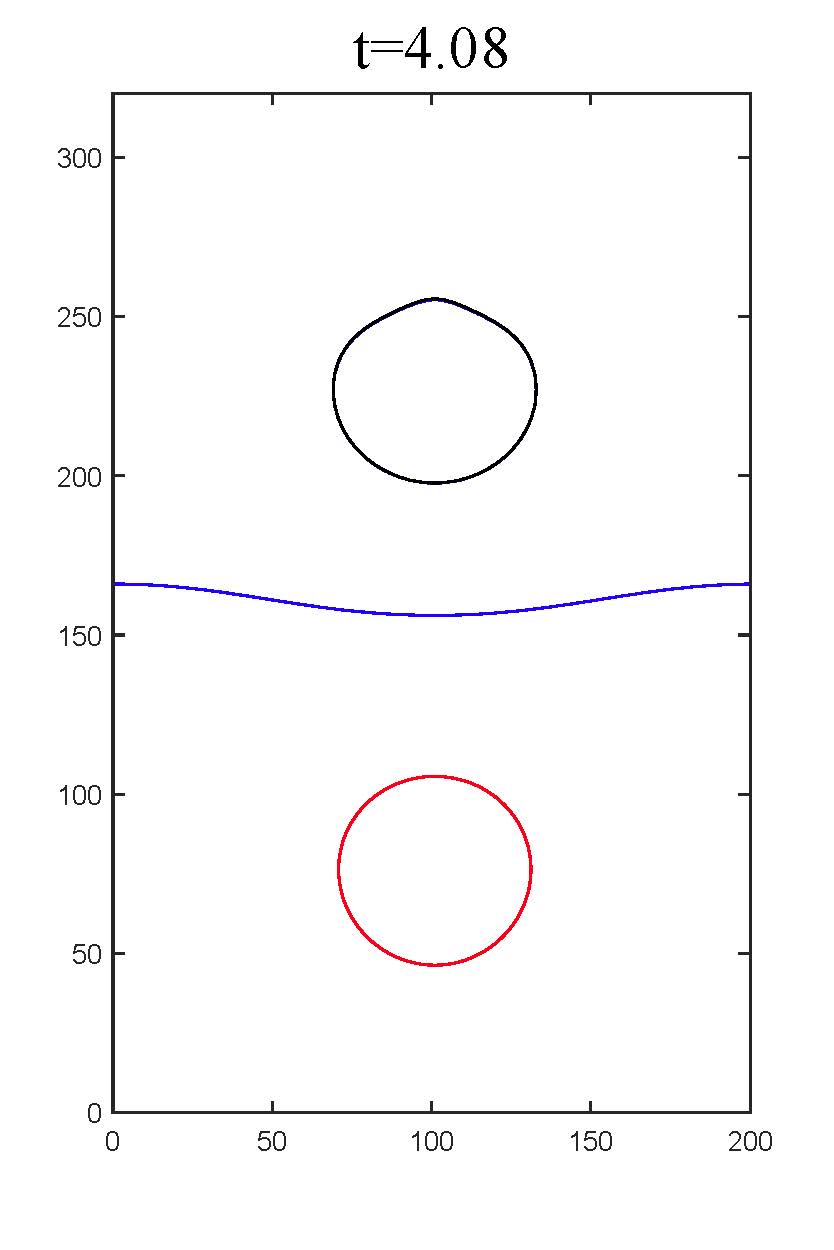}}
\subfigure[]{ \label{fig2_2e}
\includegraphics[width=1.2in]{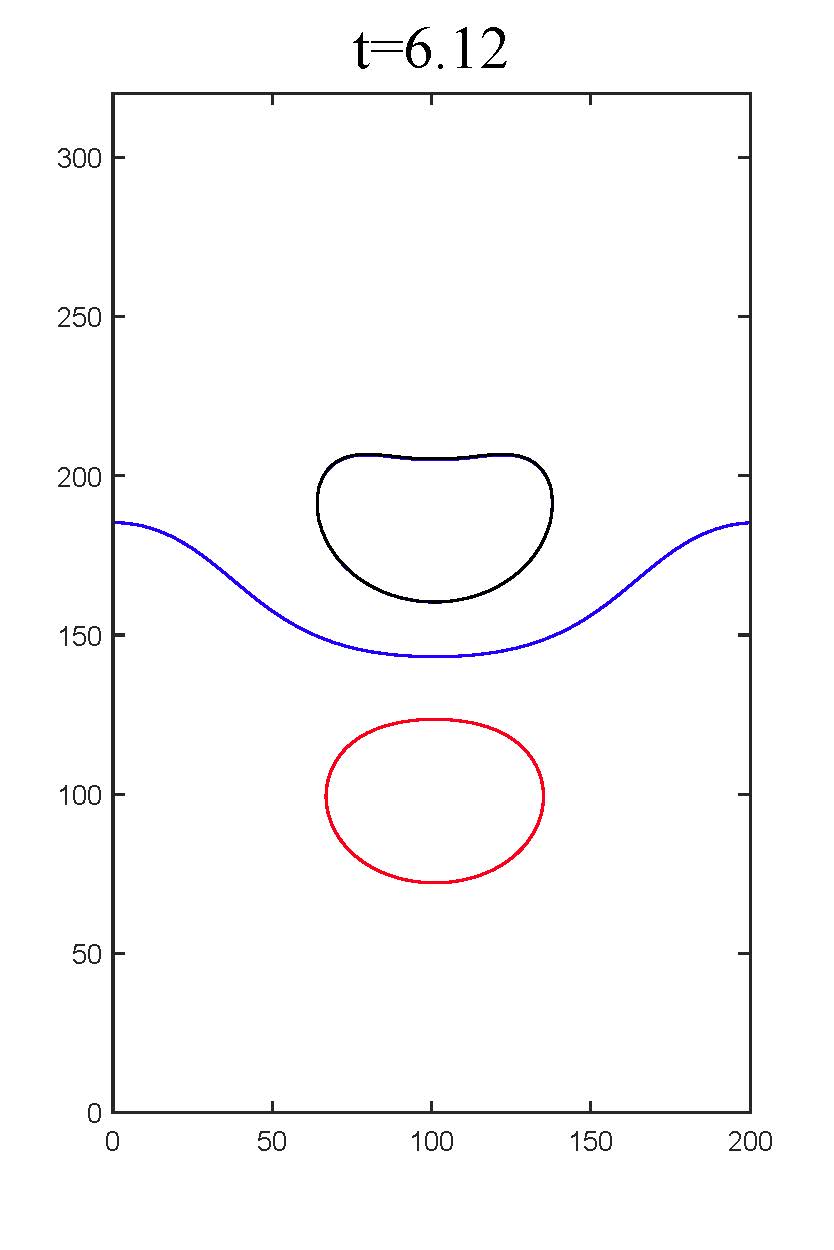}}
\subfigure[]{ \label{fig2_2f}
\includegraphics[width=1.2in]{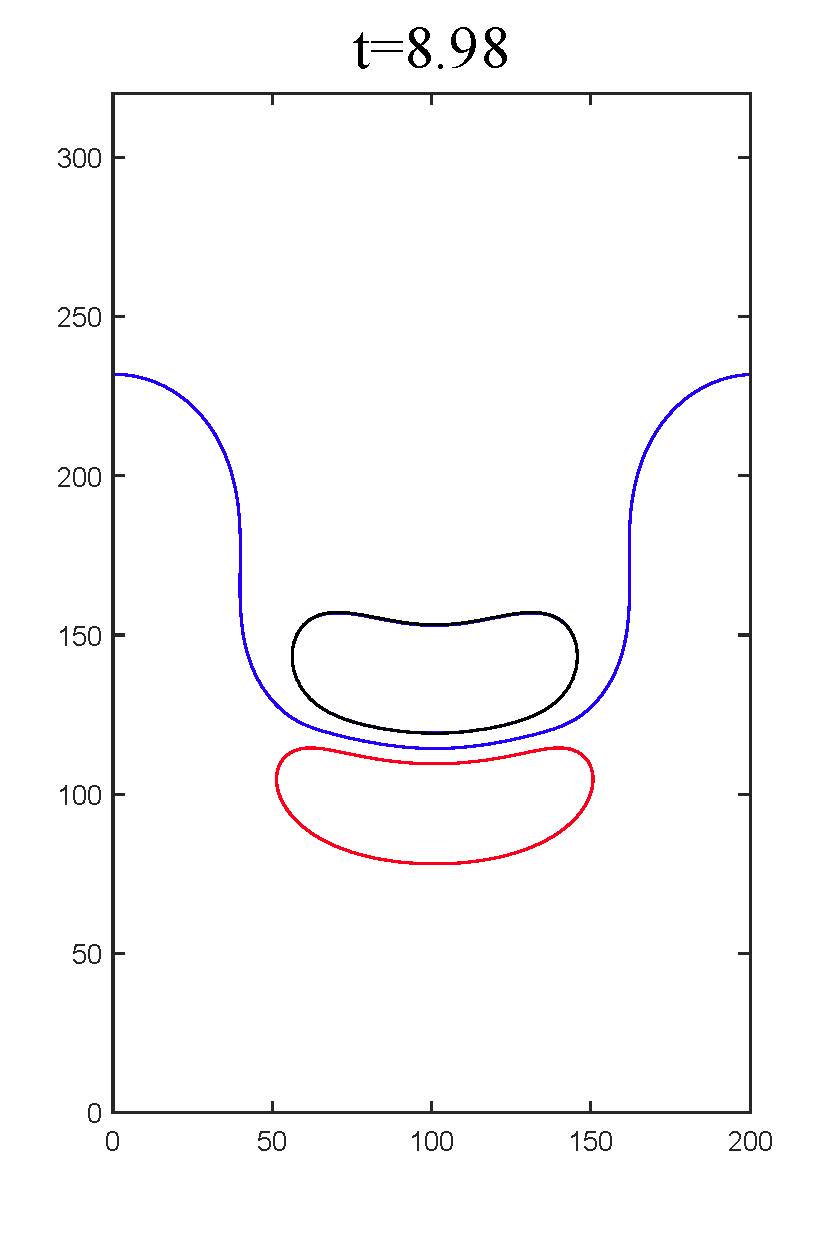}}
\subfigure[]{ \label{fig2_2g}
\includegraphics[width=1.2in]{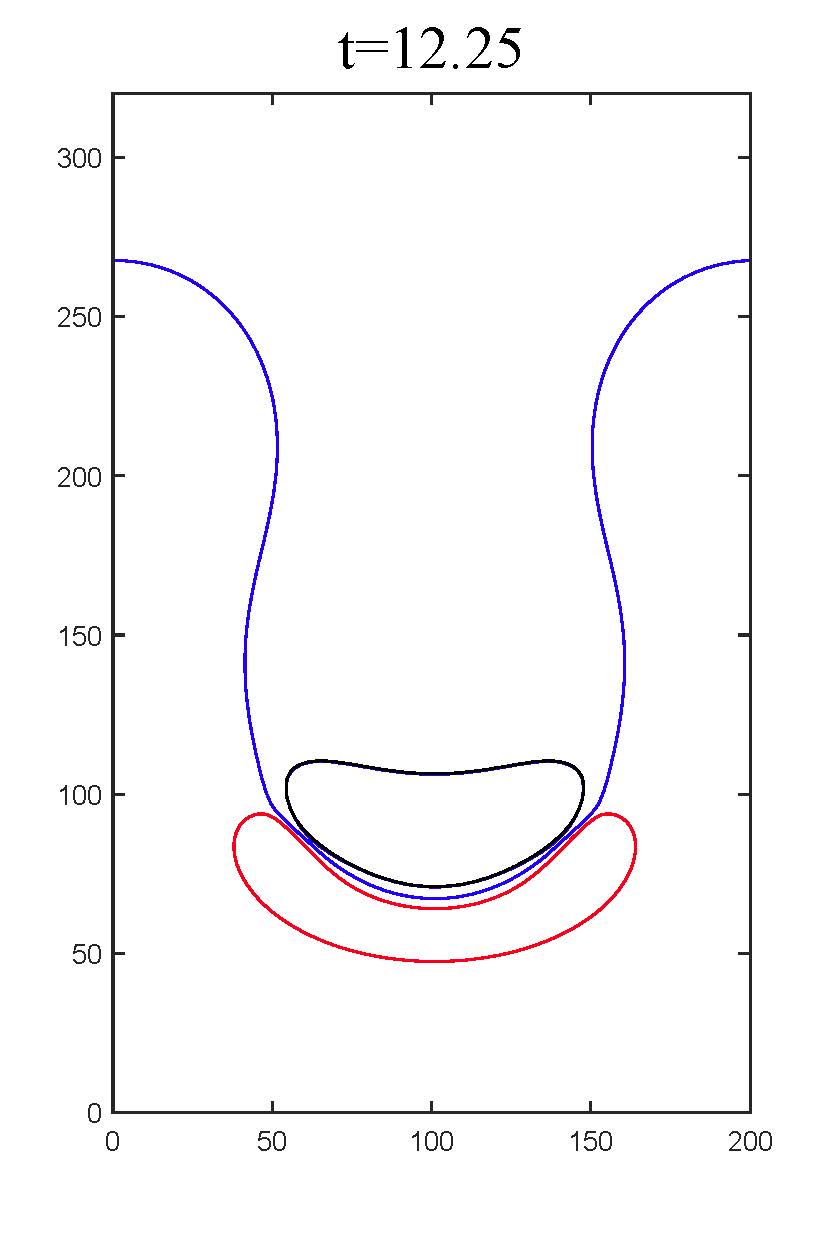}}
\caption{(Color online) Evolution of the fluid interfaces with $\sigma_{13}=0.006$, where the time has been non-dimensionalized through $t=\frac{t_l}{\sqrt{\frac{D}{g}}}$.}
\label{fig2_2}
\end{figure}

\begin{figure}[ht]
\centering
\includegraphics[width=0.7\textwidth]{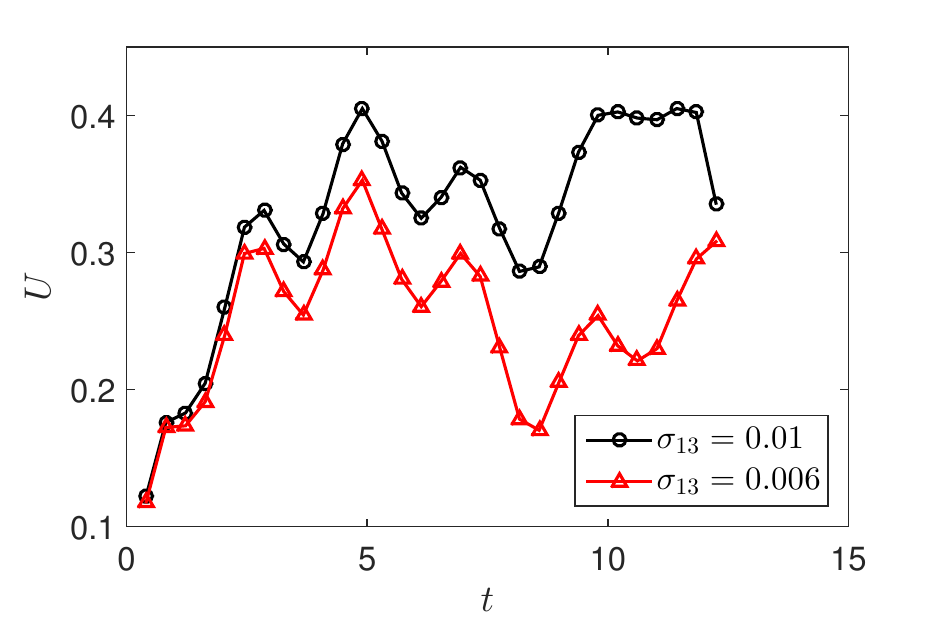}
\caption{(Color online) Comparison of the droplet front velocity with different surface tension $\sigma_{13}$.}
\label{fig2_3}
\end{figure}

We next look into the effect of the surface tension on the fluid interfaces. Here we set the surface tension between phase $1$ and phase $3$ to $\sigma_{13}=0.006$. Fig. {\ref{fig2_2}} depicts the evolution of the fluid interfaces. One can observe that at $t=8.98$, the droplet (phase $1$) in Fig. {\ref{fig2_2f}} produces a larger deformation than in Fig. {\ref{fig2_1f}}, and in Fig. {\ref{fig2_2g}}, phase $3$ doesn't break to form small droplets at $t=12.25$. To explain this, we introduce a dimensionless parameter, $Bo=\frac{\Delta \rho g D^2}{\sigma_{13}}$, which can be used to describe the ratio of droplet gravity to surface tension. In the former case, $Bo=10.8$, while in this case, $Bo=18.0$. As the $Bo$ number increases, the gravity of the droplet is more prominent, and the droplet is more susceptible to deformation (see Fig. {\ref{fig2_2f}}). This deformation causes the droplet to be subjected to greater resistance during the falling process, resulting in a slower drop of the droplet. As shown in Fig. {\ref{fig2_3}}, when $\sigma_{13}=0.006$, the falling speed of the droplet $U$ is significantly smaller than the one when $\sigma_{13}=0.01$, where $U$ has been normalized by $U=\frac{U_{l}}{\sqrt{gD}}$, and $U_l$ is the velocity in the lattice unit. Eventually, a smaller drop velocity will cause the interface dynamics of this four-phase flow problem to be less complicated.

\subsection{Dynamics of droplets and interfaces for five fluid phases}

\begin{figure}[ht]
\centering
\includegraphics[width=0.3\textwidth]{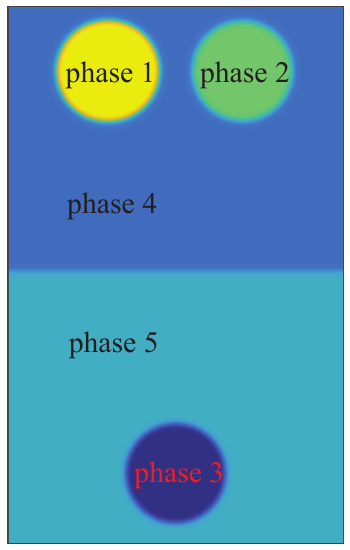}
\caption{(Color online) Initial configuration of dynamics of droplets and interfaces for five fluid phases.}
\label{fig3_1}
\end{figure}

\begin{figure}[ht]
\centering
\subfigure[]{ \label{fig3_2a}
\includegraphics[width=1.2in]{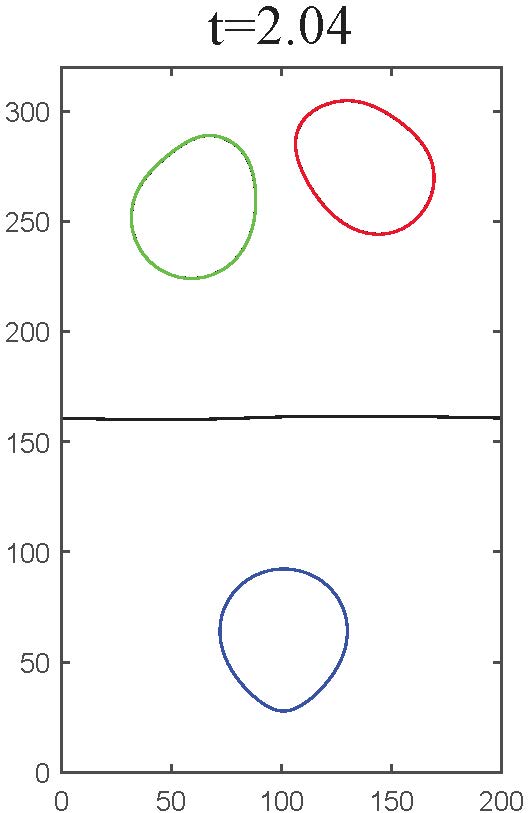}}
\subfigure[]{ \label{fig3_2b}
\includegraphics[width=1.2in]{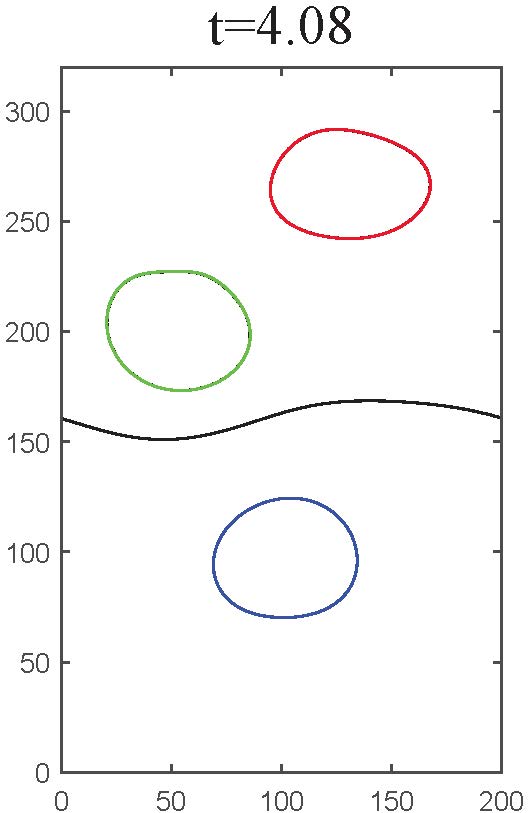}}
\subfigure[]{ \label{fig3_2c}
\includegraphics[width=1.2in]{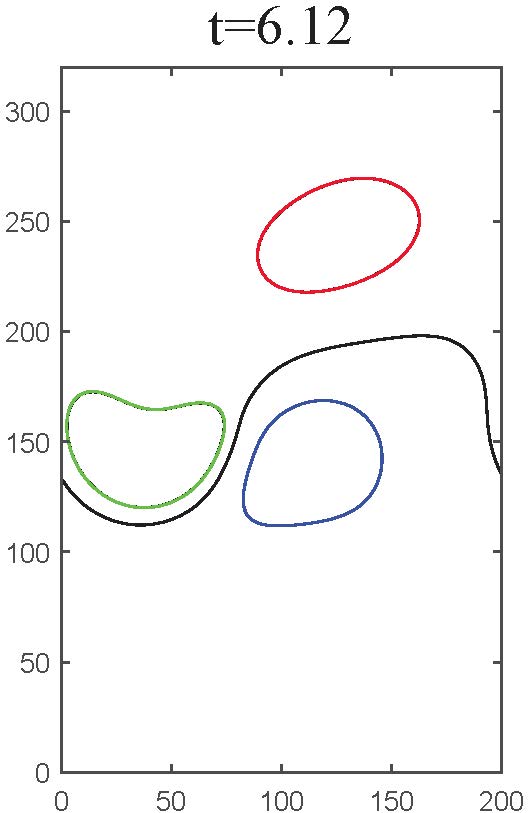}}\\
\subfigure[]{ \label{fig3_2d}
\includegraphics[width=1.2in]{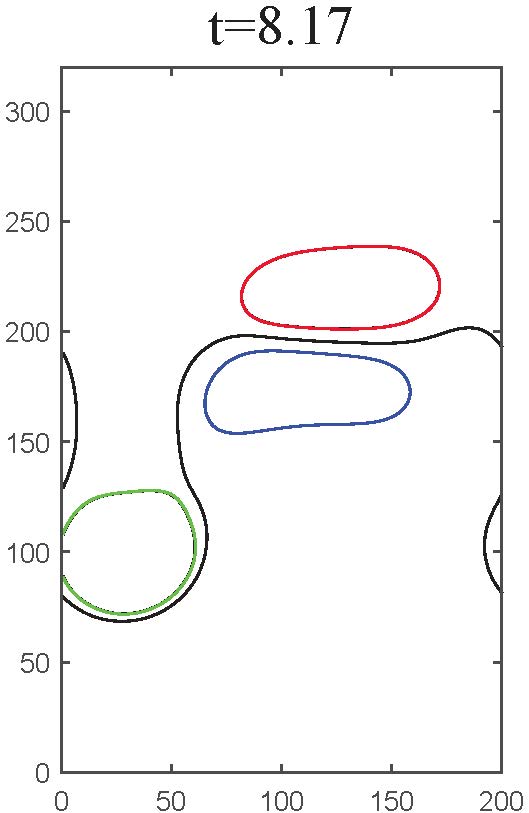}}
\subfigure[]{ \label{fig3_2e}
\includegraphics[width=1.2in]{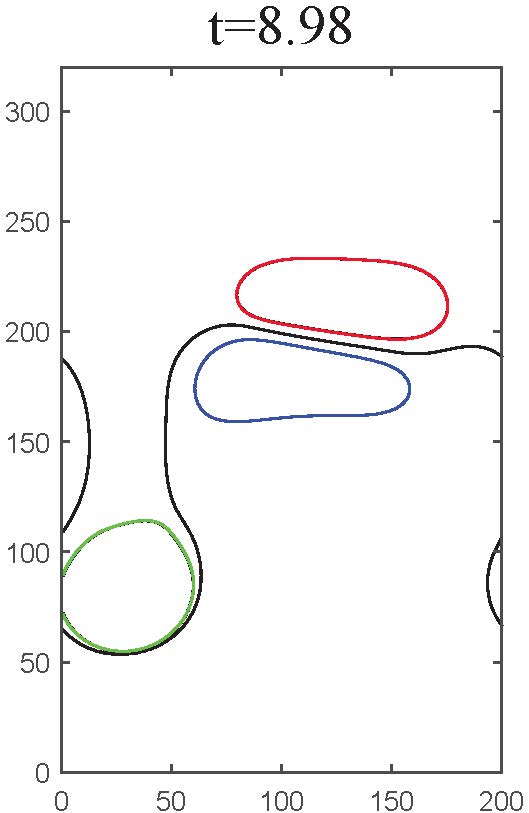}}
\subfigure[]{ \label{fig3_2f}
\includegraphics[width=1.2in]{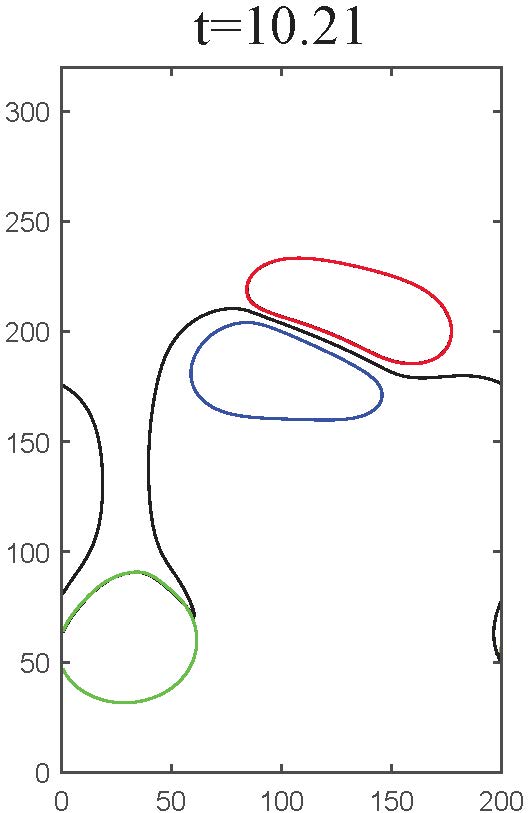}}\\
\subfigure[]{ \label{fig3_2g}
\includegraphics[width=1.2in]{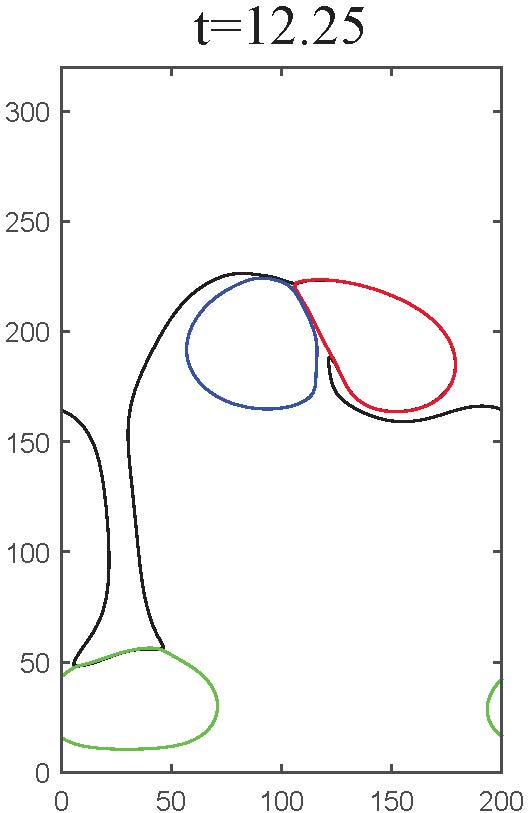}}
\subfigure[]{ \label{fig3_2h}
\includegraphics[width=1.2in]{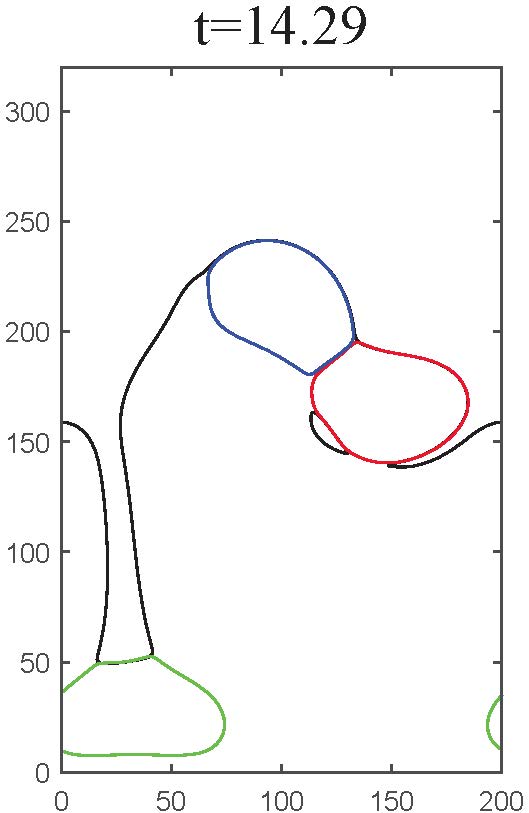}}
\subfigure[]{ \label{fig3_2i}
\includegraphics[width=1.2in]{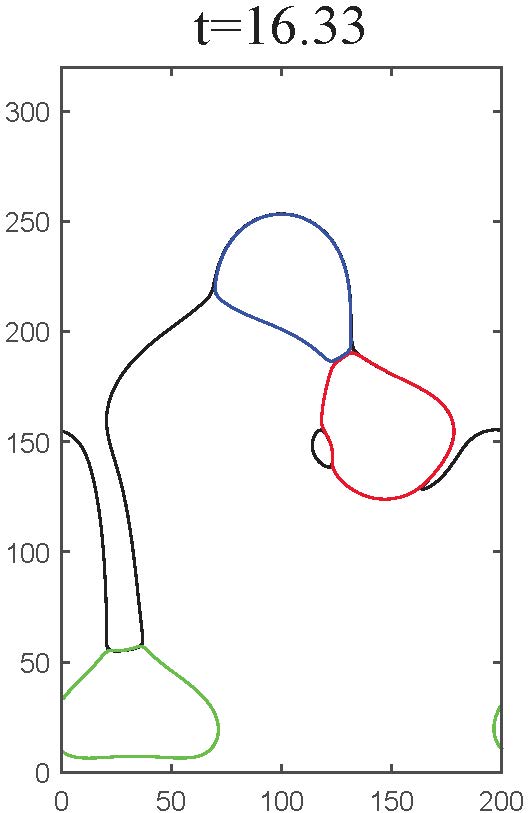}}\\
\caption{(Color online) Evolution of the fluid interfaces for five fluid phases with $\rho_1 : \rho_2  : \rho_3: \rho_4 :\rho_5=6:4:1:2:3$, where the time has been non-dimensionalized through $t=\frac{t_l}{\sqrt{\frac{D}{g}}}$.}
\label{fig3_2}
\end{figure}

In this section, to demonstrate the ability to calculate the dynamic problem involving five fluid phases using the present model, we will consider a complex problem of droplet and interface motion involving five fluid phases. The initial setup of the physical problem is shown in Fig. {\ref{fig3_1}}. Similar to the previous example, the upper part of the domain is filled with phase $4$, and the lower part is filled with phase $5$. While in this simulation, two droplets (i.e., phase $1$ and phase $2$) of the same diameter are located inside phase $4$, and a bubble of the same size, referred to as phase $3$, is located in phase $4$. Then they start moving simultaneously under gravity (buoyancy). The simulation was carried out in a uniform mesh of $NY \times NX=320 \times 200$ with periodic boundary conditions in the horizontal direction and halfway bounce-back boundary conditions in the vertical direction. The profiles of the volume fractions can be initialized by
\begin{subequations}
\begin{align}
&c_1(x,y)=0.5+0.5\tanh
\frac{R-\sqrt{(x-x_{c_1})^2+(y-y_{c_1})^2}}{\sqrt2 \eta}, \\
&c_2(x,y)=0.5+0.5\tanh
\frac{R-\sqrt{(x-x_{c_2})^2+(y-y_{c_2})^2}}{\sqrt2 \eta}, \\
&c_3(x,y)=0.5+0.5\tanh
\frac{R-\sqrt{(x-x_{c_3})^2+(y-y_{c_3})^2}}{\sqrt2 \eta}, \\
&c_5(x,y)=\left(0.5-0.5\tanh\frac{y-y_0}{\sqrt2 \eta} \right)\times[1.0-c_3(x,y)]\\
&c_4(x,y)=1.0-c_1(x,y)-c_2(x,y)-c_3(x,y)-c_5(x,y),
\end{align}
\label{eq:N39}
\end{subequations}
where $R$ is the radius of the circle drop for phase $1$, $2$ and $3$. $(x_{c_1},y_{c_1})$, $(x_{c_2},y_{c_2})$ and $(x_{c_3},y_{c_3})$ are the center of two droplets and the bubble, and $y_0=\frac{1}{2}NY$ is the initial surface position between phase $4$ and phase $5$. In this test, $R=30$, $(x_{c_1},y_{c_1})=(60,280)$, $(x_{c_2},y_{c_2})=(140,280)$, $(x_{c_3},y_{c_3})=(100,40)$. This complex five-phase problem has higher requirements for the selection of difference schemes. To further reduce the parasitic currents and guarantee the stability of present LB model, here we adopt the following isotropic mixed difference scheme \cite{Zu}
\begin{equation}
\nabla \zeta(\mathbf{x},t)=\sum_{i \ne 0}\frac{ \omega_{i}
\mathbf{c}_{i} [\zeta(\mathbf{x}+2\mathbf{c}_{i} \Delta t,t)+5\zeta(\mathbf{x}+\mathbf{c}_{i} \Delta t,t)-3\zeta(\mathbf{x},t)-\zeta(\mathbf{x}-\mathbf{c}_{i} \Delta t,t)]}{4c_s^2
\Delta t}. \label{eq:N40}
\end{equation}
Although this scheme may not conserve the mass and momentum precisely due to the discretization errors \cite{Guo2011}, due to the better stability of this scheme, we will adopt this difference scheme for five-phase flows. Actually, the mass change can be controlled within $0.1\%$ after numerical tests, which can be ignored in the simulation. The other physical parameters in our simulations are set to be $\rho_1 : \rho_2  : \rho_3: \rho_4 :\rho_5=6:4:1:2:3$, $\sigma_{ij}=0.01$, $(1 \leq i \neq j \leq 5)$, $\tau_g=\tau_1=\tau_2=\tau_3=\tau_4=0.8$, $m_1=m_2=m_3=m_4=0.1$, $\eta=\sqrt{2}$, $\beta=\sqrt{0.06}$, $g=10^{-5}$. Fig. {\ref{fig3_2}} depicts the evolution of the fluid interfaces for five fluid phases. It is found that after releasing at $t=0$, the two droplets of phase $1$ and phase $2$ fall freely in phase $4$, while the bubble of phase $3$ rise in phase $5$. Due to the higher density, phase $1$ falls faster than phase $2$, and the interface between phase $4$ and phase $5$ begins to deform due to the movement of the droplets [see Fig. \ref{fig3_2b}]. Fig. \ref{fig3_2c} shows that the droplets and the bubble approach the surface between phase $4$ and $5$. As time goes by, phase $1$ continues to move downward, causing phase $4$ to penetrate into phase $5$ to form a spike [see Figs. \ref{fig3_2d} and \ref{fig3_2e}]. At the same time, another droplet (phase $2$) and the bubble (phase $3$) are about to impact the interface between phase $4$ and phase $5$. At $t=10.21$, the droplet penetrates into phase $5$ from phase $4$. Meanwhile, phase $2$ and phase $3$ gradually become unstable, and the interface between them begins to incline to the right. Subsequently, phase $2$ continues to move upward due to the buoyancy, phase $3$ moves downward around phase $2$, and causes a small drop of phase $4$ to be trapped between phase $2$ and phase $5$.

\begin{figure}[ht]
\centering
\subfigure[]{ \label{fig3_3_1}
\includegraphics[width=1.1in]{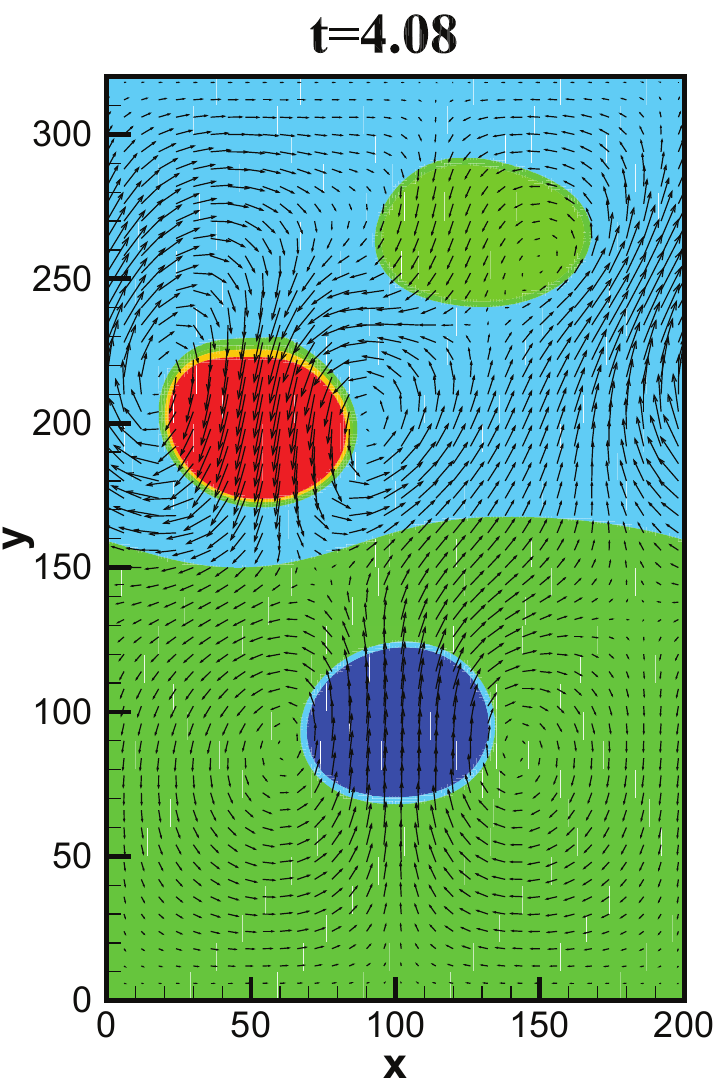}}
\subfigure[]{ \label{fig3_3_2}
\includegraphics[width=1.1in]{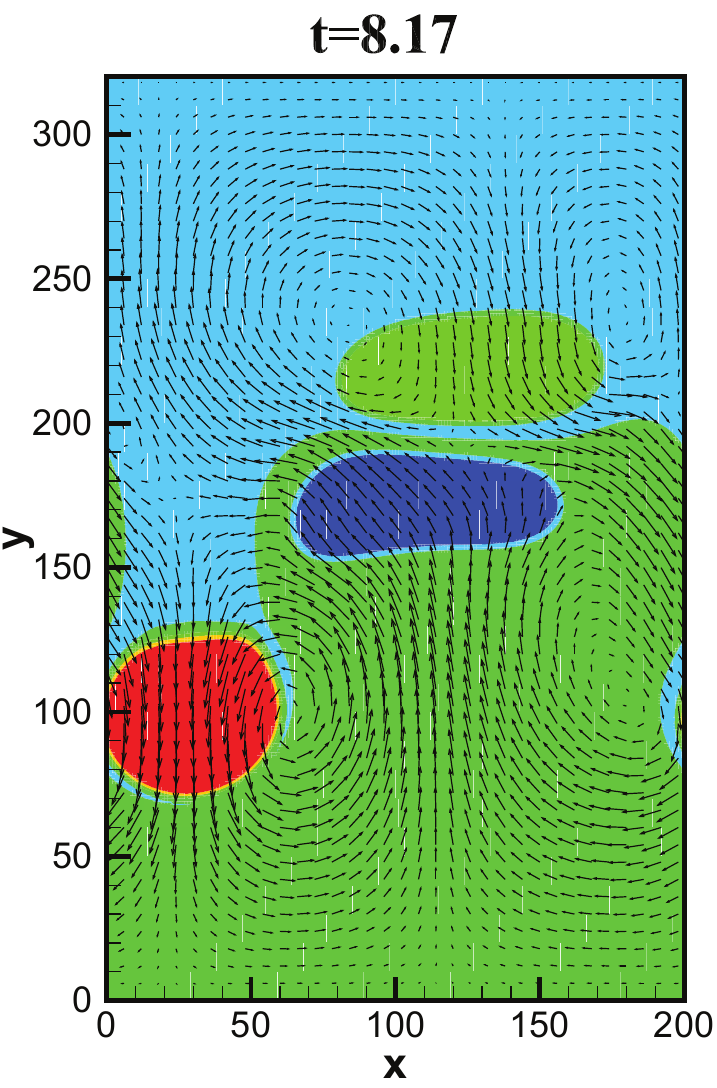}}
\subfigure[]{ \label{fig3_3_3}
\includegraphics[width=1.1in]{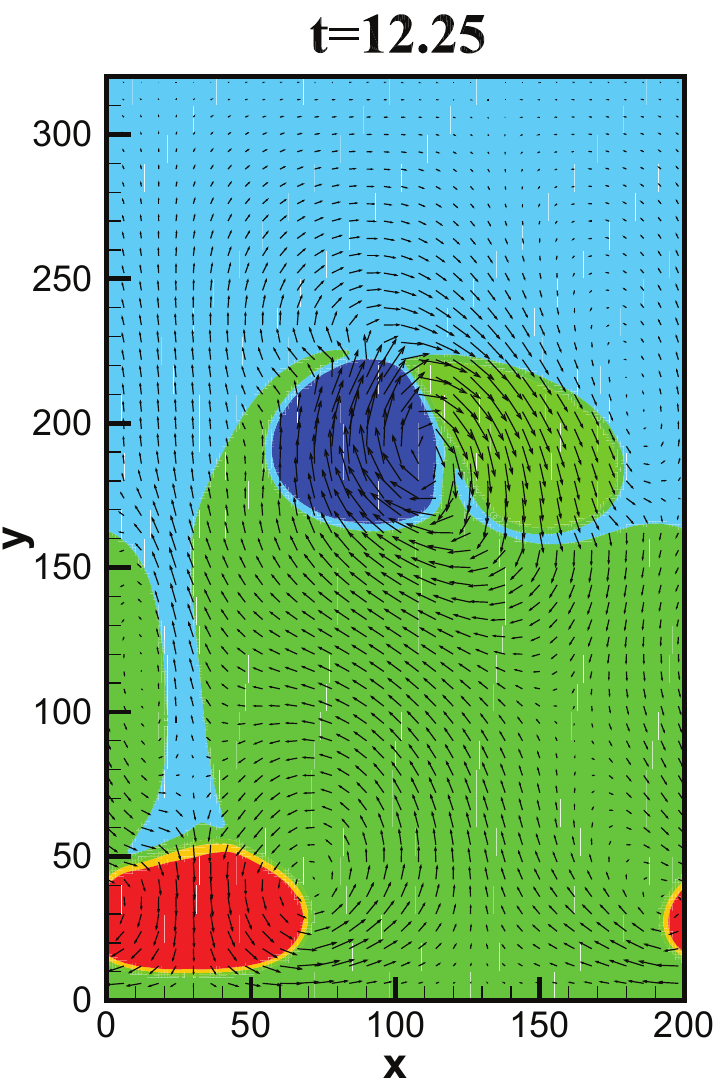}}
\subfigure[]{ \label{fig3_3_4}
\includegraphics[width=1.1in]{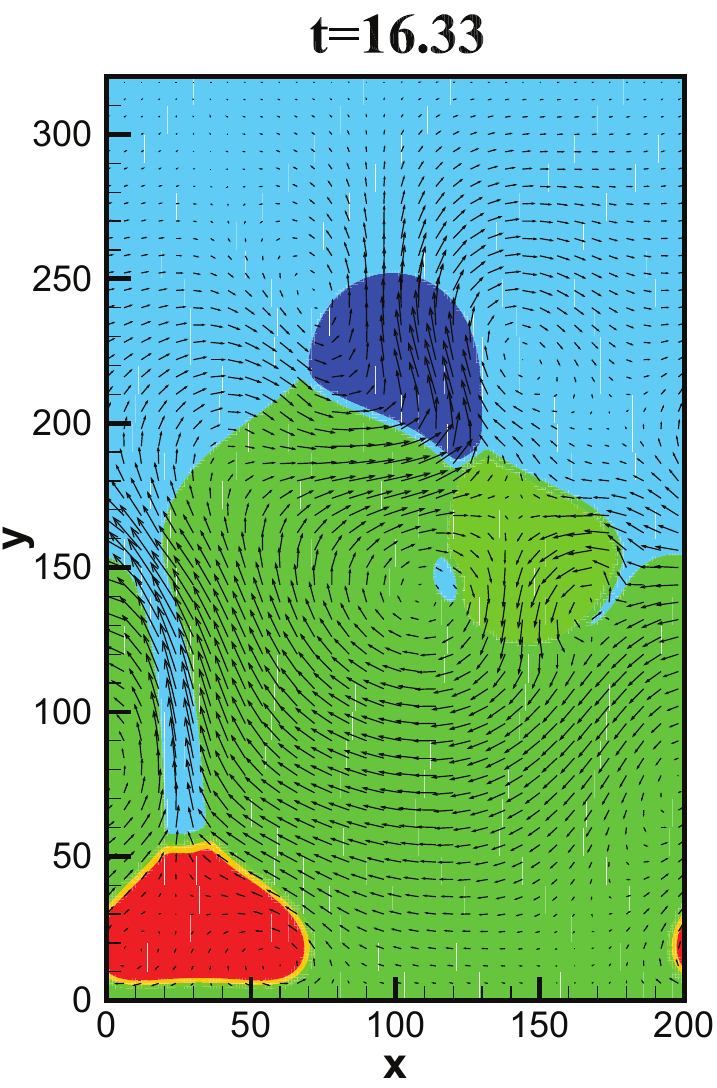}}
\caption{(Color online) Temporal sequence of snapshots of velocity fields for five fluid phase with $\rho_1 : \rho_2  : \rho_3: \rho_4 :\rho_5=6:4:1:2:3$, where the time has been non-dimensionalized through $t=\frac{t_l}{\sqrt{\frac{D}{g}}}$.}
\label{fig3_3}
\end{figure}

Fig. {\ref{fig3_3}} depicts the temporal sequence of snapshots of velocity fields at a few special moments. At $t=4.08$, the falling droplet of phase $1$ and the bubble (phase $3$) have induced a distinct velocity field. Subsequently, phase $1$ moves downward, penetrating into phase $5$ with a pair of vortices [see Fig. {\ref{fig3_3_2}}]. After that, the velocity field of phase $1$ gradually subsides [Figs. {\ref{fig3_3_3}} and {\ref{fig3_3_4}}]. On the other hand, phase $2$ and phase $3$ produce a clockwise velocity field after contact at $t=12.25$, eventually causing phase $2$ and Phase $3$ to bypass each other and then move in the opposite direction.

\begin{figure}[ht]
\centering
\subfigure[]{ \label{fig3_2a1}
\includegraphics[width=1.2in]{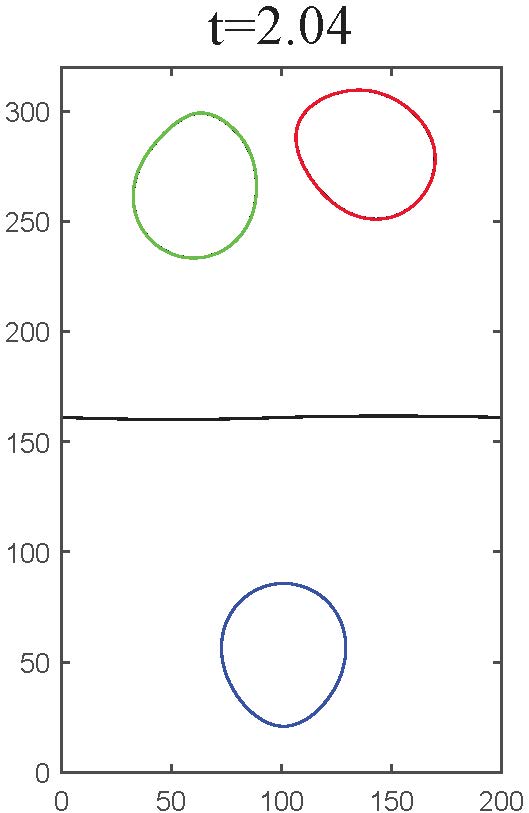}}
\subfigure[]{ \label{fig3_2b1}
\includegraphics[width=1.2in]{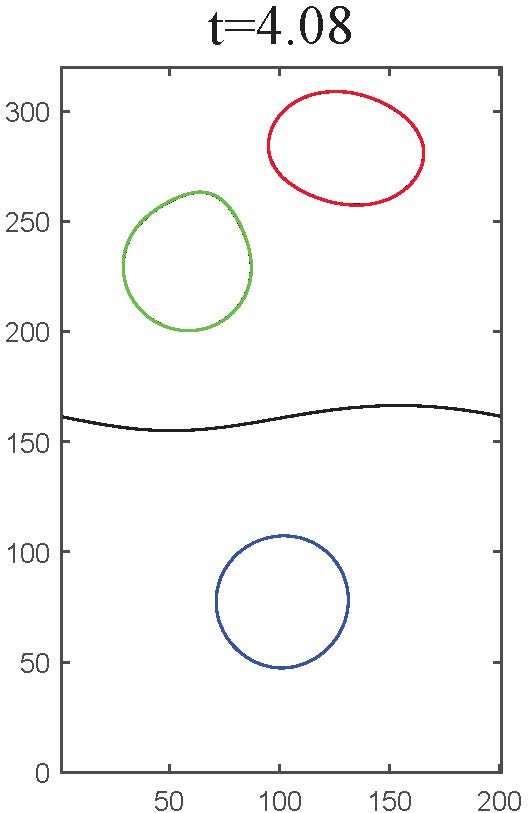}}
\subfigure[]{ \label{fig3_2c1}
\includegraphics[width=1.2in]{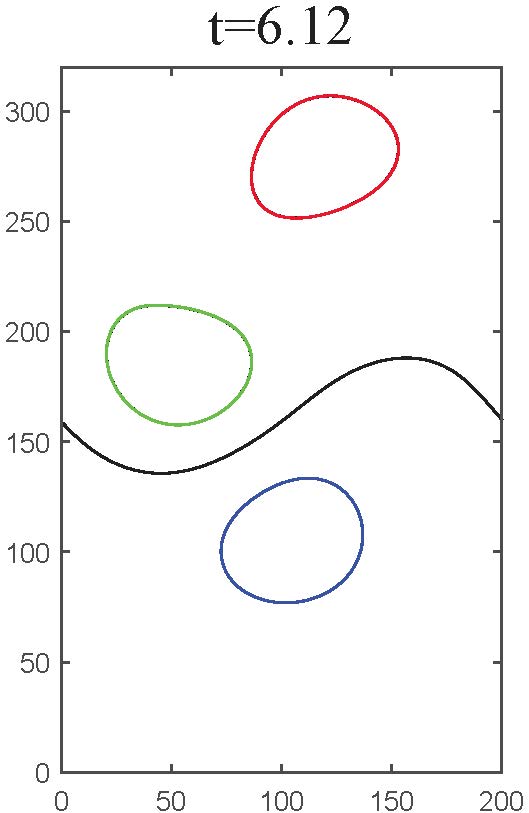}}
\subfigure[]{ \label{fig3_2d1}
\includegraphics[width=1.2in]{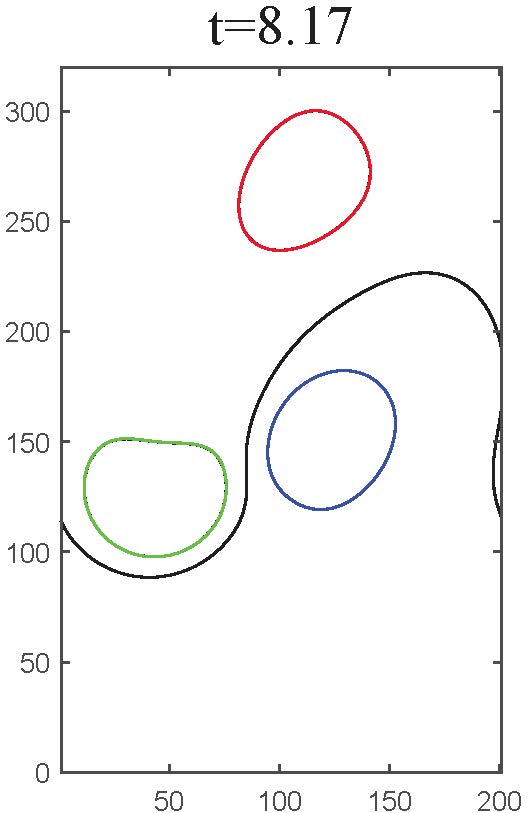}}
\subfigure[]{ \label{fig3_2e1}
\includegraphics[width=1.2in]{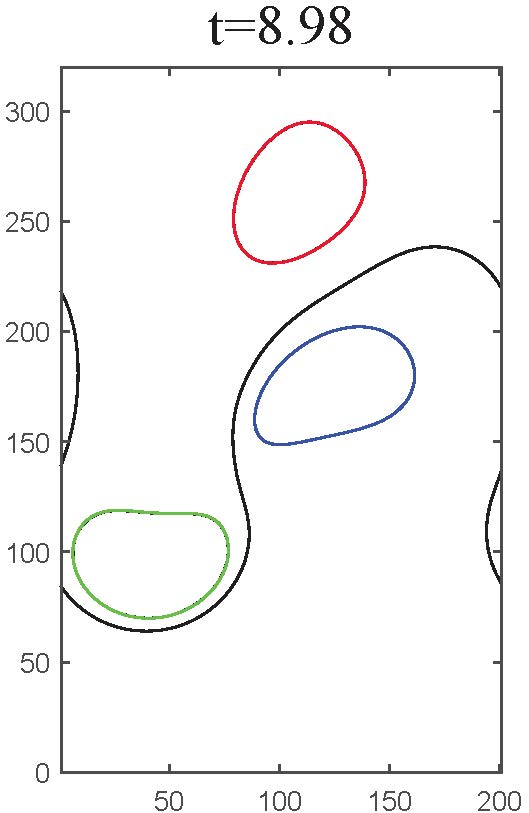}}
\subfigure[]{ \label{fig3_2f1}
\includegraphics[width=1.2in]{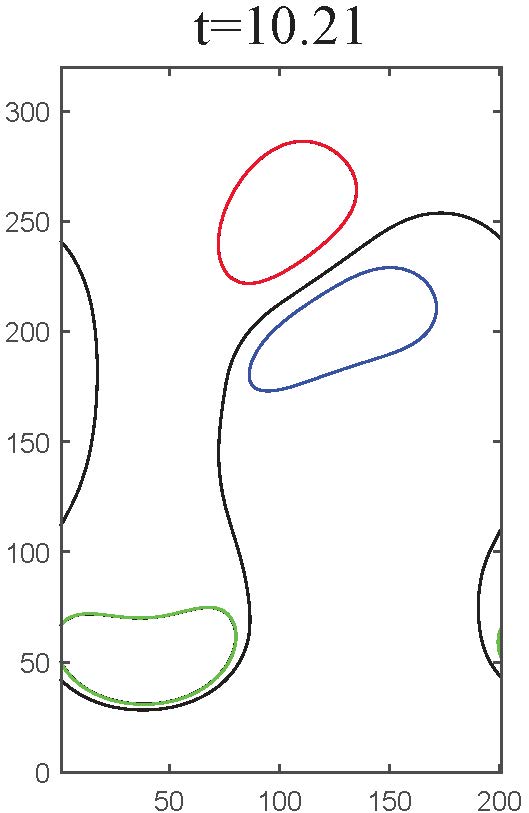}}
\subfigure[]{ \label{fig3_2g1}
\includegraphics[width=1.2in]{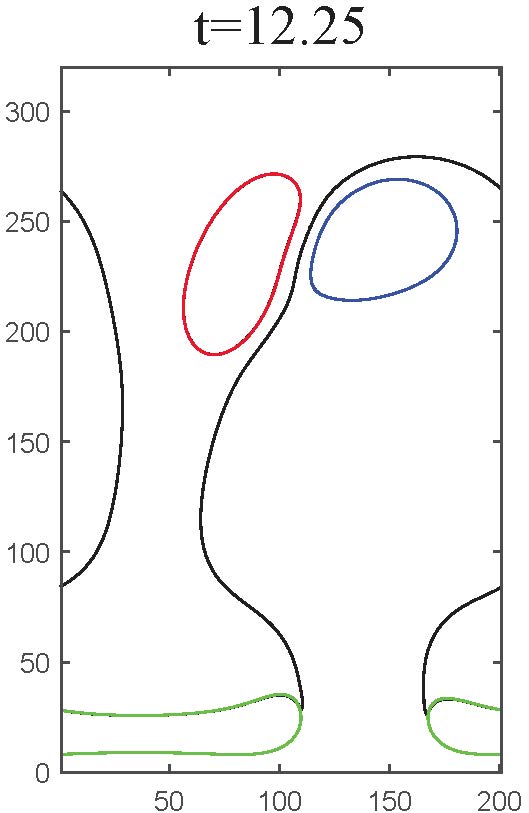}}
\subfigure[]{ \label{fig3_2h1}
\includegraphics[width=1.2in]{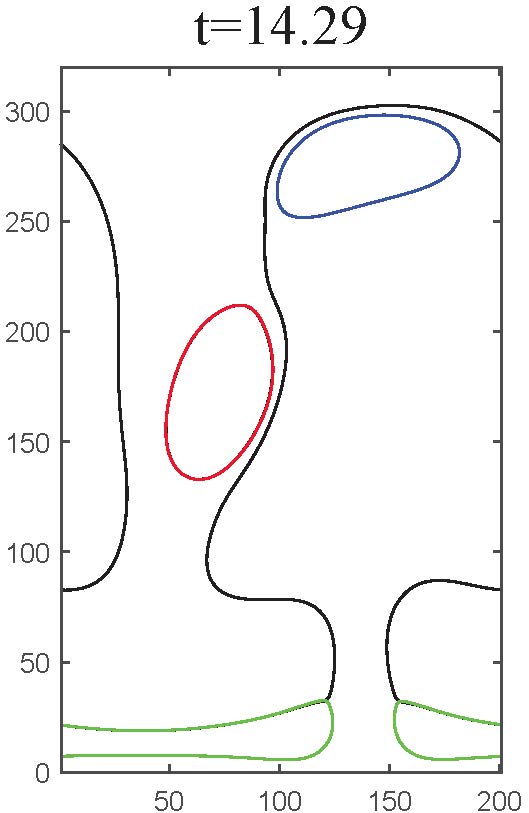}}
\subfigure[]{ \label{fig3_2i1}
\includegraphics[width=1.2in]{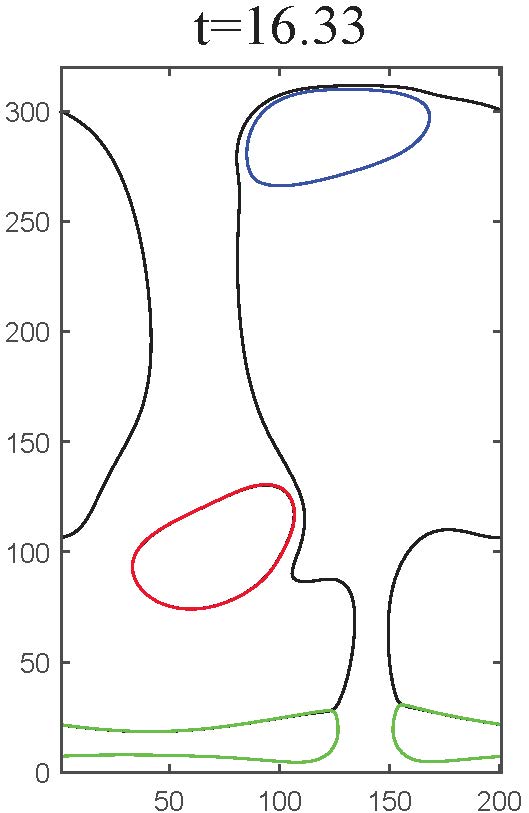}}
\caption{(Color online) Evolution of the fluid interfaces for five fluid phases with $\rho_1 : \rho_2  : \rho_3: \rho_4 :\rho_5=6:4:1:3:2$, where the time has been non-dimensionalized through $t=\frac{t_l}{\sqrt{\frac{D}{g}}}$.}
\label{fig3_4}
\end{figure}

\begin{figure}[ht]
\centering
\subfigure[]{ \label{fig3_4_1}
\includegraphics[width=1.1in]{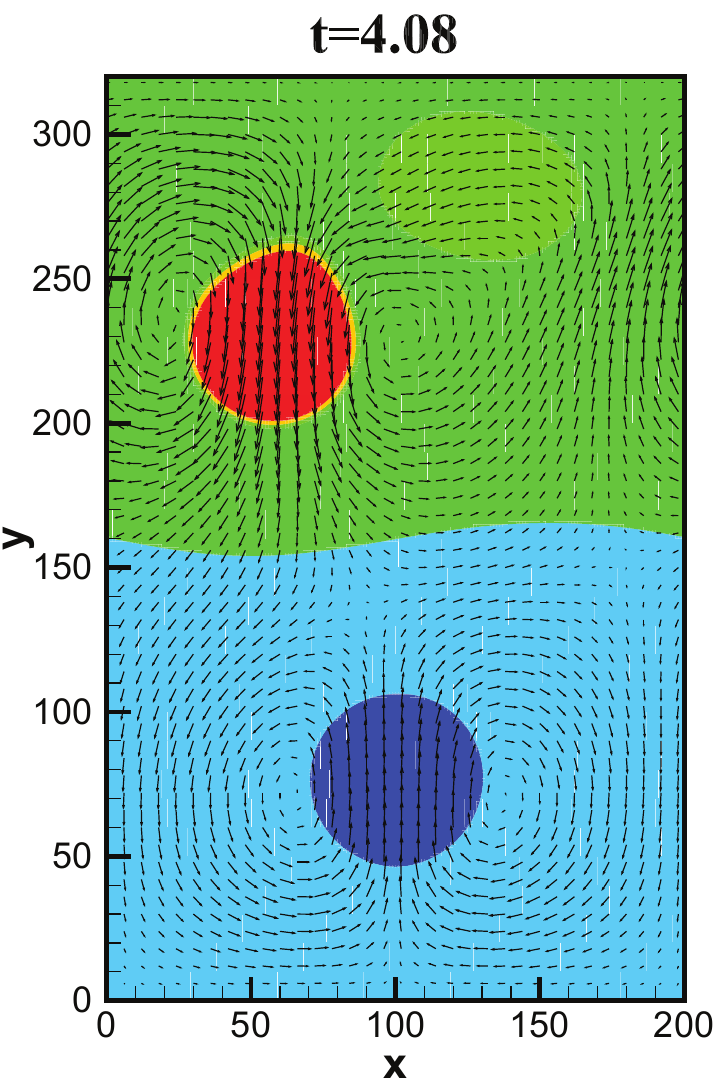}}
\subfigure[]{ \label{fig3_4_2}
\includegraphics[width=1.1in]{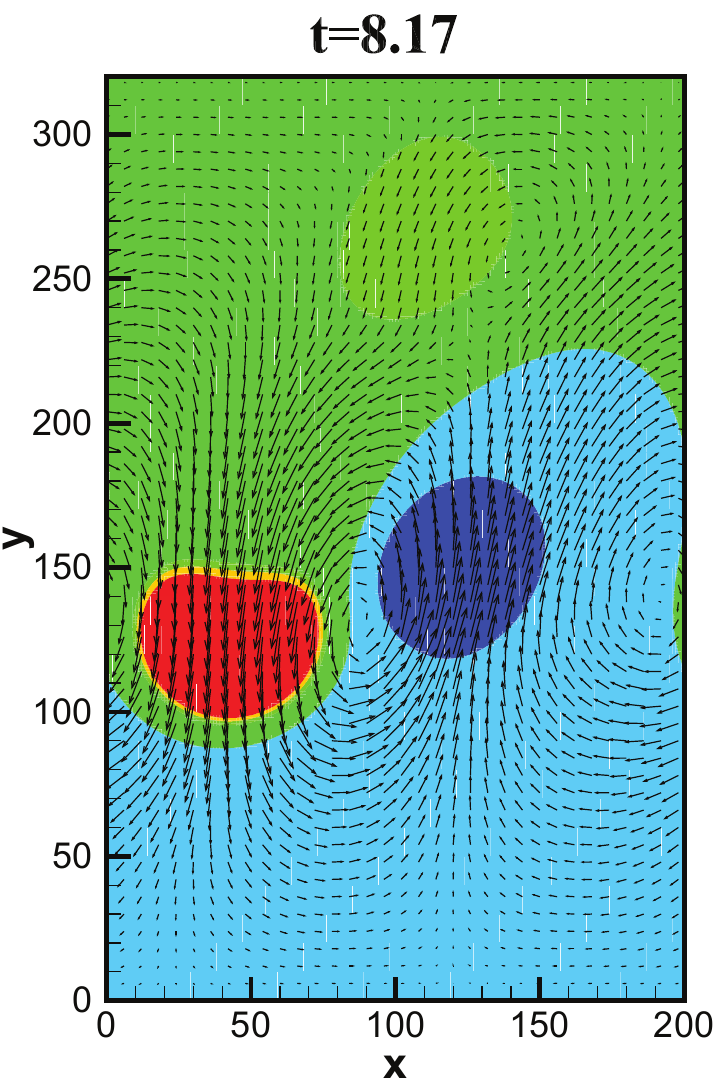}}
\subfigure[]{ \label{fig3_4_3}
\includegraphics[width=1.1in]{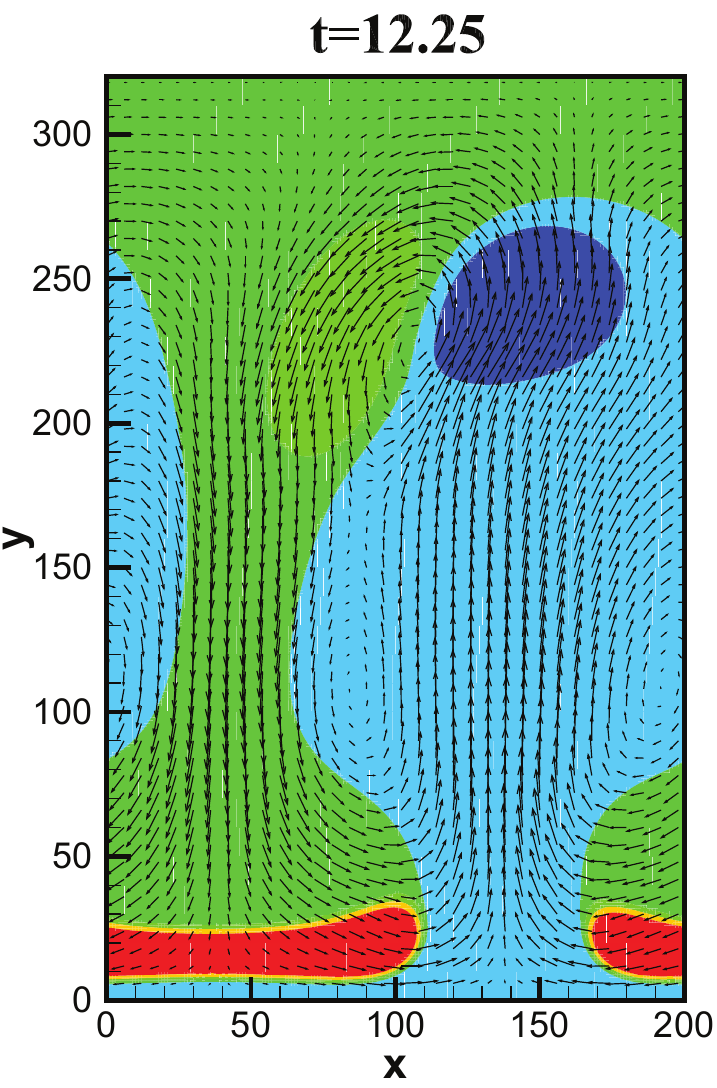}}
\subfigure[]{ \label{fig3_4_4}
\includegraphics[width=1.1in]{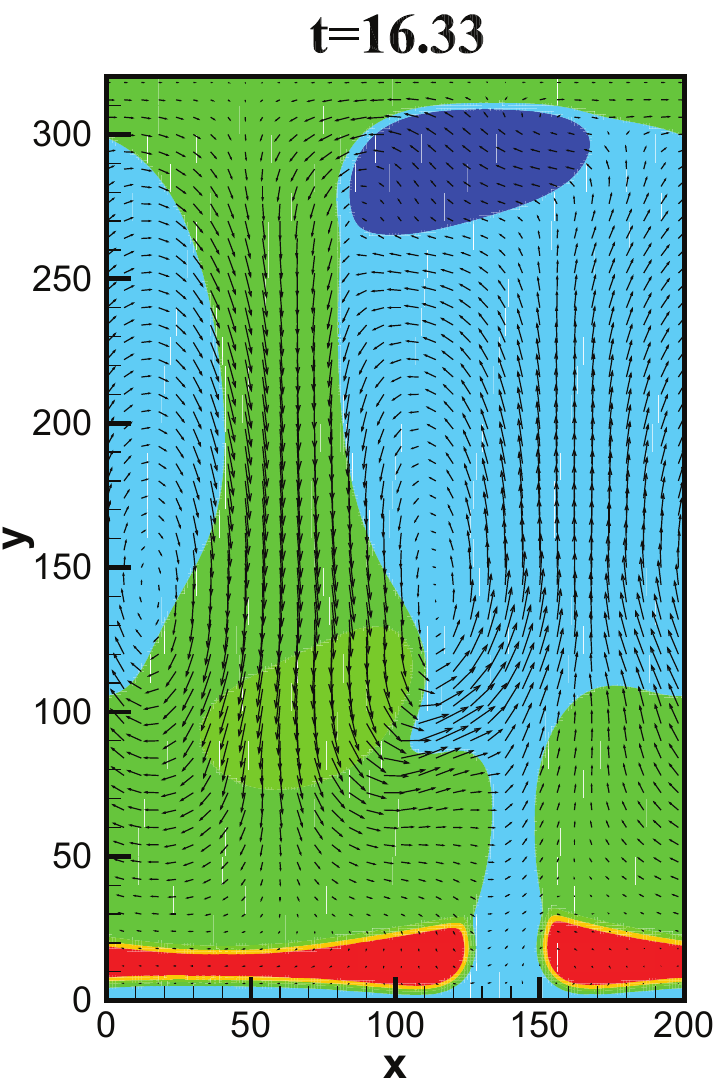}}
\caption{(Color online) Temporal sequence of snapshots of velocity fields for five fluid phase with $\rho_1 : \rho_2  : \rho_3: \rho_4 :\rho_5=6:4:1:3:2$, where the time has been non-dimensionalized through $t=\frac{t_l}{\sqrt{\frac{D}{g}}}$.}
\label{fig3_5}
\end{figure}

We next investigate the effect of density ratio. In the present case, the density ratio is set to be $\rho_1 : \rho_2  : \rho_3: \rho_4 :\rho_5=6:4:1:3:2$. As illustrated in Fig. \ref{fig3_4}, the current droplets and bubble (phases $1$, $2$ and $3$) move more slowly in contrast to the previous case, because the droplets and bubble are subject to smaller gravity (buoyancy) [Figs. \ref{fig3_2a1}-\ref{fig3_2c1}]. Another significant difference is that phase $2$ and phase $3$ bypass each other in the opposite direction to the previous example [Figs. \ref{fig3_2f1}-\ref{fig3_2h1}]. This phenomenon can be observed and explained more intuitively through the velocity field at time $t=12.25$ [Fig. \ref{fig3_4_3}]. At this moment, we find that the vortex between phase $2$ and phase $3$ is counterclockwise in the current situation, which is the opposite of the vortex direction in the previous example.  Because the density of the upper fluid (phase $4$) is larger than that of the lower one (phase $5$), the interface between phase $4$ and phase $5$ undergoes a more distinct deformation [Figs. \ref{fig3_2g1}-\ref{fig3_2i1}]. Finally, phase $1$ and phase $2$ fall to the bottom and phase $3$ rise to the top of the calculation domain.

\section{Conclusions}
In this study, a LB model based on the multicomponent phase field theory is proposed for $N$-phase flow systems ($N \geq 2$). In the proposed model, we reformulate the governing equations of the $N$-phase system proposed by Dong \cite{Dong2014}, and the new governing equations are also thermodynamically consistent. We point out that the velocity in this model is the volume-averaged mixture velocity and is divergence free. Besides, the mixing energy density coefficients in the free energy involve the interaction between different phases and can be determined by the pairwise surface tensions among the N fluids. In the present LB model, $(N-1)$ LB equations are employed to capture the interface, and another LB equation is used to solve the N-S equations. To reduce the calculation of the gradient term, a distribution function for the force term is carefully designed in the LB equation for flow field , and the governing equations can be recovered correctly through Chapmann-Enskog analysis.

The proposed model is first validated by two classical benchmark problems, including tests of static droplets and the spreading of a liquid lens, the simulation results show that the current LB model is accurate and efficient for simulating incompressible $N$-phase fluid flows. To further demonstrate the capability and performance of the LB model, we use two numerical simulations, including dynamics of droplet collision for four fluid phases and dynamics of droplets and interfaces for five fluid phases,
  to test the proposed model. The results show that the present model can be used to handle complex interactions among multiple types of fluid interfaces.


\section*{Acknowledgements}
This work is financially supported by the National Natural Science Foundation of China (Grant Nos. 51576079, 51836003), and the National Key Research and Development Program of China (Grant No. 2017YFE0100100).

\appendix

\section{\label{app:sec1}Chapman-Enskog analysis of the present model}
 In this appendix, the Chapman-Enskog (C-E) expansion is performed to demonstrate the consistency of the present LB model with hydrodynamic equations [Eqs. (\ref{eq:N19})].

 Before performing C-E expansion, the moment conditions are first given based on the expressions of the equilibrium and force distribution functions,
 \begin{subequations}
 \begin{equation}
 \begin{split}
 &\sum_k g_k^{eq}=\rho_0,\quad \sum_k \mathbf{c}_{k \alpha} g_k^{eq}=\rho u_{\alpha}, \\
 &\sum_k \mathbf{c}_{k\alpha} \mathbf{c}_{k\beta}
 g_k^{eq}=p \delta_{\alpha \beta}+\rho u_{\alpha} u_{\beta},\\
 &\sum_k \mathbf{c}_{k\alpha} \mathbf{c}_{k\beta} \mathbf{c}_{k\gamma} g_k^{eq}=\rho c_s^2 \Delta_{\alpha \beta \gamma \theta} u_{\theta}.
 \end{split}
 \label{eq:a1a}
 \end{equation}
 \begin{equation}
 \begin{split}
 &\sum_k G_k=u_{\alpha}\partial_{\alpha}\rho,\quad \sum_k \mathbf{c}_{k \alpha} G_k=F_{\alpha},\\
 &\sum_k \mathbf{c}_{k \alpha} \mathbf{c}_{k \beta} G_k=u_{\alpha} F_{\beta}+u_{\beta}F_{\alpha}+c_s^2 u_{\alpha} \partial_{\beta} \rho+c_s^2 u_{\beta} \partial_{\alpha} \rho+(c_s^2 u_{\gamma} \partial_{\gamma} \rho) \delta_{\alpha \beta}+\frac{\tilde{J}_{\alpha} u_{\beta}}{\Delta t (\tau_g-0.5)},
 \end{split}
 \label{eq:a1b}
 \end{equation}
 \label{eq:a1}
 \end{subequations}
where $\Delta_{\alpha \beta \gamma \theta}$ is given by $\Delta_{\alpha \beta \gamma \theta}=\delta_{\alpha \beta} \delta_{\gamma \theta}+\delta_{\alpha \gamma} \delta_{\beta \theta}+\delta_{\beta \gamma}\delta_{\alpha \theta}$, and Greek indices denote Cartesian spatial components.

In the C-E analysis, the time and space derivatives, as well as the force term can be expanded as
\begin{subequations}
\begin{equation}
g_k=g_k^{(0)}+\epsilon g_k^{(1)}+\epsilon^{2}g_k^{(2)}+ \cdots,
\label{eq:a2}
\end{equation}
\begin{equation}
G_k=\epsilon G_{k}^{(1)}+\epsilon^2 G_{k}^{(2)}, \label{eq:a3}
\end{equation}
\begin{equation}
\partial_t=\epsilon \partial_{t_1}+\epsilon^2 \partial_{t_2},
\partial_{\alpha}=\epsilon \partial_{1\alpha},
\label{eq:a4}
\end{equation}
\begin{equation}
F_{\alpha}=\epsilon F_{\alpha}^{(1)}+\epsilon^2 F_{\alpha}^{(2)},
\label{eq:a5}
\end{equation}
\label{eq:A2}
\end{subequations}
 where $\epsilon$ denotes a small expansion parameter. Using the Taylor
expansion to Eq. (\ref{eq:24}), one have
\begin{equation}
\Delta t D_k g_k(\textbf{x},t)+\frac{\Delta t^2}{2} D_k^2
g_k(\textbf{x},t) + \cdots=-\frac{1}{\tau_g}
\left(g_k(\mathbf{x},t)-g_k^{eq}(\mathbf{x},t)\right)+\Delta
t(1-\frac{1}{2\tau_g})G_k, \label{eq:A2A}
\end{equation}
where $D_{k}=\partial_{t}+c_{k\alpha} \partial_{\alpha}$, and
substituting Eq. (\ref{eq:A2}) into Eq. (\ref{eq:A2A}), the following multi-scale equations can be obtained,
\begin{subequations}
\begin{equation}
O(\epsilon^0):g_k^{(0)}=g_k^{eq}, \label{eq:a6}
\end{equation}
\begin{equation}
O(\epsilon^1):D_{1k}g_k^{(0)}=-\frac{1}{\tau_g \Delta
t}g_k^{(1)}+(1-\frac{1}{2\tau_g})G_k^{(1)}, \label{eq:a7}
\end{equation}

\begin{equation}
O(\epsilon^2):\partial_{t_2}g_k^{(0)}+D_{1k}g_k^{(1)}+\frac{\Delta
t}{2}D_{1k}^2 g_k^{(0)}=-\frac{1}{\tau_g \Delta t}g_k
^{(2)}+(1-\frac{1}{2\tau_g})G_k^{(2)}, \label{eq:a8}
\end{equation}
\label{eq:A3}
\end{subequations}
where $D_{1k}=\partial_{t_1}+c_{k\alpha} \partial_{1\alpha}$.

Then, the substitution of Eq. (\ref{eq:a7}) into Eq. (\ref{eq:a8})
yields
\begin{equation}
\partial_{t_2}g_k^{(0)}+(1-\frac{1}{2\tau_g})D_{1k}g_k^{(1)}+\frac{\Delta
t}{2}(1-\frac{1}{2\tau_g})D_{1k}G_k^{(1)}=-\frac{1}{\tau_g \Delta
t}g_k ^{(2)}+(1-\frac{1}{2\tau_g})G_k^{(2)}. \label{eq:a9}
\end{equation}
By summing Eq. (\ref{eq:a7}) and Eq. (\ref{eq:a7})$\times
c_{k\beta}$ over $k$, we can obtained the recovered equations at $\epsilon$ scale,
\begin{subequations}
\begin{equation}
\partial_{1\alpha} \rho u_{\alpha}=-\frac{1}{\tau_g \Delta t}
\sum_k g_k^{(1)}+(1-\frac{1}{2\tau_g})\sum_k G_k^{(1)},
\label{eq:a10}
\end{equation}
\begin{equation}
\partial_{t_1} \rho u_{\beta}+\partial_{1\alpha} (p \delta_{\alpha \beta}+\rho u_{\alpha} u_{\beta})=-\frac{1}{\tau_g \Delta t}
\sum_k c_{k\beta} g_k^{(1)}+(1-\frac{1}{2\tau_g})\sum_k c_{k\beta}
G_k^{(1)}. \label{eq:a11}
\end{equation}
\label{eq:A4}
\end{subequations}

Similarly, the recovered equations at $\epsilon^2$ scale can also be obtained from Eq. (\ref{eq:a9})
\begin{subequations}
\begin{equation}
\begin{split}
&(1-\frac{1}{2\tau_g})\left[\partial_{t_1} (\sum_k g_k^{(1)})+\partial_{1\alpha}(\sum_k c_{k\alpha}
g_k^{(1)})\right]+\frac{\Delta t}{2}(1-\frac{1}{2\tau_g})\left[\partial_{t_1} (\sum_k G_k^{(1)})+\right.
\\ & \left.\partial_{1\alpha}(\sum_k c_{k\alpha} G_k^{(1)})\right]
=-\frac{1}{\tau_g \Delta t}\sum_k g_k^{(2)}+(1-\frac{1}{2\tau_g})\sum_k G_k^{(2)},
\end{split}
\label{eq:a12}
\end{equation}
\begin{equation}
\begin{split}
&\partial_{t_2} \rho u_{\beta}+(1-\frac{1}{2\tau_g})\left[\partial_{t_1}
(\sum_k c_{k\beta} g_k^{(1)})+\partial_{1\alpha}\Lambda^{(1)}
\right]+\frac{\Delta t}{2}(1-\frac{1}{2\tau_g})\left[\partial_{t_1}
(\sum_k c_{k\beta}
G_k^{(1)})+\right.\\
& \left. \partial_{1\alpha}(\sum_k c_{k\alpha} c_{k\beta}
G_k^{(1)})\right] =-\frac{1}{\tau_g \Delta t}\sum_k c_{k\beta}
g_k^{(2)}+(1-\frac{1}{2\tau_g})\sum_k c_{i\beta} G_k^{(2)},
\end{split}
\label{eq:a13}
\end{equation}
\label{eq:A5}
\end{subequations}
where $\Lambda^{(1)}=\sum_k c_{k\alpha} c_{k\beta} g_k^{(1)}$ is the
first-order momentum flux tensor.

Summing Eq. (\ref{eq:23a}) and Eq. (\ref{eq:23a})$\times
c_{k\alpha}$ over $k$, we can obtain the following relations
\begin{subequations}
\begin{equation}
\sum_k g_k^{eq}=\sum_k g_k+\frac{\Delta t}{2}\sum_k G_k, \label{eq:a14}
\end{equation}
\begin{equation}
\sum_k c_{k\alpha} g_k^{eq}=\sum_k c_{k\alpha} g_k+\frac{\Delta t}{2} \sum_k
c_{k\alpha} G_k, \label{eq:a15}
\end{equation}
\label{eq:A6}
\end{subequations}
which can be further recast as
\begin{subequations}
\begin{equation}
\sum_k g_k^{(1)} =-\frac{\Delta t}{2} \sum_k G_k^{(1)},\sum_k
g_k^{(2)} =-\frac{\Delta t}{2} \sum_k G_k^{(2)}, \label{eq:a16}
\end{equation}
\begin{equation}
\sum_k c_{k\alpha}g_k^{(1)} =-\frac{\Delta t}{2} \sum_k c_{k\alpha}
G_k^{(1)},\sum_k c_{k\alpha}g_k^{(2)} =-\frac{\Delta t}{2} \sum_k
c_{k\alpha}G_k^{(2)}. \label{eq:a17}
\end{equation}
\label{eq:A7}
\end{subequations}
Substituting Eq. (\ref{eq:A7}) into Eqs. (\ref{eq:A4}) and (\ref{eq:A5}), one have
\begin{subequations}
\begin{equation}
\partial_{1\alpha}\rho u_{\alpha}=\sum_k G_k^{(1)},
\label{eq:a18}
\end{equation}
\begin{equation}
\partial_{t_1} \rho u_{\beta}+\partial_{1\alpha} (p \delta_{\alpha \beta}+\rho u_{\alpha} u_{\beta})=\sum_k c_{k\beta}
G_k^{(1)}, \label{eq:a19}
\end{equation}
\label{eq:A8}
\end{subequations}

\begin{subequations}
\begin{equation}
0=\sum_i G_i^{(2)},
\label{eq:a20}
\end{equation}

\begin{equation}
\partial_{t_2}\rho u_{\beta}+(1-\frac{1}{2\tau_g})\partial_{1\alpha}\Lambda^{(1)} +\frac{\Delta
t}{2}(1-\frac{1}{2\tau_g})\partial_{1\alpha}(\sum_k c_{k\alpha}
c_{k\beta} G_k^{(1)}) =\sum_k c_{k\beta} G_k^{(2)}. \label{eq:a21}
\end{equation}
\label{eq:A9}
\end{subequations}
Combining Eq. (\ref{eq:a18}) and Eq. (\ref{eq:a20}) at $\epsilon$ and $\epsilon^2$ scales with the help of Eqs. (\ref{eq:a1}) yields
\begin{equation}
\partial_{\alpha}u_{\alpha}=0.
\label{eq:a22}
\end{equation}

To derive the equation at $\epsilon^2$ scale, $\Lambda^{(1)}$ should be expressed as
\begin{equation}
\begin{split}
\Lambda^{(1)}&=-\tau_g \Delta t \left[ \partial_{t_1}\sum_k c_{k \alpha} c_{k \beta}g_k^{(0)}+\partial_{1\gamma}(\sum_k c_{k \alpha} c_{k \beta}c_{k \gamma}g_k^{(0)})-(1-\frac{1}{2\tau_g})\sum_k c_{k\alpha} c_{k\beta} G_k^{(1)} \right] \\
&=-\tau_g \Delta t \left \{\partial_{t_1} p\delta_{\alpha\beta}+\partial_{t_1}(\rho u_{\alpha}u_{\beta})+c_s^2\partial_{1\gamma}( \rho u_{\gamma}\delta_{\alpha\beta})+c_s^2 \partial_{1\gamma}[\rho(u_{\alpha \delta_{\beta \gamma}}+u_{\beta}\delta_{\alpha
\gamma})] \right. \\&\left. \quad -(1-\frac{1}{2\tau_g})\sum_k c_{k\alpha}c_{k\beta}G_k^{(1)}\right \} \\
\end{split}
\label{eq:a24}
\end{equation}
where Eq. (\ref{eq:a7}) has been used, and the term
$\partial_{t_1}(\rho u_{\alpha} u_{\beta})$ can be written as
\begin{equation}
\partial_{t_1}(\rho u_{\alpha}
u_{\beta})=u_{\alpha}F_{\beta}^{(1)}+u_{\beta}F_{\alpha}^{(1)}-(u_{\alpha}
\partial_{1\beta}p+u_{\beta}
\partial_{1\alpha}p),
\label{eq:a25}
\end{equation}
where the Eqs. (\ref{eq:a18}) and Eq. (\ref{eq:a19}) are used, and the term of $O(Ma^3)$ has been neglected.

Combining Eq. (\ref{eq:a19}) with Eq. (\ref{eq:a21}) at $\epsilon$
and $\epsilon^2$ scales, together  with Eq. (\ref{eq:a24}), Eq.
(\ref{eq:a25}) and Eq. (\ref{eq:a1b}), we have
\begin{equation}
\begin{split}
\partial_t (\rho u_{\beta})+\partial_{\alpha}(\rho
u_{\alpha}u_{\beta})=&-\partial_{\beta}p+\partial_{\alpha}\left[
\rho
\nu(\partial_{\alpha}u_{\beta}+\partial_{\beta}u_{\alpha})\right]+\epsilon
\Delta t (\tau_g-0.5)\partial_{\alpha} \left [ \partial_{t_1}p
\delta_{\alpha \beta} \right.\\ & \left.-(u_{\alpha}
\partial_{1\beta}p+ u_{\beta}  \partial_{1\alpha}p)\right ]-\partial_{\alpha}(\tilde{J}_{\alpha} u_{\beta}) +F_{\beta},
\end{split}
\label{eq:a26}
\end{equation}
with the relation of
\begin{equation}
\nu=c_s^2(\tau_g-0.5)\Delta t.
\end{equation}

In the limit of a low Mach number, the dynamic pressure is assumed to be $\delta p\sim O(Ma^2)$. As a result, Eq. ({\ref{eq:a26}}) will reduce to
\begin{equation}
\partial_t (\rho u_{\beta})+\partial_{\alpha}(\rho
u_{\alpha}u_{\beta})=-\partial_{\beta}p+\partial_{\alpha}\left[\rho \nu(\partial_{\alpha}u_{\beta}+\partial_{\beta}u_{\alpha})\right]-\partial_{\alpha}(\tilde{J}_{\alpha} u_{\beta}) +F_{\beta},
\label{eq:a27}
\end{equation}
Thus, Eqs. ({\ref{eq:a22}}) and ({\ref{eq:a27}}) clearly shows that the incompressible N-S equations can be exactly recovered from the present LB model.


\end{document}